\documentclass[aip,graphicx,amsmath,amssymb,preprintnumbers]{revtex4-1}
\usepackage{graphicx}
\usepackage{subcaption}
\usepackage[final]{changes}
\draft % marks overfull lines with a black rule on the right

\begin{document}

%\preprint{SAND2020-2836 O}

\title{QMCPACK: Advances in the development, efficiency, and application of auxiliary field and real-space variational and diffusion Quantum Monte Carlo} %Title of paper

\author{P. R. C. Kent}
\email[]{kentpr@ornl.gov}
\affiliation{Center for Nanophase Materials Sciences Division and Computational Sciences and Engineering Division, Oak Ridge National Laboratory, Oak Ridge, Tennessee 37831, USA\footnote{{Notice: This manuscript has been authored by UT-Battelle, LLC, under Contract No. DE-AC0500OR22725 with the U.S. Department of Energy. The United States Government retains and the publisher, by accepting the article for publication, acknowledges that the United States Government retains a non-exclusive, paid-up, irrevocable, world-wide license to publish or reproduce the published form of this manuscript, or allow others to do so, for the United States Government purposes. The Department of Energy will provide public access to these results of federally sponsored research in accordance with the DOE Public Access Plan (\url{http://energy.gov/downloads/doe-public-access-plan})}}}

\author{Abdulgani Annaberdiyev}
\affiliation{Department of Physics, North Carolina State University, Raleigh, North Carolina 27695-8202, USA}
%\email{aannabe@ncsu.edu}

\author{Anouar Benali}
\affiliation{Computational Science Division, Argonne National Laboratory, 9700 S. Cass Avenue, Lemont, IL 60439, USA}
%\email{benali@anl.gov}

\author{M. Chandler Bennett}
\affiliation{Materials Science and Technology Division, Oak Ridge National Laboratory, Oak Ridge, Tennessee, 37831, USA}
%\email{bennettcc@ornl.gov}

\author{Edgar Josu{\'e} Landinez Borda}
\affiliation{Quantum Simulations Group, Lawrence Livermore National Laboratory, 7000 East Avenue, Livermore, CA 94551, USA}
%\email{landinezbord1@llnl.gov}

\author{Peter Doak}
\affiliation{Center for Nanophase Materials Sciences Division and Computational Sciences and Engineering Division, Oak Ridge National Laboratory, Oak Ridge, Tennessee 37831, USA}
%\email{doakpw@ornl.gov}

\author{Hongxia Hao}
\affiliation{Department of Chemistry, University of California, Berkeley, California 94720, USA}
%\email{hh8@berkeley.edu}

\author{Kenneth D. Jordan}
\affiliation{Department of Chemistry, University of Pittsburgh, Pennsylvania 15260, USA}
%\email{jordan@pitt.edu}

\author{Jaron T. Krogel}
\affiliation{Materials Science and Technology Division, Oak Ridge National Laboratory, Oak Ridge, Tennessee 37831, USA}
%\email{krogeljt@ornl.gov}

\author{Ilkka Kyl\"anp\"a\"a}
\affiliation{Computational Physics Laboratory, Tampere University, P.O. Box 692, 33014 Tampere, Finland}
%\email{ilkka.kylanpaa@tuni.fi}

\author{Joonho Lee}
\affiliation{Department of Chemistry, Columbia University, New York, NY 10027, USA}
%\email{linusjoonho@gmail.com}

\author{Ye Luo}
\affiliation{Computational Science Division, Argonne National Laboratory, 9700 S. Cass Avenue, Lemont, IL 60439, USA}
%\email{yeluo@anl.gov}

\author{Fionn D. Malone}
\affiliation{Quantum Simulations Group, Lawrence Livermore National Laboratory, 7000 East Avenue, Livermore, CA 94551, USA}
%\email{malone14@llnl.gov} 

\author{Cody A. Melton}
\affiliation{Sandia National Laboratories, Albuquerque, New Mexico 87123, USA}
%\email{cmelton@sandia.gov}

\author{Lubos Mitas}
\affiliation{Department of Physics, North Carolina State University, Raleigh, North Carolina 27695-8202, USA}
%\email{lmitas@ncsu.edu}

\author{Miguel A. Morales}
\affiliation{Quantum Simulations Group, Lawrence Livermore National Laboratory, 7000 East Avenue, Livermore, CA 94551, USA}
%\email{moralessilva2@llnl.gov}

\author{Eric Neuscamman}
\affiliation{Department of Chemistry, University of California, Berkeley, California 94720, USA}
\affiliation{Chemical Sciences Division, Lawrence Berkeley National Laboratory, Berkeley, CA, 94720, USA}
%\email{eneuscamman@berkeley.edu}

\author{Fernando A. Reboredo}
\affiliation{Materials Science and Technology Division, Oak Ridge National Laboratory, Oak Ridge, Tennessee 37831, USA}
%\email{reboredofa@ornl.gov}

\author{Brenda Rubenstein}
\affiliation{Department of Chemistry, Brown University, Providence, RI 02912, USA}
%\email{brenda\_rubenstein@brown.edu}

\author{Kayahan Saritas}
\affiliation{Department of Applied Physics, Yale University, New Haven, CT 06520, USA}
%\email{kayahan.saritas@yale.edu}

\author{Shiv Upadhyay}
\affiliation{Department of Chemistry, University of Pittsburgh, Pennsylvania 15260, USA}
%\email{shu8@pitt.edu}

\author{Guangming Wang}
\affiliation{Department of Physics, North Carolina State University, Raleigh, North Carolina 27695-8202, USA}
%\email{gwang18@ncsu.edu}

\author{Shuai Zhang}
\affiliation{Laboratory for Laser Energetics, University of Rochester, 250 E River Rd, Rochester, NY 14623, USA}
%\email{szha@lle.rochester.edu}

\author{Luning Zhao}
\affiliation{Department of Chemistry, University of Washington, Seattle, Washington, 98195, USA}
%\email{zhaoln@uw.edu}

\date{\today}

\begin{abstract}
We review recent advances in the capabilities of the open source ab initio Quantum Monte Carlo (QMC) package QMCPACK and the workflow tool Nexus used for greater efficiency and reproducibility.
The auxiliary field QMC (AFQMC) implementation has been greatly expanded to include k-point symmetries,
tensor-hypercontraction, and accelerated graphical processing unit (GPU) support. These scaling and memory reductions greatly
increase the number of orbitals that can practically be included in AFQMC calculations, increasing accuracy. Advances in real
space methods include techniques for accurate computation of band gaps and for systematically improving the nodal surface of
ground state wavefunctions. Results of these calculations can be used to validate application of more approximate electronic
structure methods including GW and density functional based techniques. To provide an improved foundation for these calculations
we utilize a new set of correlation-consistent effective core potentials (pseudopotentials) that are more accurate than
previous sets; these can also be applied in quantum-chemical and other many-body applications, not only QMC. These advances
increase the efficiency, accuracy, and range of properties that can be studied in both molecules and materials with QMC and
QMCPACK.
\end{abstract}


\maketitle %\maketitle must follow title, authors, abstract and \pacs

\section{Introduction}\label{sec:introduction}
Quantum Monte Carlo (QMC) methods are an attractive approach for accurately computing and analyzing solutions of the
Schr\"{o}dinger equation.~\cite{BeccaQuantum2017,MartinInteracting2016,Foulkes2001} The methods form a general ab initio
methodology able to solve the quantum many-body problem, applicable to idealized models such as chains or lattices of atoms
through to complex and low-symmetry molecular and condensed matter systems, whether finite or periodic, metallic or insulating,
and with weak to strong electronic correlations. Significantly, the methods can naturally treat systems with significant
multi-reference character, and are without electron self-interaction error, which challenges many quantum chemical approaches and
density functional theory approximations, respectively. The methods continue to be able to take advantage of improvements in
computational power, giving reduced time to solution with new generations of computing. Due to these features, usage of QMC
methods for first principles and ab initio calculations is growing.

Compared to traditional deterministic approaches, QMC methods are generally distinguished by: (1) use of statistical methodologies
to solve the Schr\"{o}dinger equation. This allows the methods to not only treat problems of high-dimensionality efficiently, but
also potentially use basis, wave function, and integral forms that are not amenable to numerical integration. (2) Use of few and
well-identified approximations that can potentially be quantified or made systematically convergeable. (3) A low power scaling
with system-size, but large computational cost prefactor. (4) High suitability to large scale parallel computing owing to lower
communications requirements than conventional electronic structure methods. Scaling has been demonstrated to millions of compute
cores.\cite{KimQMCPACKJCP2018}  

Modern applications of QMC have expanded to cover many of the same systems studied by density functional theory (DFT) and quantum
chemical approaches, and in many cases also at a similar atom and electron count, although at far greater computational cost.
Besides those described in below, recent molecular applications of QMC include studies of the nature of the quadruple bond in
C$_2$\cite{Genovesenature2019}, acenes\cite{DupuyFate2018}, physisorption of water on graphene\cite{BrandenburgPhysisorption2019},
binding of transition metal diatomics\cite{SheeAchieving2019}, and DNA stacking energies\cite{QinInconsistencies2020}. Materials
applications include nitrogen defects in ZnO\cite{YuFixed2017}, excitations in Mn doped phosphors\cite{SaritasExcitation2018}, and
the singlet-triplet excitation in MgTi$_2$O$_4$\cite{BusemeyerPrediction2019}. Methodological improvements include: reducing the
sensitivity of pseudopotential evaluation\cite{Zennew2019}, extensions to include linear response\cite{MussardTime2018}, density
functional embedding\cite{Doblhoff-DierDensity2018}, excited states including geometry optimization\cite{DashExcited2019},
improved twist averaging\cite{Azadiefficient2019}, and accurate trial wavefunctions via accurate densities\cite{PerDensity2019}.
Importantly, for model systems such as the hydrogen chain, the methods can be used to benchmark themselves as well as other
many-body approaches\cite{MottaTowards2017}. This partial list of developments and applications from the last two years
alone indicates that the field is growing and maturing.

In this article we describe recent updates to the QMCPACK code and its ecosystem of wavefunction converters and workflow tools.
These updates have aimed to expand the range of systems, properties, and accuracies that can be achieved both with QMCPACK and
with QMC techniques in general. For a description of the underlying methodology we refer the reader to
Refs.~\onlinecite{KimQMCPACKJCP2018,Motta2019,BeccaQuantum2017,MartinInteracting2016,Foulkes2001}. In particular, a thorough description of real
space QMC methods is given in Section 5 of Ref.~\onlinecite{KimQMCPACKJCP2018}. For an extensive introduction to AFQMC we refer the
reader to Ref.\onlinecite{Motta2019}.

QMCPACK is a fully open source and openly
developed QMC package, with 48 coauthors on the primary citation paper~\cite{KimQMCPACKJCP2018} published in 2018 and an
additional 5 contributors since then. The main website for QMCPACK is \url{https://qmcpack.org} and the source code is currently
available through \url{https://github.com/QMCPACK/qmcpack}. QMCPACK aims to implement state of the art QMC methods, be generally
applicable, easy to use, and high-performing on all modern computers.  Since the publication of Ref.~\cite{KimQMCPACKJCP2018}, the
range of QMC calculations that are possible has been expanded by significant enhancements to the Auxiliary-Field QMC (AFQMC)
solver.  This orbitally based method is distinct from and complementary to the longer-implemented real space methods of
variational and diffusion QMC (VMC and DMC, respectively). The AFQMC implementation can fully take advantage of graphics
processing units (GPUs) for a considerable speedup and, unlike the real-space methods, can also exploit k-point symmetries. It
shares the same workflow tool, Nexus, which helps simplify and ease application of all the QMC methods by new users as well as aid
in improving reproducibility of complex multi-step research investigations. To our knowledge this is currently the only AFQMC code
designed for large scale research calculations that is open source. To help guarantee the future of the code, it is undergoing
rapid development and refactoring to target the upcoming Exascale architectures as part of the U.S. Exascale Computing
Project\cite{ExascaleSkin2020}, which also entails major updates to the testing, validation and maintainability.

The electronic structure and quantum chemical codes that QMCPACK is interfaced to for trial wavefunctions has been expanded to
include Qbox\cite{GygiArchitecture2008}, PySCF\cite{PYSCF}, Quantum Espresso\cite{Gianozzi2009}, Quantum Package\cite{QP2019}, and
GAMESS\cite{SchmidtGeneral1993}. Additional codes such as NWCHEM\cite{ValievNWChem2010} can be interfaced straightforwardly.

In the following, we first review in Section~\ref{sec:opendev} the open development principles of QMCPACK. In
Section~\ref{sec:workflow} we discuss updates to the Nexus workflow package. This integrates entire research electronic structure
workflows for greater productivity and reproducibility than by-hand invocation of individual calculations.  Due to the
infeasibility of performing QMC calculations for general systems using an all electron approach, use of effective core potentials
(ECPs), or pseudopotentials, is essential. To improve the accuracy obtainable we have developed a new approach and set of
``correlation consistent'' ECPs. These can be used in all ab initio calculations, not only QMC, and are described in
Section~\ref{sec:ecps}. Advances in the AFQMC implementation are described in Section~\ref{sec:afqmc}. Turning to real-space QMC
methods in Section~\ref{sec:gsnodes}, algorithms and multiple determinant trial wavefunctions can now be used to obtain improved
ground state energies as well as band gaps in solid-state materials. As a result, it is now possible to begin to test the accuracy
of the nodal surfaces that have long been used in these calculations. Finally, in Section~\ref{sec:applications}, we give three
applications: first, application to non-valence anions, which challenge all electronic structure and quantum chemical techniques.
Second, application to excitations of localized defects in solids~\ref{sec:excited_defect}. Third, the ability to obtain the
momentum distribution has recently been improved, motivated by recent experiments on VO$_2$. A summary is given in
Section~\ref{sec:summary}.

\section{Open Development and Testing}\label{sec:opendev}
Fully open source development is an important core value of the QMCPACK development team. Besides improving the quality of the
software, anecdotally it also improves the on-boarding experience for new users. While the developers of many electronic structure
packages now practice some degree of open development, QMCPACK has seen very significant benefits from this in the last few years.
We expect other packages would also benefit from full adoption and therefore give details here.

QMCPACK is an open source package, with releases and the latest development source code available through
\url{https://github.com/QMCPACK/qmcpack}. QMCPACK is written in C++14, with MPI parallelization between compute nodes, OpenMP
threading used for multicore parallelism. CUDA is used for NVIDIA accelerators. Options to support CUDA, complex valued
wavefunctions, and to adjust the numerical precision used internally are currently compile time options. 

Besides adoption of a distributed source code control system, we have found that
development productivity can be further increased by adoption of code reviews and continuous integration (testing). To maximize
the efficiency of both contributors and reviewers and shorten the development cycle of new features, work-in-progress pull
requests are encouraged for early engagement in the process. The early review allows guidance to be given, e.g.\
are the
algorithms clear enough to other developers and are the coding guidelines being followed. At the same time, continuous integration
is applied to the proposed code change. This process routinely catches cases that developers may not have considered or tested against, e.g.\ the
complex-valued build of QMCPACK or accelerated GPU support, that are compile time options. This period of comment while the work is being completed also helps
advertise the work to other developers and minimizes risk of duplicated work. Our experience strongly suggests that this process reduces bugs, reduces potential
developer's effort, and saves reviewer's time compared to a late engagement with an unexpected pull request. All the discussions
around the code change become archived searchable documentation and potential learning materials.

Testing of QMCPACK has been significantly expanded. Two years ago, QMCPACK had limited unit, integration and performance testing
categories: unit tests that run quickly on individual components; integration tests that exercise entire runs; performance tests
for monitoring relative performance between code changes. However, due to the stochastic nature of QMC, as the number of tests and
build combinations increased it became impractical to run the integration tests long enough to obtain a statistically reliable
pass/fail: the smallest (shortest) integration test set currently takes around one hour to execute on a 16 core machine, and must
necessarily suffer from occasional statistical failures. Thus, a new category of tests was needed for quickly examining full QMC
execution with a reproducible Monte Carlo trajectory. The new deterministic integration tests are modified QMC runs with only a
few steps, very few Monte Carlo walkers, and fixed random seeds for absolute reproducibility. All the major features of QMCPACK
are covered by this new of category tests. Running all the unit and deterministic integration tests takes approximately one minute
which is fast enough for iterative development and fast enough to be used in continuous integration. This fast to run set of tests
facilitates significant changes and refactoring of the application which otherwise would be far more difficult to test and
unlikely to be attempted by non-experts without long experience with the codebase. All
the deterministic tests are accompanied by longer running statistical tests that can be used to verify a new implementation when
changes alter a previous deterministic result. Combinations of these tests are run automatically on a
nightly basis and report to a public dashboard \url{https://cdash.qmcpack.org}. At the time of writing, around 25 different
machine and build combinations are used to run around 1000 labeled tests each, and most of these cover multiple features.

Improving source code readability is critical for both new and experienced developers. In the past, misleading variable or class
names and confusing function names have confused developers and resulted in subtle bugs, e.g.\ due to similarly named functions, only one of which updates internal state in the Monte Carlo algorithm. For this reason, coding standards including naming conventions have been added in the manual and are enforced on
newly contributed codes. Existing codes are updated to follow the standards as they need other modifications. Automatic source
formatting is also applied with the help of the clang-format tool. Concomitantly, both developer sections in the manual and source
code documentation are significantly expanded.

As a result of the above changes, new contributors with a basic theoretical background can connect source code with textbook
equations with much less difficulty than in the past. These efforts are clearly bringing long term benefit to QMCPACK and
hopefully can be transferred to other scientific applications as well.

\section{Improving QMC Workflows with Nexus}\label{sec:workflow}

QMC techniques are progressing from methods under research towards more routine application.  In this transition,
usability of QMC becomes an important factor. A mature, usable computational method transfers
responsibility for correct execution from users to the code.  Major factors determining overall usability include: ease of
requesting a desired result (in the form of input), robustness of the code in obtaining the desired result, and complexity of the
overall calculation process.  All of these contribute to the effort required by the user to obtain desired
results. In essence, higher required effort translates directly into lower productivity of the user base. Lower productivity
in turn risks a lower overall adoption rate and thus blunts overall impact of the method.  It is therefore important to seek to understand and
minimize barriers to the practical use of QMC.

To illustrate the complexity of the QMC calculation process, we describe below a basic but realistic
sequence of calculations (a scientific workflow) that is required to obtain a final fixed node DMC total energy per formula unit
for a single crystalline solid with QMCPACK.  In this workflow, we suppose that self-consistent (SCF) and
non-self-consistent (NSCF) calculations are performed with Quantum Espresso\cite{Gianozzi2009} and wavefunction optimization
(OPT), variational Monte Carlo (VMC), and diffusion Monte Carlo calculations are performed with QMCPACK.  SCF/NSCF calculations
might be performed on a workstation or a few nodes of a cluster, VMC/OPT calculations on a research-group sized cluster ($\sim$30
nodes), and DMC on high performance computing resources ($\sim$1000 nodes).

\begin{enumerate}
\item{Converge DFT orbitals with respect to plane-wave energy cutoff (4--6 SCF calculations ranging from 300 to 800 Ry for the energy cutoff).}

\item{Converge B-spline orbital representation with respect to B-spline mesh spacing (1 NSCF, $\sim$5 VMC calculations in a small supercell over a series of finer mesh spacings).}

\item{Converge twist grid density ($\sim$5 VMC calculations in a small supercell for a series of increasingly dense supercell Monkhorst-Pack twist grids).}

\item{Determine best optimization process ($\sim$6 optimization (OPT) calculations in a small supercell over varying input parameters and e.g. Jastrow forms).}

\item{Obtain fixed node DMC total energy ($\sim$3 NSCF, $\sim$3 OPT and $\sim$9 DMC calculations, 3 successively smaller timesteps for timestep extrapolation, 3 successively larger supercells for finite size extrapolation).}
\end{enumerate}

This basic workflow process is to be compared with the much reduced complexity for obtaining a single converged total energy for
DFT, which typically requires only a single input file and single program execution to perform a single SCF calculation for the
final energy.  The complexity intrinsic to the basic workflow translates into a large degree of effort on the part of the user and
limits the accessibility of the method for new users or for experienced users pursuing ambitious projects comprised of a large
number of DMC calculations. 

Scientific workflow tools make the QMC process more accessible in multiple ways: (1) bringing the constellation of electronic
structure codes needed to produce a single QMC result under a single framework, (2) reducing the number of inputs required to
request a desired result to a single user-facing input file, (3) reducing overall complexity by abstracting the execution process,
(4) minimizing the direct effort required to execute the workflow process by assuming the management of simulation execution and
monitoring from the user.  Workflow tools have been applied with significant benefit to related electronic structure methods such
as DFT\cite{Ortiz2009,Hachmann2011,Curtarolo2012,Jain2013} and also to QMC\cite{KonkovQMC2019,Saritas2017}.

The Nexus workflow automation system\cite{Krogel2016Nexus} was created to realize these advantages for users of QMCPACK.  Nexus is
a Python-based, object oriented workflow system that can be run on a range of target architectures.  Nexus has been used
successfully on simple workstations and laptops, small group or institutional computing clusters, university level high
performance computing centers in the U.S. and internationally, and Leadership Computing Facilities supported by the U.S.
Department of Energy. Nexus has been used in a growing number of QMC studies involving QMCPACK and its uptake by new users is
high.  

Nexus abstracts user's interactions with each target simulation code that are components of a desired simulation workflow.  Access
to each respective code is enabled through single function calls that only require the user to specify a reduced set of important
input parameters.  Each function call resembles a small input block from a standard input file for an electronic structure code.
Taken together, a sequence of these blocks comprises a new meta-input file that represents the data flow and execution pattern of
the underlying simulation codes as a combined workflow.  

Nexus assumes the responsibility of initiating and monitoring the progress of each simulation job in the workflow.  Nexus
generates expanded input files to each code based on the reduced inputs provided by the user.  It also generates job submission
files and monitors job execution progress via a
lightweight polling mechanism.  Apart from direct execution of
each workflow step, Nexus also automates some tasks that previously fell to users.  One example is that Nexus selects the best
wavefunction produced during the non-linear statistical optimization process employed by QMCPACK and automatically passes this
wavefunction to other calculations (such as diffusion Monte Carlo), which require it. 

In the future, additional productivity gains might be realized with Nexus by further abstracting common workflow patterns.  For
example, convergence studies for orbital parameters (k-points, mesh-factors, source DFT functional) often follow similar patterns
which could be encapsulated as simple components for users.  Additionally, more of the responsibility for obtaining desired
results, e.g.\ total energies to a statistically requested tolerance, could be handled by Nexus through algorithms that create and
monitor dynamic workflows.

\section{Effective Core Potentials}\label{sec:ecps}
\subsection{Introduction}
All-electron (AE) QMC calculations become inefficient and eventually infeasible with increasing atomic number $Z$ since the
computational cost grows roughly as~\cite{Ceperley1986,Hammond1987} $Z^6$. Since our primary interest is in valence properties,
pseudopotentials and/or effective core potentials (ECP) are commonly employed to eliminate the atomic cores leading to
valence-only effective Hamiltonians. Unfortunately, the existing tables and ECP generating tools have proved to exhibit somewhat
mixed fidelity to the true all-electron calculations, especially in high accuracy QMC studies. In order to overcome this
limitation, we have proposed and constructed a new generation of valence-only Hamiltonians called correlation consistent ECPs
(ccECP)~\cite{Bennett2017,Bennett2018,Annaberdiyev2018,Wang2019}. The key feature of this new set is the many-body construction of
ccECPs from the outset, in particular: (i) we have emphasized and put upfront the accuracy of many-body valence spectra
(eigenvalues and eigenstates) as a guiding principle in addition to the well-known norm conservation/shape consistency principles;
(ii) we have opted for simplicity, transparency and eventual wide use, in addition to offering several choices of core sizes or
even smoothed-out all-electron nuclear Coulomb potentials; (iii) we have used a set of tests and benchmarks such as molecular
bonds over a range of distances in order to extensively probe for the quality and transferability of the ccECPs; (iv) we have established
reference data sets for the {\em exact/nearly-exact} atomic total energies, kinetic energies, as well as single-reference and
multi-reference fixed-node DMC energies. At present, this covers elements H-Kr with subsequent plans to fill the periodic table. 

\subsection{ccECP atomic and molecular properties.}
The construction of ccECPs builds in electron correlations obtained from the accurate coupled-cluster singles doubles with
perturbative triples (CCSD(T)) method. By doing so, ccECPs achieve
very high accuracy and enjoy spectral properties on the valence subspace that are in close agreement with the scalar relativistic all-electron (AE) Hamiltonian. 
The agreement is often within chemical accuracy over a large range of atomic excitations and ionizations that
often spans hundreds of eV energy windows. Molecular properties such as binding energies in multiple geometries, equilibrium bond
lengths, and vibrational frequencies were also considered in the development, mostly examining oxides, hydrides, and homonuclear
dimers. Especially, compressed bond length properties were given priority as this corresponds to high-pressure applications and
probes for the proper behavior of the valence charge in the core region. These atomic and molecular tests provide a direct and
comprehensive comparison of ccECP and other core approximations such as BFD\cite{BFD_2008}, STU\cite{STU_1987},
eCEPP\cite{TN_2017}, CRENBL\cite{CRENBL_1985}, SBKJC\cite{SBKJC_1984}, UC (uncorrelated, self-consistent, all-electron core), and
ccECP.S (optimization including only atomic spectrum). Here we illustrate some of these results for selected cases.

\begin{figure*}[!htbp]
\centering
\begin{subfigure}{0.5\textwidth}
\includegraphics[width=\textwidth]{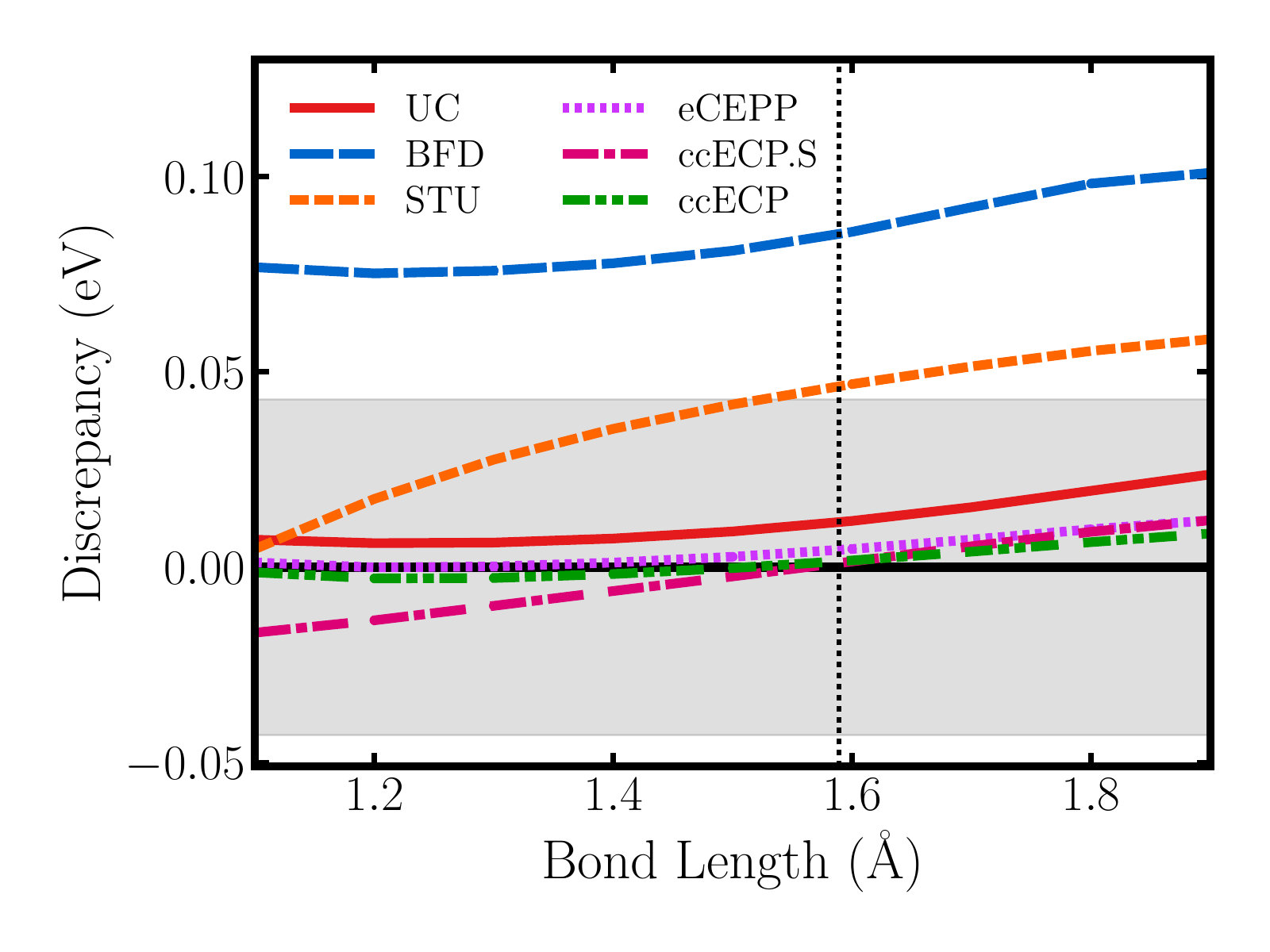}
\caption{FeH binding curve discrepancies.}\label{fig:FeH}
\end{subfigure}%
\begin{subfigure}{0.5\textwidth}
\includegraphics[width=\textwidth]{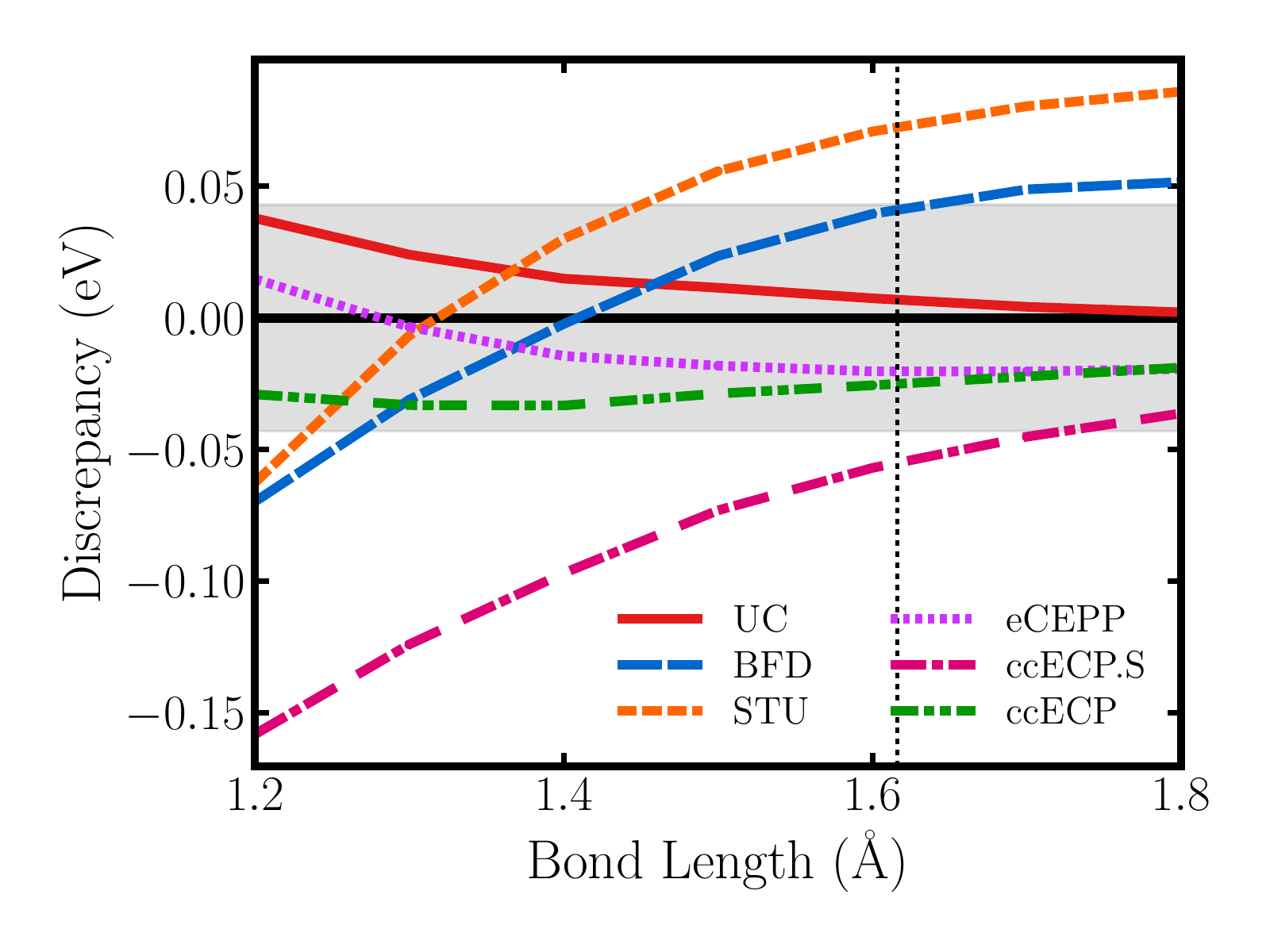}
\caption{FeO binding curve discrepancies}\label{fig:FeO}
\end{subfigure}
\begin{subfigure}{0.5\textwidth}
\includegraphics[width=\textwidth]{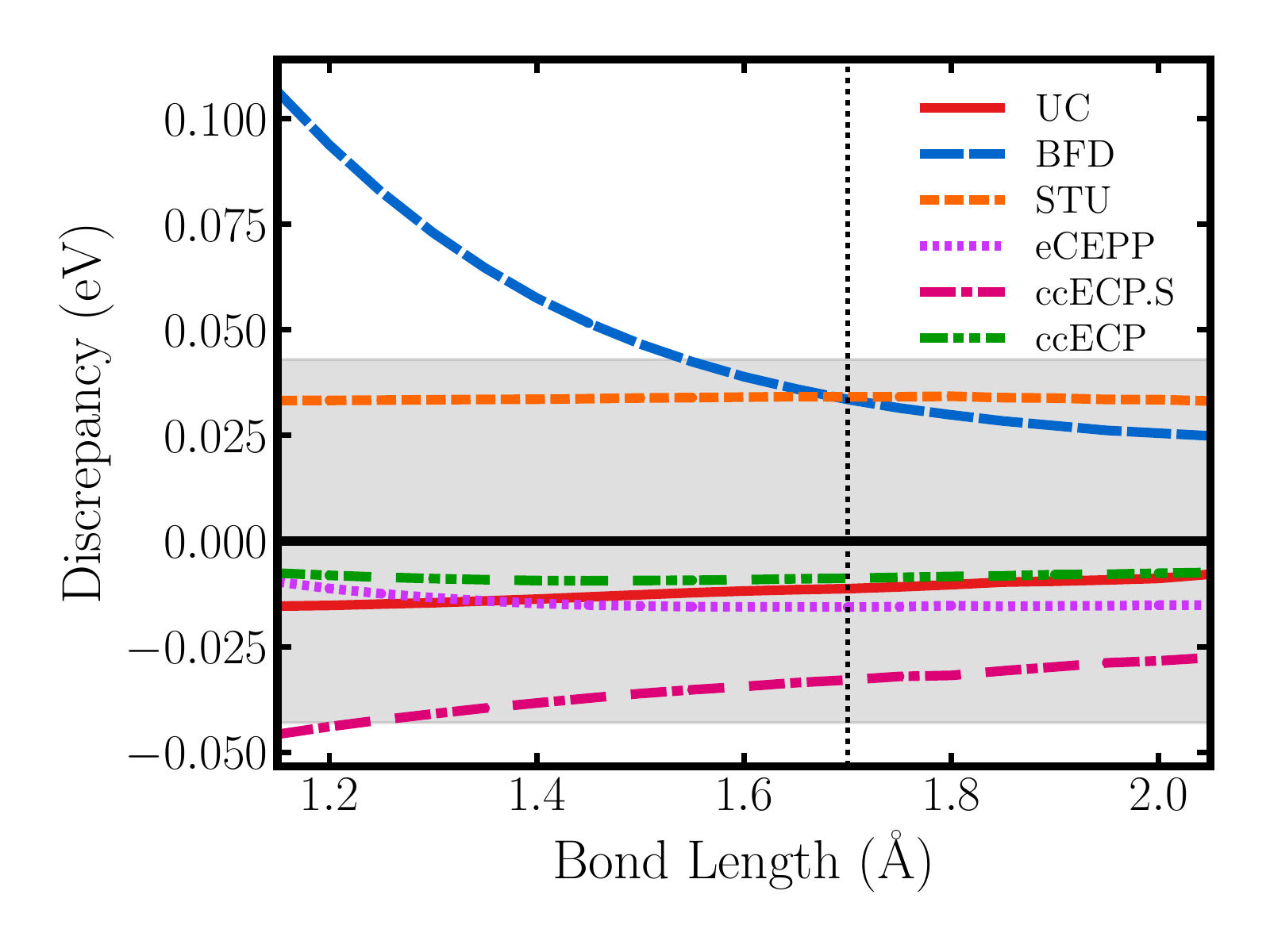}
\caption{VH binding curve discrepancies}\label{fig:VH}
\end{subfigure}%
\begin{subfigure}{0.5\textwidth}
\includegraphics[width=\textwidth]{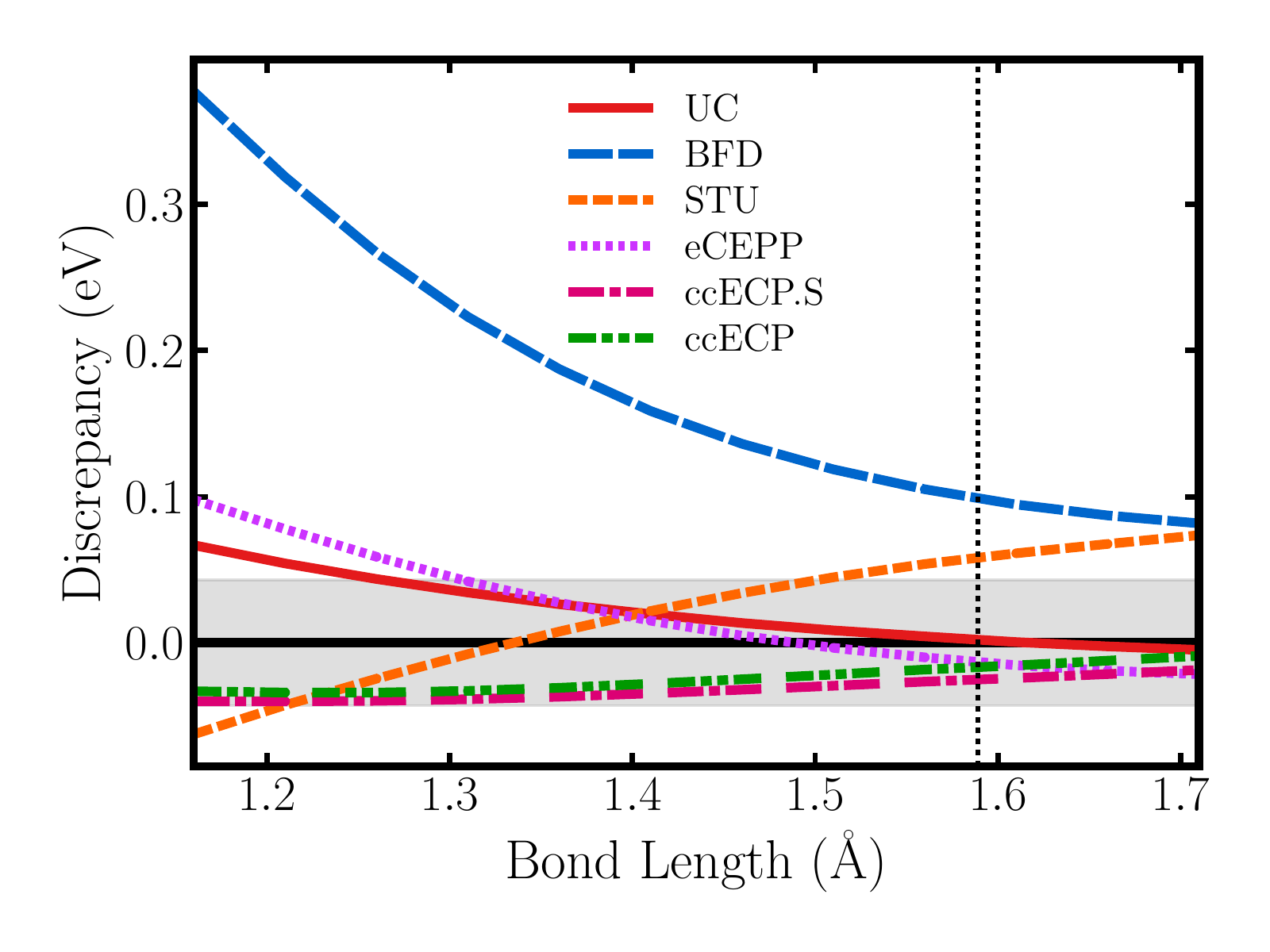}
\caption{VO binding curve discrepancies}\label{fig:VO}
\end{subfigure}%
\caption{Binding energy discrepancies for (a) FeH, (b) FeO, (c) VH, and (d) VO molecules. The binding curves are relative to scalar relativistic AE CCSD(T) binding curve. The shaded region indicates a discrepancy of chemical accuracy in either direction.
The ccECPs are the only valence Hamiltonians that are consistently within the shaded region of chemical accuracy including short bond lengths which are relevant for high pressures.
Reproduced from Ref.~\onlinecite{Annaberdiyev2018}, with the permission of AIP Publishing.}\label{fig:Fe_mols}
\end{figure*}

\begin{figure*}[!htbp]
\centering
\begin{subfigure}{0.5\textwidth}
\includegraphics[width=\textwidth]{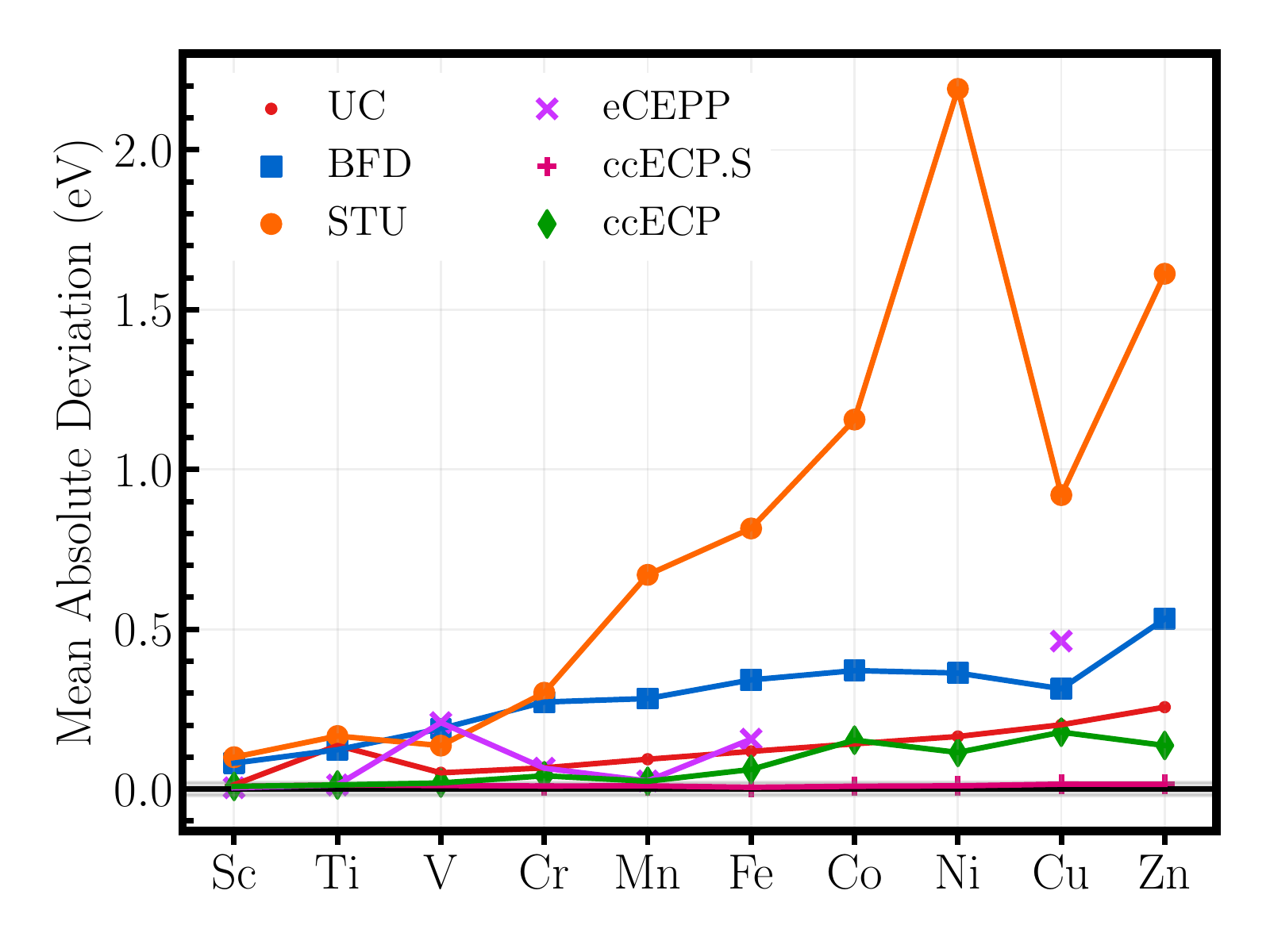}
\caption{}\label{fig:3d-spectrum}
\end{subfigure}%
\begin{subfigure}{0.5\textwidth}
\includegraphics[width=\textwidth]{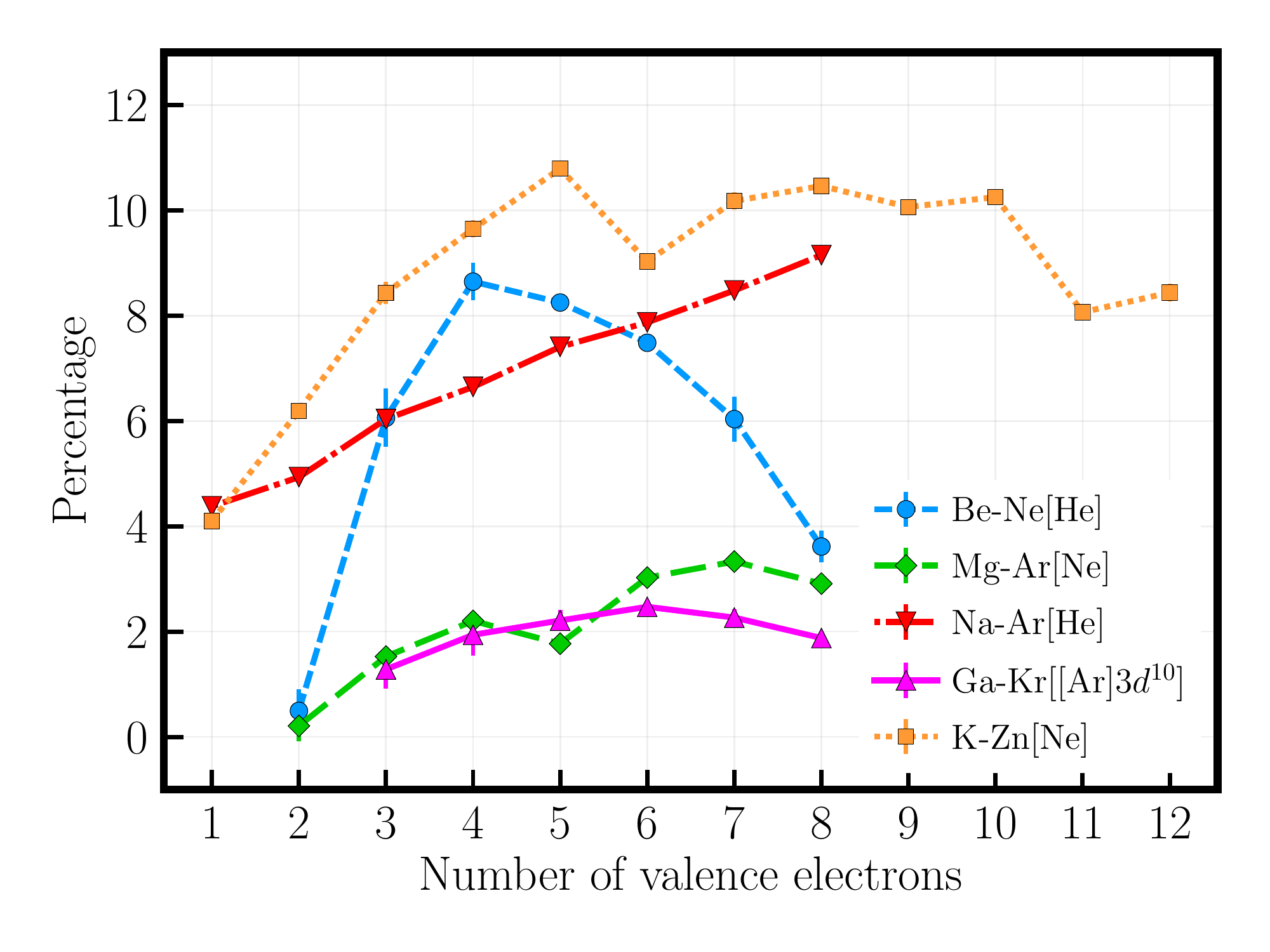}
\caption{}\label{fig:fn_bias}
\end{subfigure}
\caption{(a) MADs for $3d$ TM benchmark states from bare [Ne] core up to low-lying neutral excitations and the anionic state. (b)
Fixed-node DMC biases ($\epsilon$) as a percentage of the correlation energy for ccECP pseudo atoms: $100\epsilon/|E_{corr}|$.
T-moves\cite{CasulaLocality2006} and single-reference trial functions were used in calculations with the exception of Be, B, and C with two-reference form
to account for the significant $2s-2p$ near-degeneracy. Fig. \ref{fig:3d-spectrum} reproduced from Ref. \onlinecite{Annaberdiyev2018}, with the permission of AIP Publishing.
%Fig. \ref{fig:fn_bias} reproduced from Ref. \onlinecite{annaberdiyev_accurate_2019}.
Fig. \ref{fig:fn_bias} is adapted with permission from Ref. \onlinecite{annaberdiyev_accurate_2019}. Copyright 2020 American Chemical Society.
}
\label{atomic-property}
\end{figure*}

\begin{figure*}[!htbp]
\centering
\begin{subfigure}{0.5\textwidth}
\includegraphics[width=\textwidth]{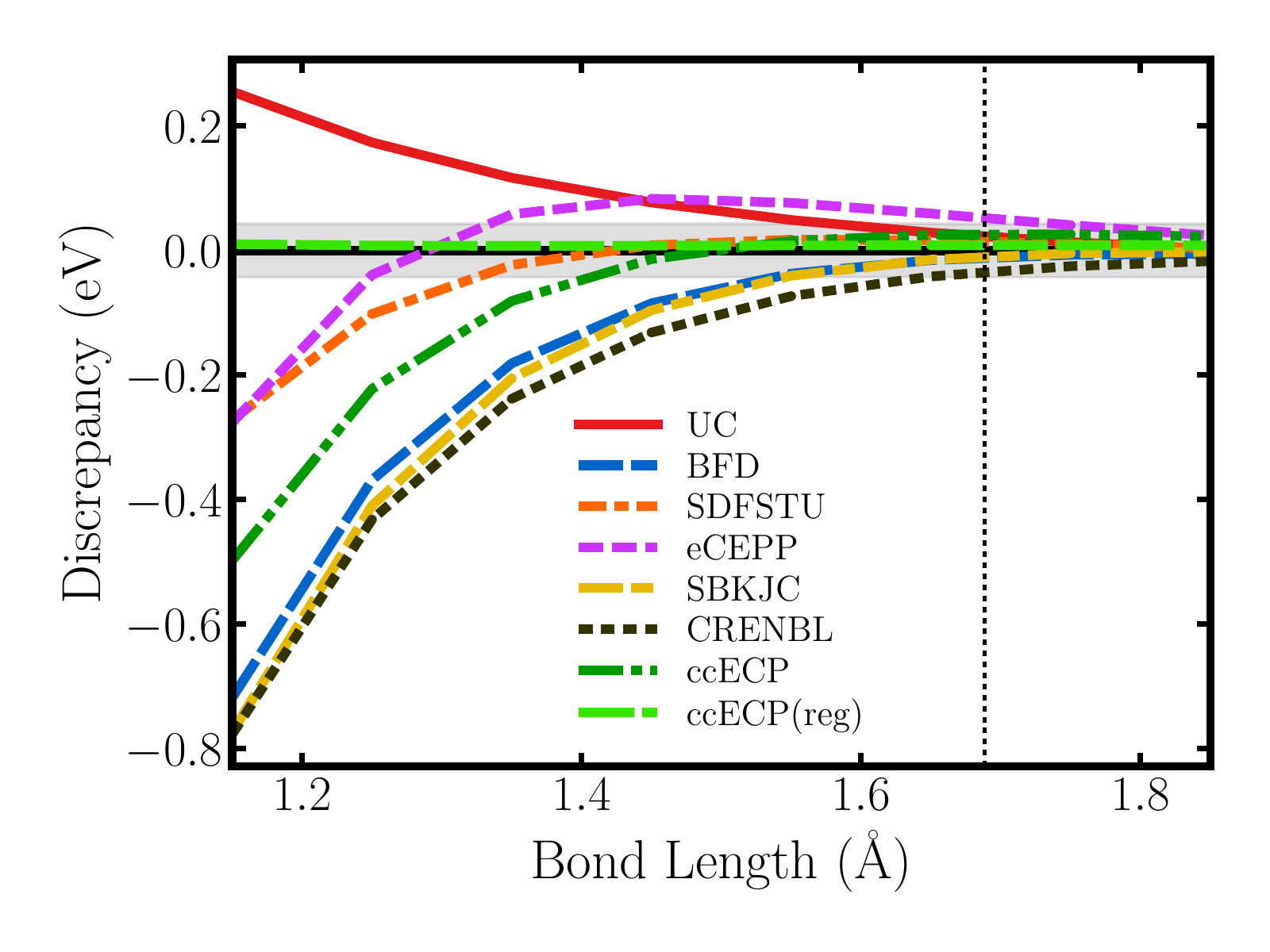}
\caption{LiO binding curve discrepancies}\label{fig:LiO}
\end{subfigure}%
\begin{subfigure}{0.5\textwidth}
\includegraphics[width=\textwidth]{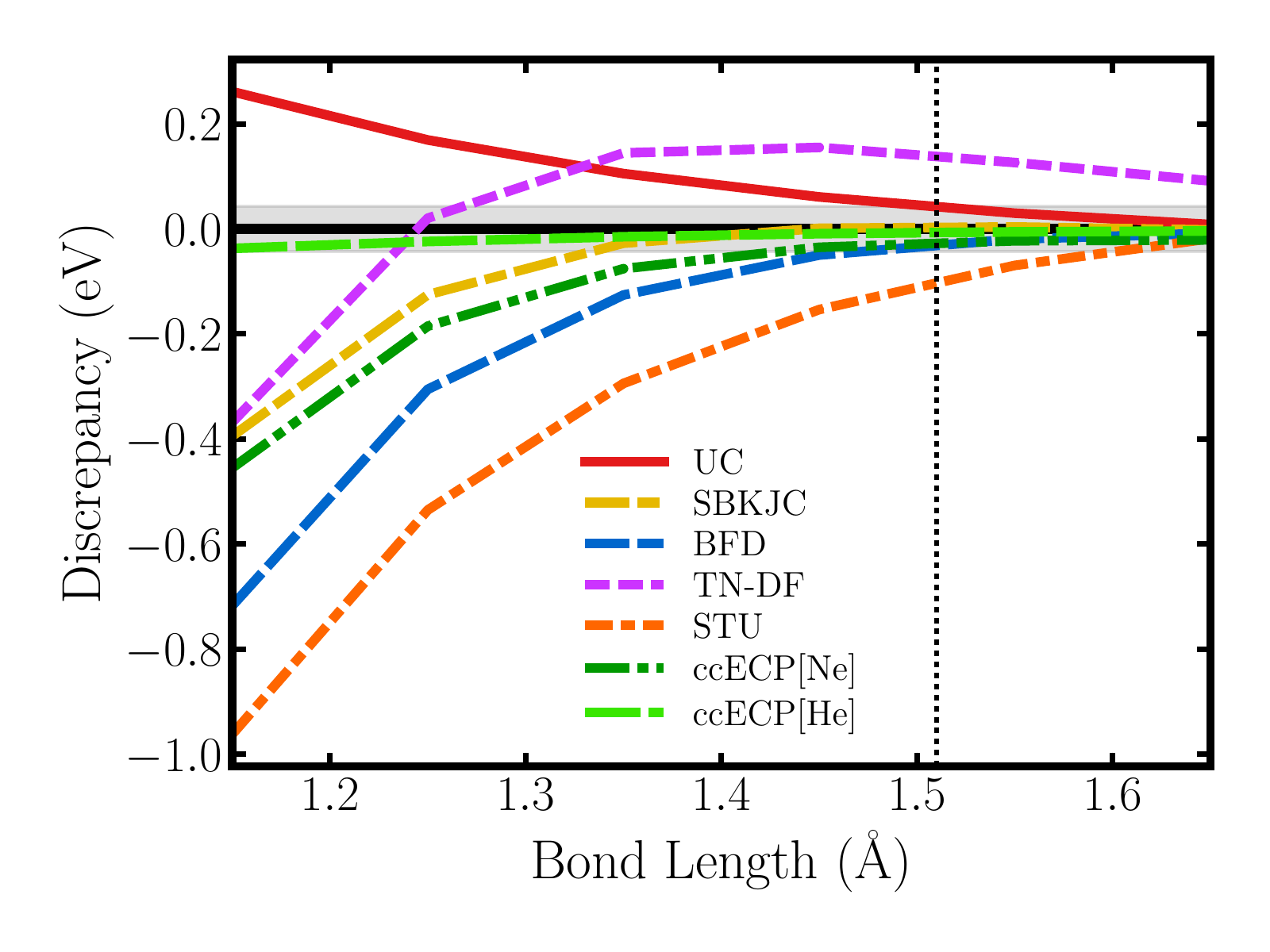}
\caption{SiO binding curve discrepancies}\label{fig:SiO}
\end{subfigure}
\caption{Binding energy discrepancies for (a) LiO and (b) SiO molecules. The uncorrelated core (UC) results,
plus those for many different effective core potentials are given
relative to scalar relativistic all electron (AE)
CCSD(T) binding curve. The shaded region indicates a discrepancy of chemical accuracy in either
direction. Details are given in Refs.~\onlinecite{Wang2019,Bennett2018}. Reproduced from Ref. \onlinecite{Wang2019} and Ref. \onlinecite{Bennett2018}, with the permission of AIP Publishing.
}
\label{fig:small-core-mols}
\end{figure*}

Figure~\ref{fig:Fe_mols} shows the molecular binding energy discrepancies for FeH, FeO, VH, and VO molecules relative to all-electron
CCSD(T) where we observe that some previous ECPs display significant errors. In addition, Table~\ref{tab:mol_mean}, lists a more
comprehensive comparison by tabulating the average of mean absolute deviations (MAD) of molecular binding properties relative to
all-electron CCSD(T) for all $3d$ transition metal (TM) molecules. Similarly, Figure~\ref{fig:3d-spectrum} presents the MAD of a
large valence spectrum for all $3d$ TM atoms. In both atomic and molecular tests, we see that ccECP achieves smaller or on par average errors
with regard to the other ECPs.
In addition, Fig.~\ref{fig:Fe_mols} shows these improvements to be consistent for different  elements and varying geometries. Hence, we believe that ccECP
accomplishes the best accuracy compromise for atomic spectral and molecular properties. Furthermore, ccECPs are provided with smaller cores than
conventionally used ones in some cases where large errors were observed. This includes Na-Ar with [He] core and H-Be with
softened/canceled Coulomb singularity at the origin (ccECP(reg)). Selected molecular test results for these are shown in
Figure~\ref{fig:small-core-mols}.

For reference, we also provide accurate total and kinetic energies for all ccECPs~\cite{annaberdiyev_accurate_2019} using methods
such as CCSDT(Q)/FCI (FCI, full configuration interaction) with DZ-6Z extrapolations to estimate the complete basis set limit. This data, for instance, is useful in the
assessment of fixed-node DMC biases. Figure~\ref{fig:fn_bias} shows the summary of single-reference (HF) fixed-node DMC errors for
ccECP pseudo atoms. 
 
\begin{table*}[!htbp]
\centering
\begin{tabular}{c|cccccc}
\hline
{} &               UC &              BFD &              STU &            eCEPP &          ccECP.S &            ccECP \\
\hline
$D_e$(eV)             & 0.0063(40) & 0.0590(41) & 0.0380(41) & 0.0163(45) & 0.0240(40) & 0.0104(40) \\
$r_e$(\AA)            & 0.0012(13) & 0.0064(13) & 0.0026(13) & 0.0019(15) & 0.0027(13) & 0.0010(13) \\
$\omega_e$(cm$^{-1}$) &   2.2(5.8) &  10.4(5.9) &   4.6(5.9) &   3.9(6.9) &   6.4(5.8) &   2.9(5.8) \\
$D_{diss}$(eV)        &  0.021(41) &  0.145(41) &  0.036(41) &  0.032(46) &  0.054(40) &  0.016(41) \\
\hline
\end{tabular}
\caption{Average MADs of binding parameters for various core approximations with respect to AE data for $3d$ TM hydride and oxide molecules. All parameters were obtained using Morse potential fit. The parameters shown are dissociation energy $D_e$, equilibrium bond length $r_e$, vibrational frequency $\omega_e$ and binding energy discrepancy at dissociation bond length $D_{diss}$. Reproduced from Ref. \onlinecite{Annaberdiyev2018}, with the permission of AIP Publishing.}\label{tab:mol_mean}
\end{table*}

\subsection{ccECP Database and Website}
\begin{figure}[!htbp]
    \centering
    \includegraphics[width=0.6\textwidth]{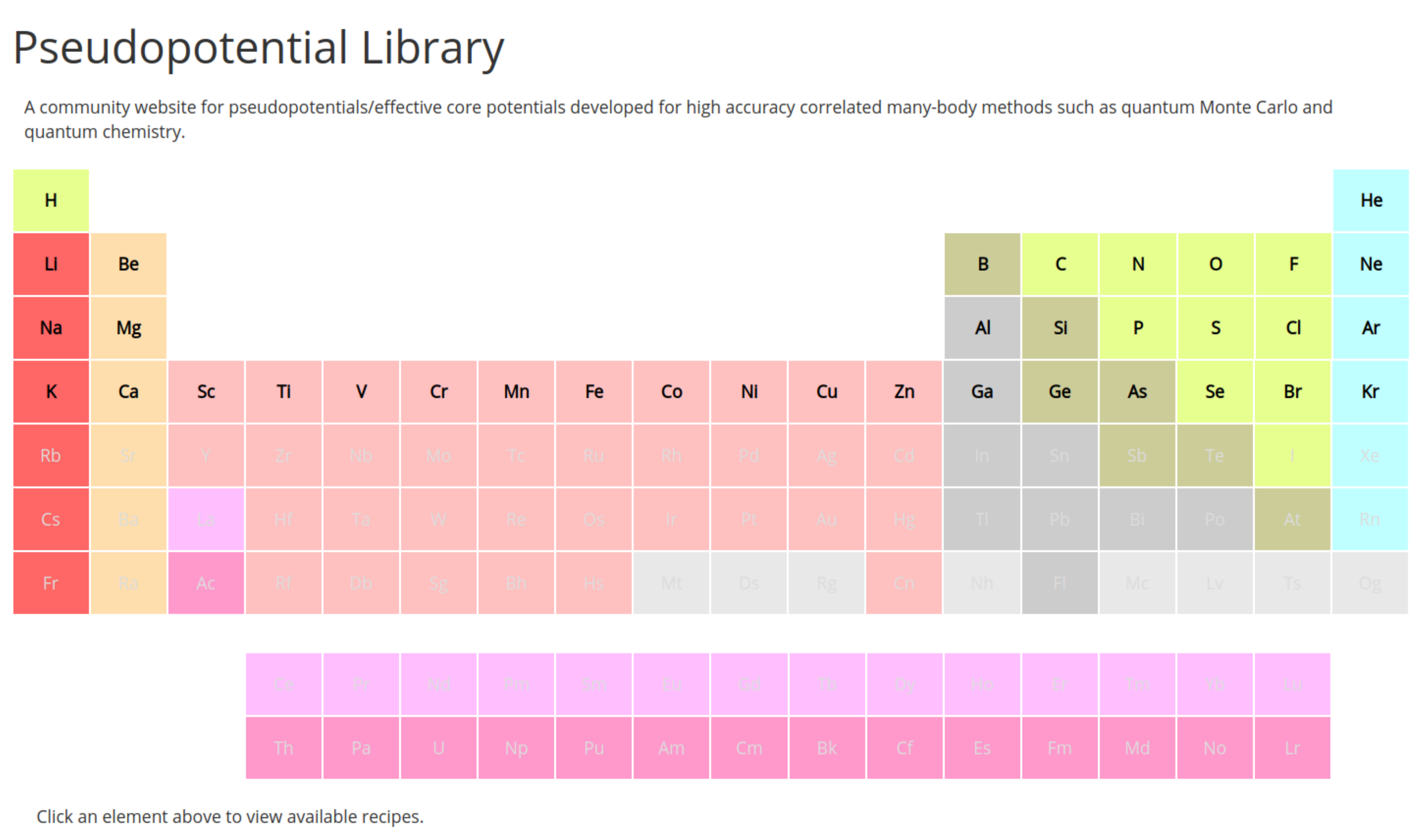}
    \caption{Pseudopotential Library, https://pseudopotentiallibrary.org.\label{fig:ecplib_web}}
\end{figure}

 In order to facilitate the use ccECPs, we have provided basis sets and a variety of ECP formats available at
\url{https://pseudopotentiallibrary.org}, shown in Figure~\ref{fig:ecplib_web}. Each ccECP is presented in a quantum chemistry
format for direct use in various codes, including \textsc{Molpro}, \textsc{GAMESS}, \textsc{NWChem}, and \textsc{PySCF} which uses
the \textsc{NWChem} format. We also provide an XML format which can directly be used in \textsc{QMCPACK}.

In addition to the ccECPs themselves, we have also provided basis sets appropriate for correlated calculations in each code
format. Specifically, we have provided Dunning style~\cite{dunning_basis_2002} correlation consistent basis sets from the DZ to
6Z, and in most cases have also provided an augmented version. For use in solid state applications using a plane wave basis, we
have also transformed the semi-local potentials into fully nonlocal Kleinman-Bylander potentials~\cite{Kleinman1982} using the
Unified Pseudopotential Format. This allows the ccECPs to be directly used in codes such as \textsc{Quantum Espresso}. A report
file is included giving detailed information about the quality of the Kleinman-Bylander version of the potential and recommended
plane wave energy cutoff energies.

\subsection{Status and future developments} The ccECP table and construction principles aim at improved account of systematic
errors built into effective valence Hamiltonians in a wide variety of correlated calculations (see encouraging feedback so
far~\cite{WangTingWang2019, zhou_diffusion_2019}). Further effort is focused on adapting ccECPs for efficient calculations with
plane wave basis set (ccECPpw versions). This requires modifying the deep potentials of the late $3d$ elements Fe-Zn in
particular. The goal is to enable calculations with plane wave cutoffs not exceeding $\sim$ 600-800 Ry. Plans for the near future
involve ccECPs for selected $4d$ and $5d$ elements that include a number of technologically important elements and require
explicit treatment of the spin-orbit interactions. Additional improvements such as core polarization and relaxation corrections
can be added per specific, application driven needs. Further plans include seeking feedback from the electronic structure
community, collecting the data for reference and validation as well as adjustments per need for use in a broad variety of {\it ab
initio} approaches. 

\section{Auxiliary Field Quantum Monte Carlo}\label{sec:afqmc}
The latest version of QMCPACK offers a now mature implementation of the phaseless auxiliary-field quantum Monte Carlo (AFQMC)
method\citep{Zhang_phaseless,Motta2019} capable of simulating both molecular\citep{borda_nomsd,LeeSymmetryBreaking2020} and solid state
systems\citep{zhang_nio,malone_isdf,lee_2019_UEG}. AFQMC is usually formulated as an orbital-space approach in which the
Hamiltonian is represented in second-quantized form as
\begin{align}
    \hat{H} &= \sum_{ij} h_{ij} \hat{c}^{\dagger}_{i}\hat{c}_{j} + \frac{1}{2}\sum_{ijkl}v_{ijkl}  \hat{c}^{\dagger}_{i}\hat{c}^{\dagger}_{j}\hat{c}_{l}\hat{c}_{k}+E_{II},\label{eq:hamil}\\
            &= \hat{H}_1 + \hat{H}_2 + E_{II},
\end{align}
where $\hat{c}^{\dagger}_{i}$  and $\hat{c}_{i}$ are the fermionic creation and annihilation operators, $h_{ij}$ and
$v_{ijkl}$ are the one- and two-electron matrix elements and $E_{II}$ is the ion-ion repulsion energy. Key to an efficient
implementation of AFQMC is the factorization of the 4-index electron-repulsion integral (ERI) tensor $v_{ijkl}$, which is
essential for the Hubbard-Stratonovich (HS) transformation\citep{hubbard_strat,zhang_cpmc}.

QMCPACK offers three factorization approaches which are appropriate in different settings. The most generic approach implemented
is based on the modified-Cholesky
factorization\citep{modified_chol_1,modified_chol_2,modified_chol_3,purwanto_cholesky,purwanto_downfolding_jctc} of the ERI
tensor:
\begin{equation}
    v_{ijkl} = V_{(ik),(lj)} \approx \sum_n^{N_\mathrm{chol}} L_{ik}^n L^{*n}_{lj},
\end{equation}
where the sum is truncated at $N_{\mathrm{chol}} = x_c M$, $x_c$ is typically between $5$ and $10$, $M$ is the number of basis
functions and we have assumed that the single-particle orbitals  are in general complex. The storage requirement is thus naively
$\mathcal{O}(M^3)$ although sparsity can often be exploited to keep the storage overhead manageable (see
Table~\ref{tab:afqmc_comp}). Note that QMCPACK can accept any 3-index tensor of the form of $L_{ik}^n$ so that alternative
density-fitting based approaches can be used. Although the above approach is efficient for moderately sized molecular and
solid-state systems, it is typically best suited to simulating systems with fewer than 2000 basis functions.

To reduce the memory overhead of storing the three-index tensor we recently adapted the
tensor-hypercontraction\citep{thc_1,thc_2,thc_3} (THC) approach for use in AFQMC\citep{malone_isdf}. Within the THC approach we
can approximate the orbital products entering the ERIs as
\begin{equation}
    \varphi^{*}_i(\mathbf{r})\varphi_k(\mathbf{r}) \approx \sum_\mu^{N_\mu} \zeta_\mu(\mathbf{r}) \varphi^*_i(\mathbf{r}_\mu)\varphi_k(\mathbf{r}_\mu),\label{eq:orb_prod}
\end{equation}
where $\varphi_i(\mathbf{r})$ are the one-electron orbitals and $\mathbf{r}_\mu$ are a set of specially selected interpolating
points, $\zeta_\mu(\mathbf{r})$ are a set of interpolating vectors and $N_\mu = x_\mu M$. We can then write the ERI tensor as a
product of rank-2 tensors
\begin{equation}
    v_{ijkl} \approx \sum_{\mu\nu} \varphi^{*}_i(\mathbf{r}_\mu)\varphi_k(\mathbf{r}_\mu) M_{\mu\nu} \varphi^{*}_j(\mathbf{r}_\nu)\varphi_l(\mathbf{r}_\nu)\label{eq:4ix_thc},
\end{equation}
where
\begin{equation}
    M_{\mu\nu} = \int d\mathbf{r}d\mathbf{r}' \zeta_\mu(\mathbf{r})\frac{1}{|\mathbf{r}-\mathbf{r}'|}\zeta^{*}_\nu(\mathbf{r}')\label{eq:mmat}.
\end{equation}
To determine the interpolating points and vectors we use the interpolative separable density fitting (ISDF)
approach~\citep{Ying_ISDF,ISDF_LinLin,ISDF_CVT}. Note that the storage requirement has been reduced to $\mathcal{O}(M^2)$. For
smaller system sizes the three-index approach is preferred due to the typically larger THC prefactors determined by $x_\mu \approx
15$ for propagation and $x_\mu \approx 10$ for the local energy evaluation. The THC approach is best suited to simulating large
supercells, and is also easily ported to GPU architectures due to its smaller memory footprint and use of dense linear algebra.
Although the THC-AFQMC approach has so far only been used to simulate periodic systems, it is also readily capable of simulating
large molecular systems using the advances from Ref.~\onlinecite{lee2019systematically}.

Finally, we have implemented an explicitly $k$-point dependent factorization for periodic systems\citep{motta_kpoint}
\begin{equation}
    V_{(i\mathbf{k}_k+\mathbf{Q}k\mathbf{k}_k),(l\mathbf{k}_lj\mathbf{k}_l-\mathbf{Q})} \approx \sum_n^{n_\mathrm{chol}^{\mathbf{Q}}} L_{ik,n}^{\mathbf{Q}\mathbf{k}_k} L^{\mathbf{Q}\mathbf{k}_l*}_{lj,n},
\end{equation}
where now $i$ runs over the number of basis functions $(m)$ for $k$-point $\mathbf{k}_i$ in the primitive cell,
$\mathbf{Q}=\mathbf{k}_i-\mathbf{k}_k +\mathbf{G} = \mathbf{k}_l-\mathbf{k}_j+\mathbf{G}'$ is the momentum transfer vector
(arising from the conservation of crystal momentum) and $\mathbf{G},\mathbf{G}'$ are reciprocal lattice vectors. Although
explicitly incorporating $k$-point symmetry reduces the scaling of many operations and the storage requirement by a factor of
$1/N_k$ (see Table.~\ref{tab:afqmc_comp}), perhaps the most significant advantage is that it permits the use of batched dense
linear algebra and is thus highly efficient on GPU architectures. Note that the THC and $k$-point symmetric factorization can be
combined to simulate larger unit cells and exploit $k$-point symmetry, however this has not been used to date. We compare the
three approaches in Table~\ref{tab:afqmc_comp} and provide guidance for their best use.

\begin{table}[h!]
    \centering
    \begin{tabular}{lccccc}
        \hline
        Method & Memory & Propagation & Energy & Setting & GPU\\
        \hline
        Dense 3-index & $x_c M^3$ & $\mathcal{O}(N M^2)$ & $\mathcal{O}(x_c N^2 M^2)$ & $M \le 1000$ & Yes \\
        Sparse 3-index & $s x_c M^3$ & $\mathcal{O}(N M^2)$ & $\mathcal{O}(N^2 M^2)$ & $M \le 2000$ & No \\
        %Sparse Three-index&
        THC & $x_\mu^2 M^2$ & $\mathcal{O}(N M^2)$ & $\mathcal{O}(x_\mu^2 N M^2)$  & $M \le 4000$ & No \\
        $k$-point & $x_c m^3 N_k^2$ & $\mathcal{O}(NM^2)$ & $\mathcal{O}(x_c m^2 n^2 N_k^3)$ & $N_k m \le 6000$ & Yes\\
        \hline
    \end{tabular}
    \caption{Comparison in the dominant scaling behavior of different factorization approaches implemented in QMCPACK. We have included a sparsity factor $s$ which can reduce the computational cost of the three-index approach significantly. For example, in molecular systems the memory requirement is asymptotically $\mathcal{O}(M^2)$ in the atomic orbital basis, whilst for systems with translational symmetry the scaling is in principle identical to that of the explicitly $k$-point dependent factorization (i.e. $s\le 1/N_k$) although currently less computationally efficient. We also indicate the current state of GPU support for the different factorizations available in QMCPACK. The THC factorization will be ported to GPUs in the near future. Note that by using plane waves the scaling of the energy evaluation and propagation can be brought down to $\mathcal{O}(N^2 M \log M)$ and $\mathcal{O}(NM\log M)$ respectively. This approach essentially removes the memory overhead associated with storing the ERIs at the cost of using a potentially very large plane-wave basis set\citep{suewattana_pw_afqmc,ma_multiple_proj}. This plane wave approach is not yet available in QMCPACK.\label{tab:afqmc_comp}}
\end{table}

\begin{figure}[h!]
    \centering
    \includegraphics{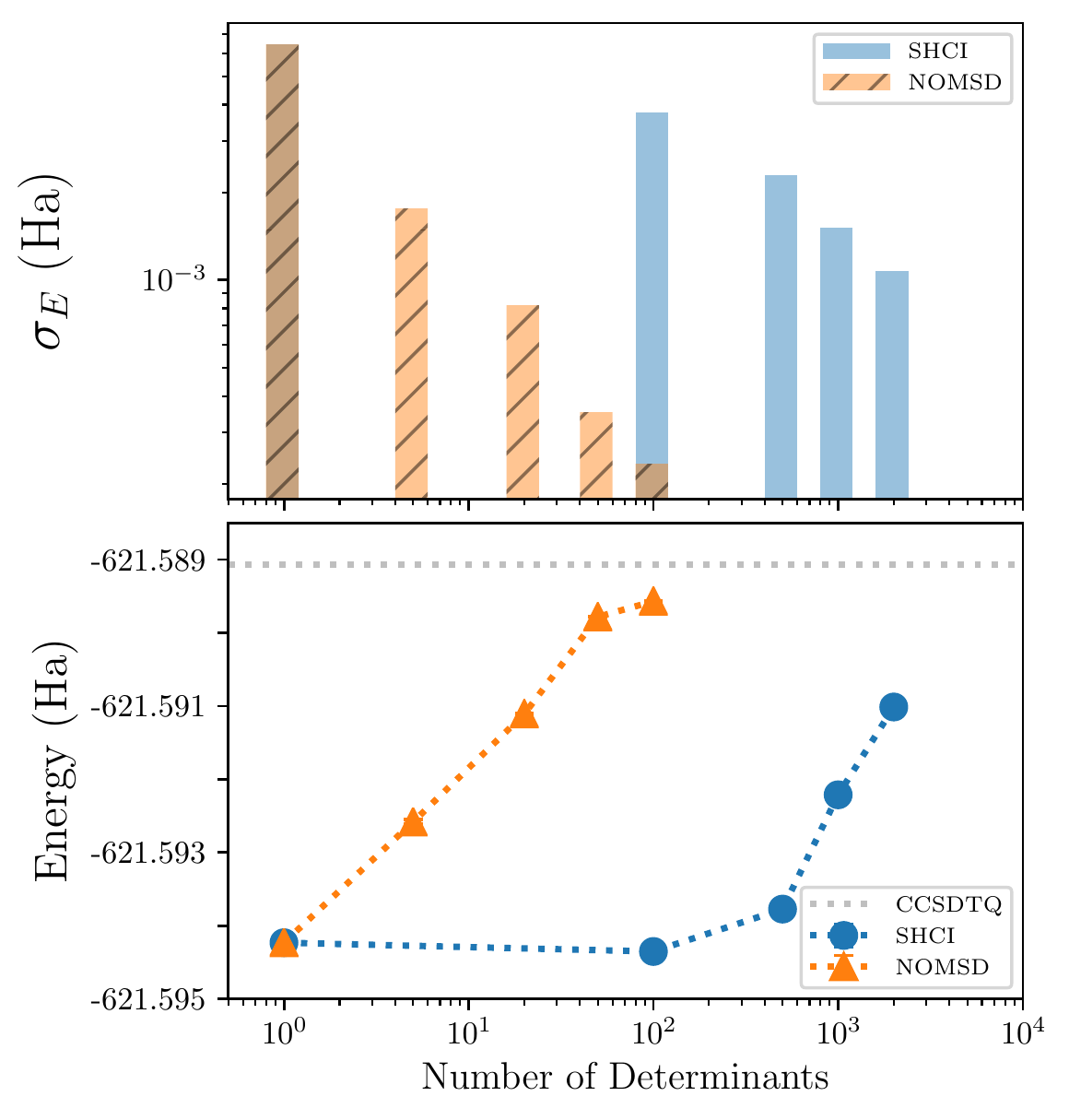}
    \caption{Comparison in the performance of selected heath-bath configuration interaction (SHCI) and NOMSD as trial wavefunctions in AFQMC calculations of NaCl in the cc-pVDZ basis set at its equilibrium bond length. The top panel demonstrates that smaller NOMSD expansions are necessary to reduce the standard deviation in the energy estimator ($\sigma_E$) compared to SHCI trial wavefunctions. The bottom panel shows that the total energy converges more rapidly with determinant number when using a NOMSD trial wavefunction, where the horizontal dashed line is the coupled cluster singles, doubles, triples and quadruples (CCSDTQ) result. The SHCI and NOMSD wavefunctions were generated using the DICE\citep{sharma_dice_1,holmes_dice_2} and PHFMOL\citep{Jimenez_phf,scuseria_phf,scuseria_phf_grad} packages respectively. The CCSDTQ result was computed using the Aquarius package\citep{SolomonikAquarius2014}. Reproduced from Ref.\onlinecite{borda_nomsd}\label{fig:afqmc_msd_conv}}
\end{figure}

In addition to state-of-the-art integral factorization techniques, QMCPACK also permits the use of multi-determinant trial
wavefunction expansions of the form
\begin{equation}
    |\psi_T\rangle = \sum_I^{N_D} c_I |D_I\rangle.
\end{equation}
We allow for either orthogonal configuration interaction expansions where $\langle D_I | D_J\rangle = \delta_{IJ}$ and also for
non-orthogonal multi Slater determinant expansions (NOMSD) where  $\langle D_I | D_J\rangle = S_{IJ}$. Orthogonal expansions from
complete active space self-consistent field (CASSCF) or selected CI methods allow for fast overlap and energy evaluation through
Sherman-Morrison based techniques, and thus do not typically incur a significant slowdown. However, they often require a large
number of determinants to converge the phaseless error. NOMSD expansions do not benefit from fast update techniques, but often
require orders of magnitude fewer determinants than their orthogonal counterparts to achieve convergence in the AFQMC total
energy\citep{borda_nomsd} (see Fig.~\ref{fig:afqmc_msd_conv}).

\begin{figure}[h!]
    \centering
    \includegraphics[scale=0.75]{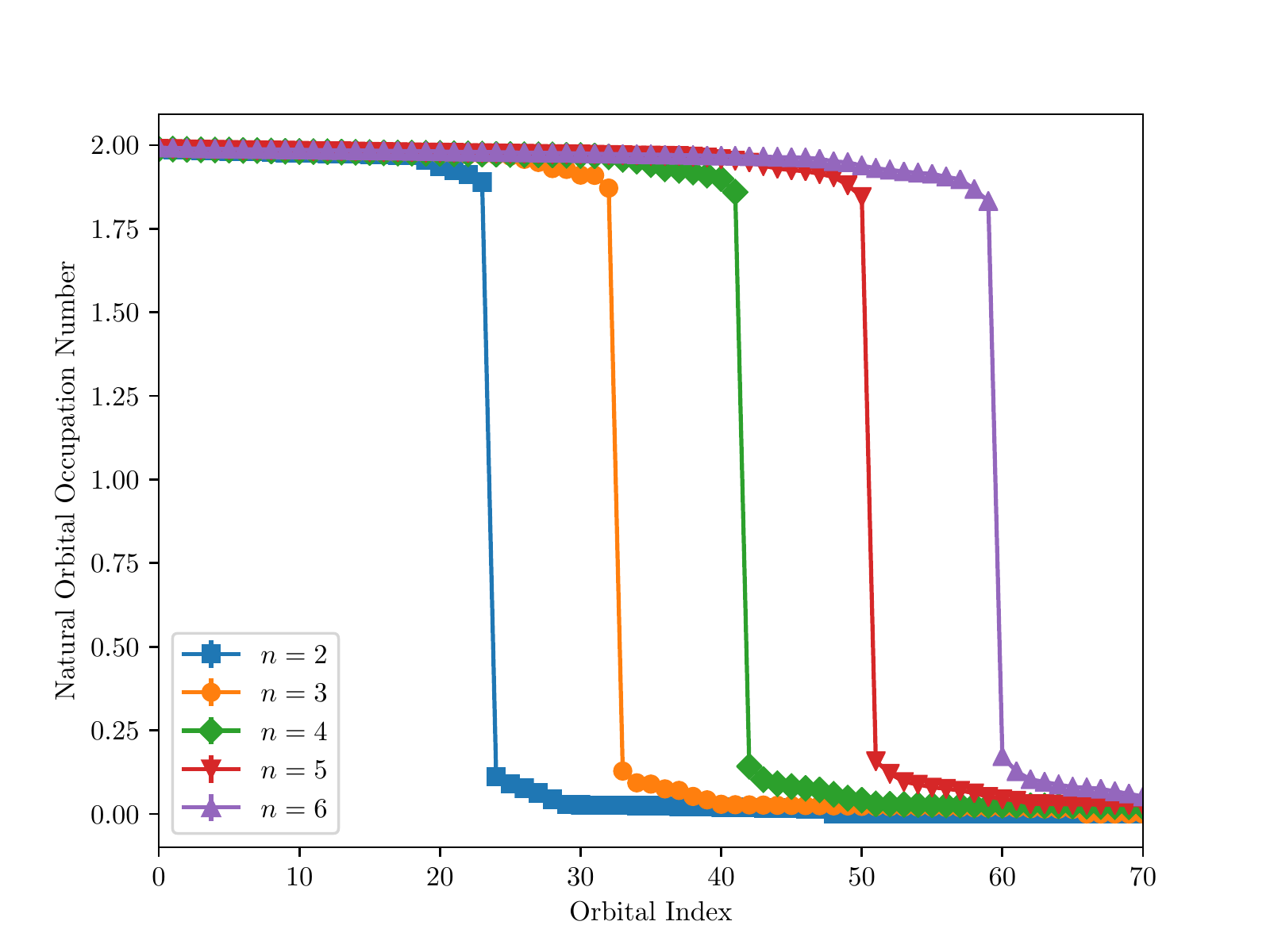}
    \caption{Phaseless AFQMC natural orbital occupation numbers computed from the back-propagated 1RDM for the $n$-acenes (C$_2$H$_4$C$_{4n}$H$_{2n}$) in the STO-3G basis set. Geometries are taken from Ref.~\onlinecite{lee_ccvb}. Error bars are plotted but are smaller than the symbol size.\label{fig:afqmc_noons}}
\end{figure}

QMCPACK also permits the evaluation of expectation values of operators which do not commute with the Hamiltonian using the back
propagation method\citep{zhang_cpmc,purwanto_back_prop,motta_back_prop}. In particular, the back-propagated one-particle reduced
density matrix (1RDM) as well components or contracted forms of the two-particle reduced density matrix are available. As an
example we plot in Fig.~\ref{fig:afqmc_noons} the natural orbital occupation numbers computed from the back-propagated phaseless AFQMC
1RDM.

Tools to generate the one- and two-electron integrals and trial wavefunctions for molecular and solid state systems are also
provided through the afqmctools package distributed with QMCPACK. To date these tools are mostly dependent on the PySCF software
package\citep{PYSCF}, however we provide conversion scripts for FCIDUMP formatted integrals, as well as simple Python routines to
convert factorized integrals or trial wavefunctions provided from any source to our internal HDF5 based file format. Detailed
tutorials on how to run AFQMC in QMCPACK are also provided. Nexus (Sec.\ref{sec:workflow}) can be used to drive the process of
mean-field calculation, wavefunction conversion, and AFQMC calculation. We are using this integration to perform a study of the relative strengths of AFQMC and real space QMC methods.

Over the next year we plan to extend the list of observables available as well as complete GPU ports for all factorization and
wavefunction combinations. In addition we plan to implement the finite temperature AFQMC
algorithm\citep{zhang_ftafqmc_99,rubenstein_ft,he_sconsis_ft,liu_ab_2018,he_continuum_prl}, and spin-orbit Hamiltonians with
non-collinear wavefunctions. We will also release our ISDF-THC factorization tools and our interface to Quantum
Espresso\citep{Gianozzi2009}. We hope our open-source effort will enable the wider use of AFQMC in a variety of challenging
settings. 

\section{Towards systematic convergence of real-space QMC calculations}\label{sec:gsnodes}
The key factor in reaching high accuracy using QMC is the choice of trial wave function $\Psi_T$. For all-electron DMC
calculations the nodes of the wavefunction are the only factor in determining the error in the computed energy, while the bulk of
the wavefunction affects the statistical efficiency and timestep error of the calculation. For calculations involving
pseudopotentials, high accuracy of the trial wavefunction is also needed around the atomic cores to minimize the approximations in
evaluating the non-local energy.  Single Slater determinant (SD) wavefunctions built with Hartree-Fock or Kohn-Sham orbitals
supplemented by a Jastrow correlation factor generally give good results, e.g. Ref. \onlinecite{DubeckyNoncovalent2017}, and are used almost
exclusively today in solid-state calculations.

Reaching systematic convergence of the trial wavefunction and its nodal surface for general systems has been a key challenge for
real space QMC methods since their invention. Besides increasing accuracy in calculated properties, this is also required to
remove the starting point dependence and allow use of QMC where all potential sources of trial wavefunction are unreliable. In 2008 this was performed for first row atoms and diatomic
molecules Ref.\cite{ToulouseFull2008}, and improved algorithms are aiding calculations on larger
systems\cite{AssarafOptimizing2017}. For general systems with many electrons, the overall challenge remains. Furthermore, if the
wavefunction is to be used in DMC, commonly used optimization techniques
only optimize the nodal surface indirectly by improving the VMC energy and/or variance. Minimization of the objective function is
therefore not guaranteed to minimize the fixed-node energy. Consistently high accuracy wavefunctions
are also needed around atomic cores to minimize the locality error in pseudopotential evaluation, posing a challenge for trial
wavefunction optimization with a large number of coefficients. 

One possible step along the way would be to optimize all the orbital coefficients in a single determinant wavefunction, but due to
the limited flexibility in describing the $(3N-1)$ dimensional nodal surface this protocol can not give exact nodes for general
systems. This approach could represent a useful starting point independent step, while keeping a simple form for the trial
wavefunction. Other possibilities for improving the trial wavefunction while retaining simplicity include techniques such as
backflow and iterative backflow\cite{TaddeiIterative2015,HolzmannOrbital2019}, and antisymmetrized geminal product wavefunctions
(e.g. Ref.~\onlinecite{CasulaCorrelated2004}). However, more flexible and complex trial wavefunctions are required to achieve systematic
convergence of the nodal surface and to approach exact results for general systems.

The most straightforward method to improve the quality of the trial wavefunction nodes in a convergeable manner is to increase its
complexity via a multi-Slater determinant (MSD) or configuration-interaction (CI) expansion:

\begin{equation}
    \Psi_{T} = \sum_{i=1}^{M}c_i D_i^{\uparrow}D_i^{\downarrow}e^J
\end{equation} 

where $\Psi_T$ is expanded in a weighted ($c_i$) sum of products of up and down spin determinants $D_i$, and $J$ is the Jastrow
correlation factor. In the limit of a full configuration interaction calculation in a sufficiently large and complete basis set,
this wavefunction is able to represent the exact wavefunction. However, direct application of configuration interaction is
prohibitively costly for all but the smallest systems, because very large numbers of determinants are usually required. To speed application, an efficient selection procedure for the determinants
is needed. This can be combined with efficient algorithms for evaluating the wavefunction in
QMC.\cite{NukalaFast2009,Scuseria2011,Scuseria2012}

\subsection{Ground state calculations}

Multiple variants of selected Configuration Interaction (sCI) methods have recently demonstrated significant success at reaching
high accuracy for ground state and excited states of molecular systems with tractable computational cost. Within the class of sCI
methods, the CIPSI\cite{Huron1973} method has proven to be practical in providing high accuracy wavefunctions for QMC for both
molecular systems and for solids\cite{SCEMAMA2019,Scemama2018,Chien2018,Li2018,Caffarel2016,Scemama2014,Giner2013}. sCI methods
enable unbiased construction of the trial wavefunction using only a single threshold parameter and therefore avoid the
complexities of, for example, CASSCF techniques which require expert selection of the active space. CIPSI algorithms are
implemented in the Quantum Package 2.0 code\cite{QP2019} and fully interfaced with QMCPACK and Nexus. 

For systems where CIPSI can be fully converged to the FCI limit and reliably extrapolated to the basis set limit, QMC is not
required, but for any reasonable number of electrons, QMC can be used to further improve the convergence. The wavefunctions
produced from CIPSI can be used either directly, in which case the nodal error is determined by the CIPSI procedure, or used to
provide an initial selection of determinants whose coefficients are subsequently reoptimized in presence of a Jastrow function, or
used within DMC where the projection procedure will improve on the CIPSI wavefunction. This procedure is equally applicable to
solids as well as molecules, provided k-points and their symmetries are fully implemented.

In the following, we illustrate these techniques by application to molecular and solid-state lithium fluoride. In both cases, we
use Linear Combination of Atomic Orbitals (LCAO) and different Gaussian basis set sizes to generate the trial wavefunctions. CIPSI
energies refer to the variational energy corrected with the sum of energies from second order perturbation
theory ($PT_2$) of each determinant, i.e. $E+PT_2$ , at convergence in energy with the number of determinants. Since the sizes of both systems are small
enough to reach CIPSI convergence with, for the largest case, less than 5M determinants, for the DMC calculations, the
coefficients of the determinants are not reoptimized in presence of a Jastrow function. The cost of DMC with a CIPSI trial wavefunction scales as $\sqrt(N)*(Var_{Ratio})^2$ where N is the number of determinants in the expansion and $Var_{Ratio}$ is the ratio between the variance of a system at 1 determinant and the same system at N determinants ($Var_{Ratio}=\frac{Var_{Ndets}}{Var_{1det}}$). In the case of the molecular systems, $Var_{Ratio}$ varies between 0.7 (for cc-pcVDZ and $N=1.6M$) and 0.8 (cc-pcVQZ and $N=5.2M$) or an increase of cost ranging from 620 and 1500 times the cost of a single determinant DMC run. In the case of the solid, $Var_{Ratio}$ varies between 0.83 (for cc-pVDZ and $N=700k$) and 0.56 (cc-pcVQZ and $N=9M$) or an increase of cost ranging from 570 and 940 times the cost of a single determinant DMC run. The main difference in the change of cost between the molecular system and solid system is the use of ECPs, reducing significantly the variance of the calculations for both DMC(SD) and DMC(CIPSI)        

\subsection{Molecular Lithium Fluoride}
Lithium Fluoride is a small molecule for which the multi-determinant expansion and therefore trial wavefunction can be fully converged to the FCI limit using the CIPSI method. Moreover, CCSD(T) calculations of the Vertical Ionization Potential (VIP) are feasible and it's experimental value is known\cite{Berkowitz1962}, providing reliable reference data. Care has to be taken in the comparison to experimental values, as experimental measurements include  intrinsic uncertainties and environmental parameters such as temperature. These effects, including zero-point motion, are not included in our study and might explain the remaining discrepancies between our calculations and the experimental ionization energy value given by Berkowitz et al. \citet{Berkowitz1962} 

\begin{figure}[h]
\centering
        \includegraphics[scale=1.0,angle=0]{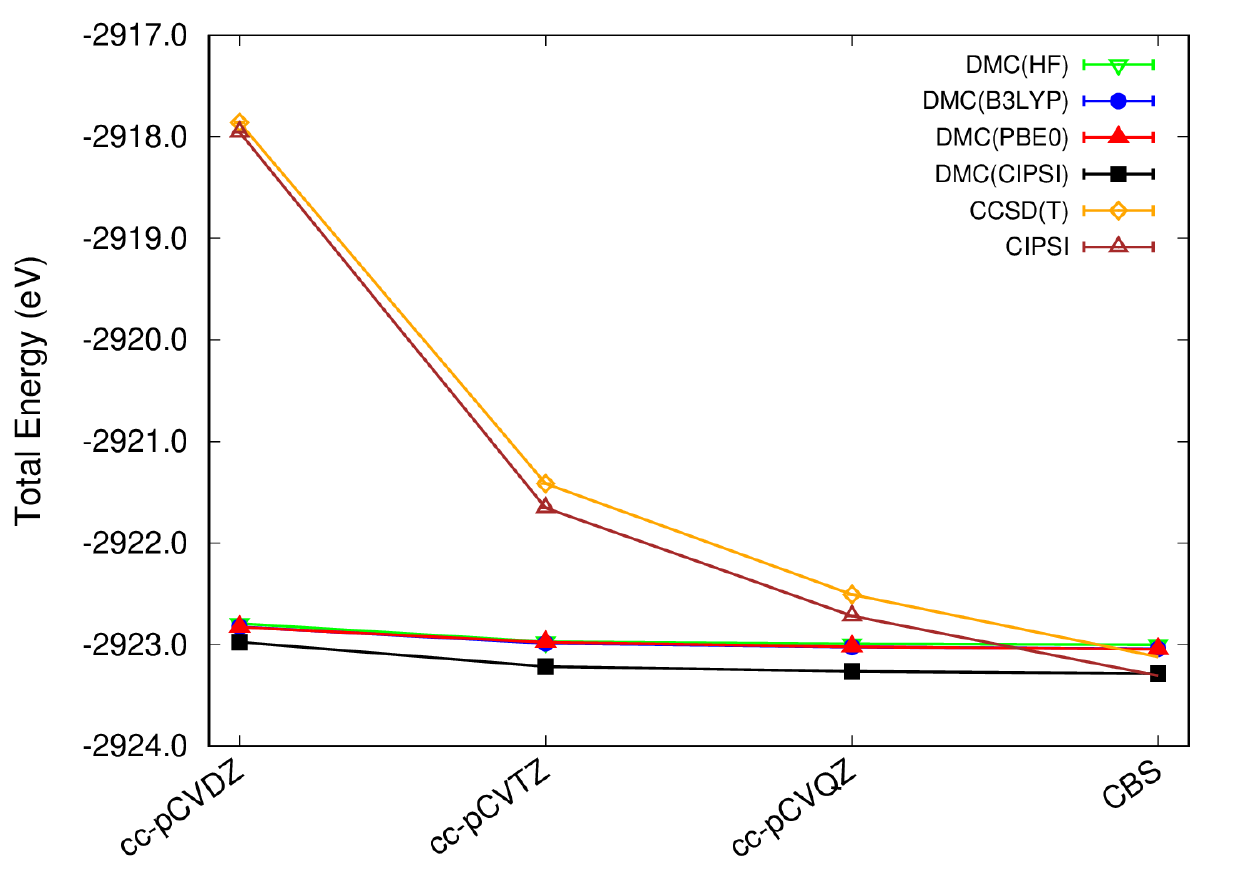}
        \caption{Ground state total energies (eV) of molecular LiF for DMC(HF), DMC(B3LYP), DMC(PBE0), DMC(CIPSI) CCSD(T) and
                converged CIPSI using cc-pCVNZ basis sets where N=D,T,Q and extrapolate to the CBS limit. All single-determinant DMC curves are on top of each
                other.}
        \label{fig:LiF}
\end{figure}

While the total energies at each basis set of ground state calculations and cation calculations are different, their trends are identical and we therefore only show figures representing the ground state. Figure-\ref{fig:LiF} shows the ground state DMC total energies of LiF,  computed using various trial wavefunctions;
single-determinant such as Hartree-Fock (HF), DFT's PBE0 and B3LYP hybrid functionals and  multi-determinant using the converged
CIPSI trial wavefunction. CCSD(T) and CIPSI energies are added to the figure for reference and all calculations are performed for
3 basis-sets increasing in size (cc-pCVNZ, N=D,T,Q) and extrapolated to the complete basis-set limit (CBS). The trends of CCSD(T) and CIPSI total energies are in agreement with each other to the CBS limit. CIPSI calculations recover more correlation energy $\sim$0.24eV for the ground state and $\sim$0.13 eV for the
cation. This is to be expected as CCSD(T) includes singles, doubles and perturbative triples excitations while CIPSI wavefunction
includes up to 9th order excitations with more than $70\%$ describing quadruple excitations ($10\%$ describing higher order
excitations) for both ground and cation states. Interestingly, at the CBS limit, CIPSI total energy converges to the same limit as
the DMC(CIPSI) energy while the CCSD(T) converges to the same energy as the SD-DMC energies. \deleted{(xxxPAUL: I did not expect this at all! we could try to expand but this is probably not the place. Please advise)}

\begin{figure}[h]
\centering
\includegraphics[scale=1.0,angle=0]{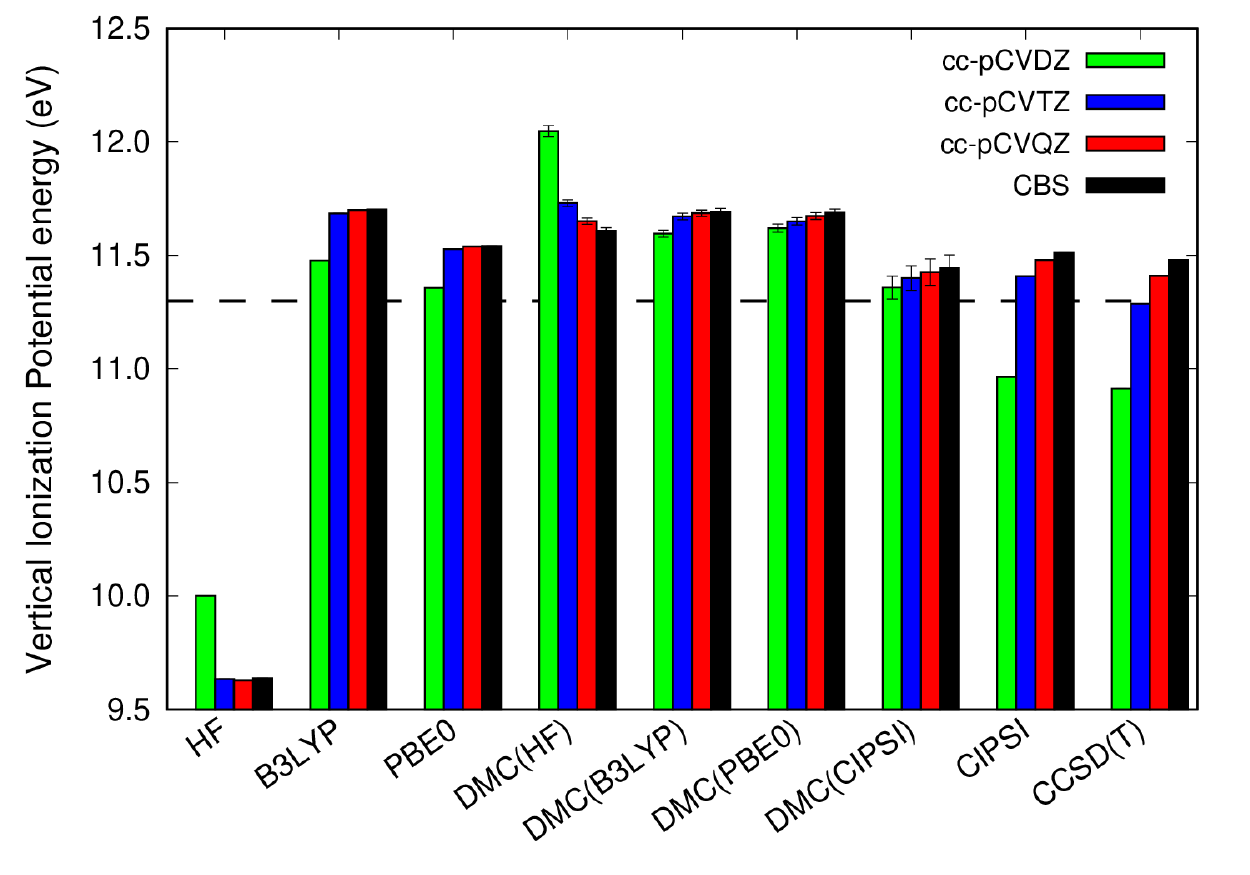}
\caption{Vertical ionization potential of LiF using different methods and trial wavefunctions. The dashed line corresponds experiment.\cite{Berkowitz1962}}
\label{fig:LiF-VIP}
\end{figure}

At the DMC level of theory, the dependence on basis-set is rather weak; less than $\sim$10 meV in the worst case. In the LiF
molecular case, the nodal surface of all tested single-determinant trial wavefunction are within error bars of each other, meaning
they are essentially the same. Such weak dependence on the starting method and on the basis set are a a significant advantage and
strength of the method when compared to other methods such as sCI or even AFQMC. The use of CIPSI-based trial wavefunctions in DMC
allows the recovery of $~0.24eV$ for the ground state and $~0.5eV$ for the cation. This difference underlines the different
sensitivity of the nodal surface to excited and charged states. The vertical ionization potential (VIP),
$E_{VIP}=E_{cation}-E_{ground}$, DMC performed using CIPSI wavefunctions shows almost no dependency to the basis set size,
Fig.~\ref{fig:LiF-VIP} while DMC(CIPSI), CIPSI and CCSD(T) are in perfect agreement at the CBS limit, demonstrating good error compensation for the latter method.

\subsection{Solid-state Lithium Fluoride}
Solid LiF is a face centered cubic material with a large gap, used mainly in electrolysis for his role in facilitating the
formation of an Li-C-F interface on the carbon electrodes\cite{Aigueperse2000}. The purpose of this example is to demonstrate
basis set effect and the systematic convergence of DMC energy with the number of determinants (a paper
demonstrating convergence to the thermodynamic limit is in preparation). We simulated a cell of ${(LiF)}_2$ (4 atoms per cell) at
the Gamma point using correlation consistent electron-core potentials (ccECP)\citet{Bennett2017,Wang2019} described in
Sec-\ref{sec:ecps} and the cc-pVDZ, cc-pVTZ and cc-pVQZ basis set associated with the ccECPs.  PySCF, Quantum Package and QMCPACK
are able to simulate all shapes of cells with both real and complex wavefunctions, corresponding to any possible k-point. In this
case, running at Gamma point is simply for convenience.

\begin{figure}[h]
\centering
\includegraphics[scale=1.0,angle=0]{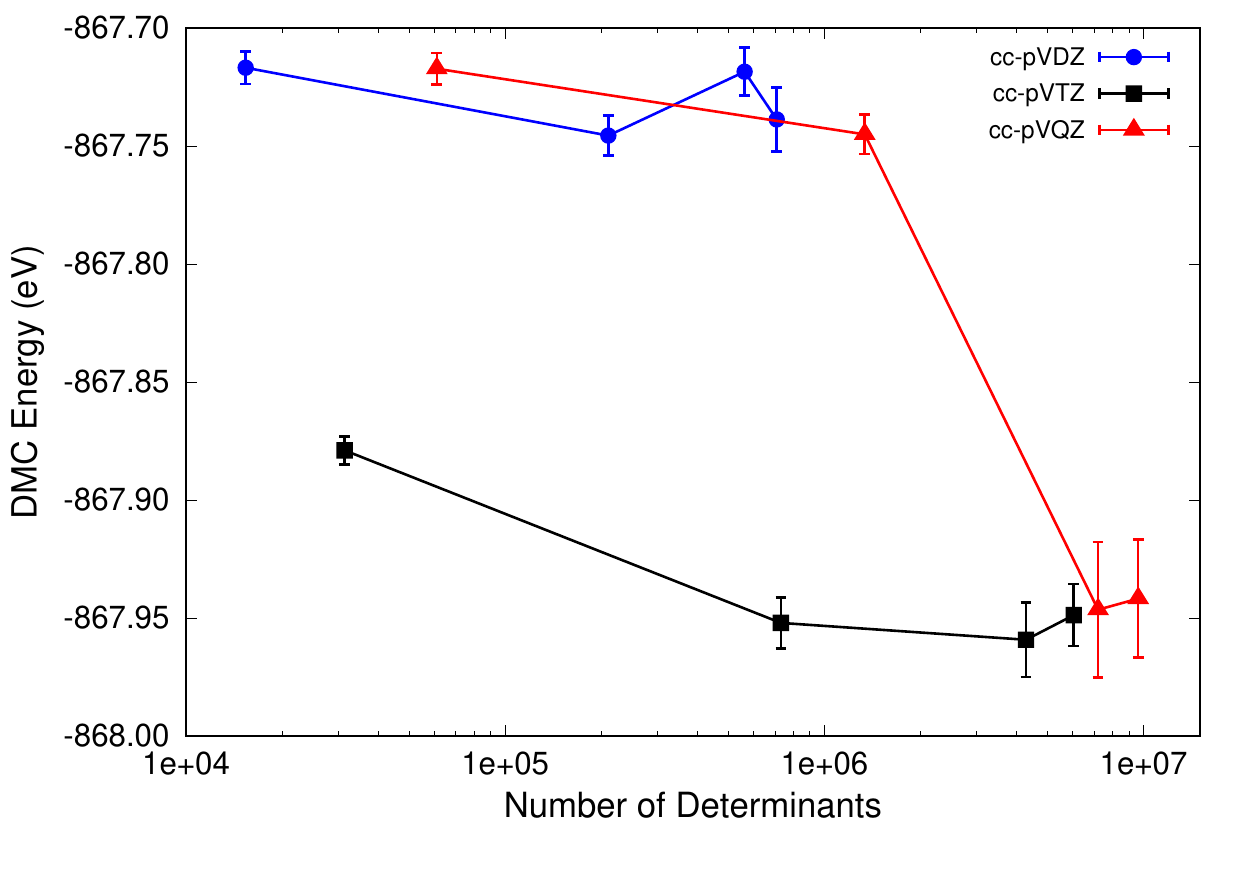}
\caption{Convergence of DMC energies of solid LiF using for different basis sets with respect to the number of determinants.}
\label{fig:LiF-PBC-DMC}
\end{figure}

For such a small simulation cell, it is possible to convergence the sCI wavefunction to the FCI limit with a reasonable number of
determinants, as can  be seen in Fig-\ref{fig:LiF-PBC-DMC}. The number of determinants needed to reach approximate convergence
remains important: around 700K in cc-pVDZ, 6M in cc-pVTZ and 9M in cc-pVQZ. Similarly to the molecular case, in the
converged energies for the cc-pVTZ and cc-pVQZ basis sets are in agreement, indicating that the basis set is sufficiently
convergence. Interestingly, in the cc-pVTZ case the DMC energy converges significantly faster with the number of determinants
(700K instead of 6M). The slower convergence of the cc-pVQZ curve indicates that important determinants describing relevant static
correlations are introduced late in the selection process. Using natural orbitals  or in general, an improved choice of orbitals
or selection scheme could accelerate the convergence. 

\subsection{Solid-state band gap calculations}
\label{sec:bandgaps}
The band gap of a solid is a critical and fundamental property of a material to predict accurately. QMC calculations for solids
have traditionally used completely independent calculations for ground and excited states and single determinant calculations.
This approach can be accurate but it relies on good error cancelation between the calculated total energy for each state, making
the selection of consistently accurate trial wavefunctions critical. Improved methods are needed to enforce good error cancelation
including approaches that can be systematically converged to give in-principle exact results.

As discussed above, convergent wavefunctions and energies can be constructed using sCI techniques. However, even for small
primitive cells with relatively uncorrelated electronic structures, this approach quickly requires millions of determinants making
it expensive to apply today. We have developed theories, methods and implementations to obtain the band-edge wavefunctions around
the fundamental gap and their relative energies efficiently and to a high accuracy. Error cancelation is built into the
methodology so that simpler trial wavefunctions are effective and the scheme is substantially more efficient to apply.
Surprisingly, for the systems examined so far, only single and double excitations need be considered to obtain accurate band gaps,
even using the simple VMC method. This makes the technique comparatively cheap to apply.
  
To compute the optical band gaps of insulators and semi-conductors we use the energy difference of optimized wave functions that
describe the valence band maximum (VBM) and the conduction band minimum (CBM). Optimizations use the recently developed excited
state variational principle,\cite{Zhao2016,Shea2017,Robinson2017}
\begin{equation}
    \label{eqn:es_var}
    \Omega\left(\omega, \Psi\right)=\frac{\left<\Psi|\omega-H|\Psi\right>}{\left<\Psi|{\left(\omega-H\right)}^2|\Psi\right>}=\frac{\omega-E}{{\left(\omega-E\right)}^2+\sigma^2}
\end{equation}
whose global minimum is not the ground state but the eigenstate with energy immediately above the chosen value $\omega$, which
could be placed within the band gap to target the first excited state and thus predict the optical gap. QMCPACK evaluates $\Omega$
via variational Monte Carlo (VMC) method to avoid explicit dealing with the $H^2$ term, and minimizes it using the linear method.
For ground state, we include the closed-shell determinant built from Kohn-Sham (KS) orbitals, plus all single-particle-hole
excitations, which represents the leading-order terms of orbital rotation that transforms KS orbitals to the ones that minimizes
$\Omega$ in the presence of the Jastrow factor. For the excited state, we include all the single-particle-hole excitations as in
Bethe-Salpeter equation (BSE)\cite{Rubio2002} methods as well as selected double excitations to capture the re-polarization of the
electron cloud in the vicinity of the exciton. We use the variance of the wavefunctions as a proxy for accuracy, and by varying
the number of determinants choose ground and excited states with consistent variance.

\begin{figure}[h]
\centering
\includegraphics[scale=0.5,angle=0]{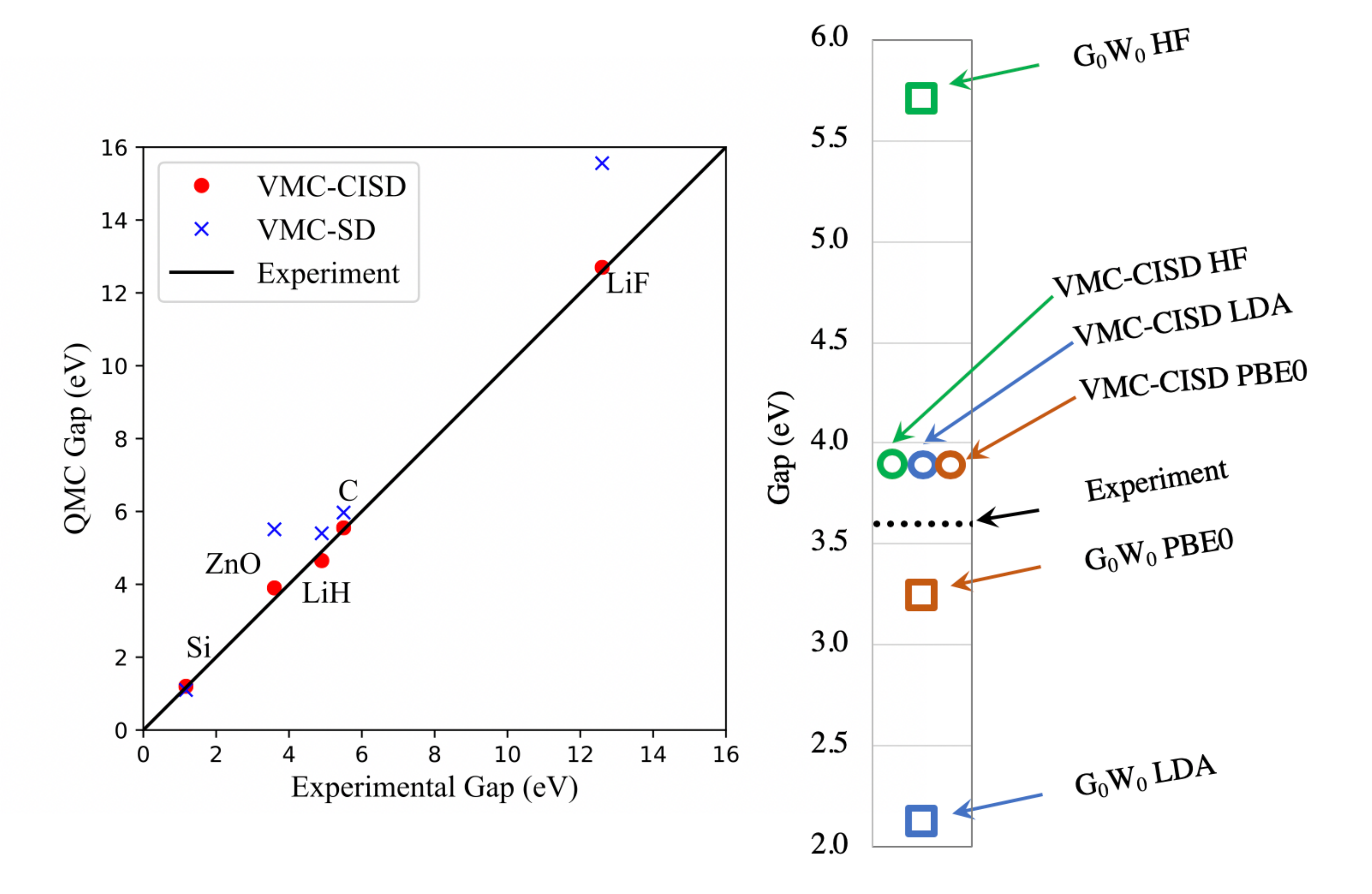}
\caption{Left: VMC optical gap predictions plotted against experimental results. The VMC wavefunctions are constructed via a
configuration interaction singles doubles expansion (VMC-CISD) or with single determinants (VMC-SD). Right: Comparison of optical gap of ZnO between
G$_0$W$_0$ using one-particle starting points that employ different fractions of exact exchange with VMC results based on the same
starting points. VMC data from Zhao,
\cite{Zhao2019}
G$_0$W$_0$ results are from Fuchs\cite{Fuchs2007} and experimental data is from
Lauck.\cite{Lauck2006}}\label{fig:money}
\end{figure}

We have used this approach in Ref.~\onlinecite{Zhao2019} to study optical gaps of a variety of solids ranging from small-gap semi-conductors to large gap
insulators and compare our results to the commonly used GW approach based on many-body perturbation theory (MBPT). As detailed in
the supplemental information \added{of} Ref.~\onlinecite{Zhao2019} the band gaps were extrapolated using calculations on 8, 16 and 24 atom supercells.
Figure~\ref{fig:money} shows that the predicted optical gaps are in excellent agreements with experimental values and the mean
absolute deviation (MAD) is just 3.5\%, compared to MADs more than twice this large for the optical gaps obtained by subtracting
the known exciton binding energy from G$_0$W$_0$ and self-consistent GW gaps. These calculations were able to run of departmental
level computing and did not require supercomputers due to the use of VMC and moderate sized supercells utilized. 
\added{Remaining errors that have yet to be investigated include the size of the CI expansion and the limited correlation energy obtained with VMC, the pseudopotentials, and residual finite size error. DMC may further improve the VMC results.}

In order to
further show the method's advantage, we performed a thorough analysis of zinc oxide, a material that is particularly challenging
for MBPT.\cite{Louie2010} As shown in Figure~\ref{fig:money}, the perturbative nature of G$_0$W$_0$ makes its prediction highly
sensitive to the amount of exact exchange included in the DFT reference. As G$_0$W$_0$ assumes a zeroth-order picture, in which
electronic excitations are simple particle-hole transitions between the one-particle eigenstates of DFT, such a sensitivity
indicates the break down of this assumption and then G$_0$W$_0$ becomes unreliable. On the other hand, our VMC approach is
designed to be insensitive to the DFT
choice for two reasons: (1) its ability to include approximate orbital relaxation counteracts the shortcomings of the DFT
orbitals. (2) unlike G$_0$W$_0$ it does not require orbital energies as input. From Figure~\ref{fig:money} we do find that its
prediction to optical gap is both accurate and independent to the choice of DFT functionals. 

\subsection{Summary}
Using DMC as a post-sCI method is very promising to systematically improve molecular and solid-state calculations beyond the single-determinant picture. It converges faster the sCI methods used on their own. From a practical perspective,
PySCF, Quantum Package and QMCPACK are fully interfaced with each other through the Nexus workflow automation system. The
necessary multi-step workflow to run the above examples is fully implemented. In the case of solids, Nexus can automatically
manage finite-size scaling calculations by setting the size of the super-cells, the number of twists angles, and drive PySCF,
Quantum Package, and QMCPACK appropriately and automatically.

\section{Applications}\label{sec:applications}
To illustrate the application of recent developments in QMCPACK, in Section~\ref{sec:anions} we give an example of using real space QMC to study non-valence anions, which are particularly challenging systems. Section~\ref{sec:excited_defect} gives an example of computing the excited states of localized defects, which is a challenge for all electronic structure methods. In Section~\ref{sec:manybody}
we give an example of computing the momentum-distribution and Compton profile from real space methods. The necessary estimators
have recently been specifically optimized for these tasks.

\subsection{Applications of DMC to Non-valence Anions}\label{sec:anions}
In addition to valence anions, molecules and clusters can posses non-valence anions in which the binding of the excess electron is dominated by a combination of long-range electrostatics and long-range dispersion-type correlation effects. 
The best known class of non-valence anions are dipole-bound anions, in which the binding of the excess electron is driven by the dipole field of the neutral.\cite{Fermi1947,Wallis1960,Crawford1967,Turner1968,Simons1987,Jordan2003}
Non-valence anions, regardless of the nature of the long-range interaction responsible for the electron binding, are challenging to treat using traditional electronic structure methods due to the large, highly diffuse basis sets required. 
Both DMC and AFQMC methods have been demonstrated to be useful in characterizing dipole-bound anions.\cite{Xu2010,Hao2018}
Non-valence anions in which electron correlation effects dominate the binding of the excess electron pose an additional challenge, namely, by definition, they do not bind the excess electron in the Hartree-Fock (HF) approximation.
In fact, HF calculations on such excess electron systems collapse on to the neutral molecule leaving the excess electron in a discretized continuum orbital.\cite{Voora2017}
Hence, methods that start from the HF wave function, e.g., MP2 and CCSD(T),\cite{Raghavachari1989} also fail to bind the excess electron.
Many-body methods such as equation-of-motion coupled cluster (EOM-CC),\cite{EOMCC1,EOMCC2,EOMCC3,EOMCCSDT} have proven successful in treating these species, but still face the problem of requiring very large basis sets.

This raises the question of whether DMC calculations using a single Slater determinant trial wave function to define the nodal surface can accurately describe non-valence correlation-bound (NVCB) anions.
To investigate this, we have undertaken DMC calculations on a (H$_2$O)$_4$ model with the monomers arranged so that the net dipole is zero.\cite{Voora2017}
This model has been studied previously using EOM-CC methods.
In the present work, we focus on a geometry at which the excess electron does not bind in the HF approximation, but for which EOM-EA-CCSDT\cite{EOMCCSDT} calculations using the aug-cc-pVDZ basis set\cite{accd1,accd2} augmented with a 7s7p set of diffuse functions located at the center of the molecular cluster give an electron binding energy (EBE) of 174 meV.\cite{Voora2017}

\begin{figure*}[!ht]
  \includegraphics[width=\textwidth]{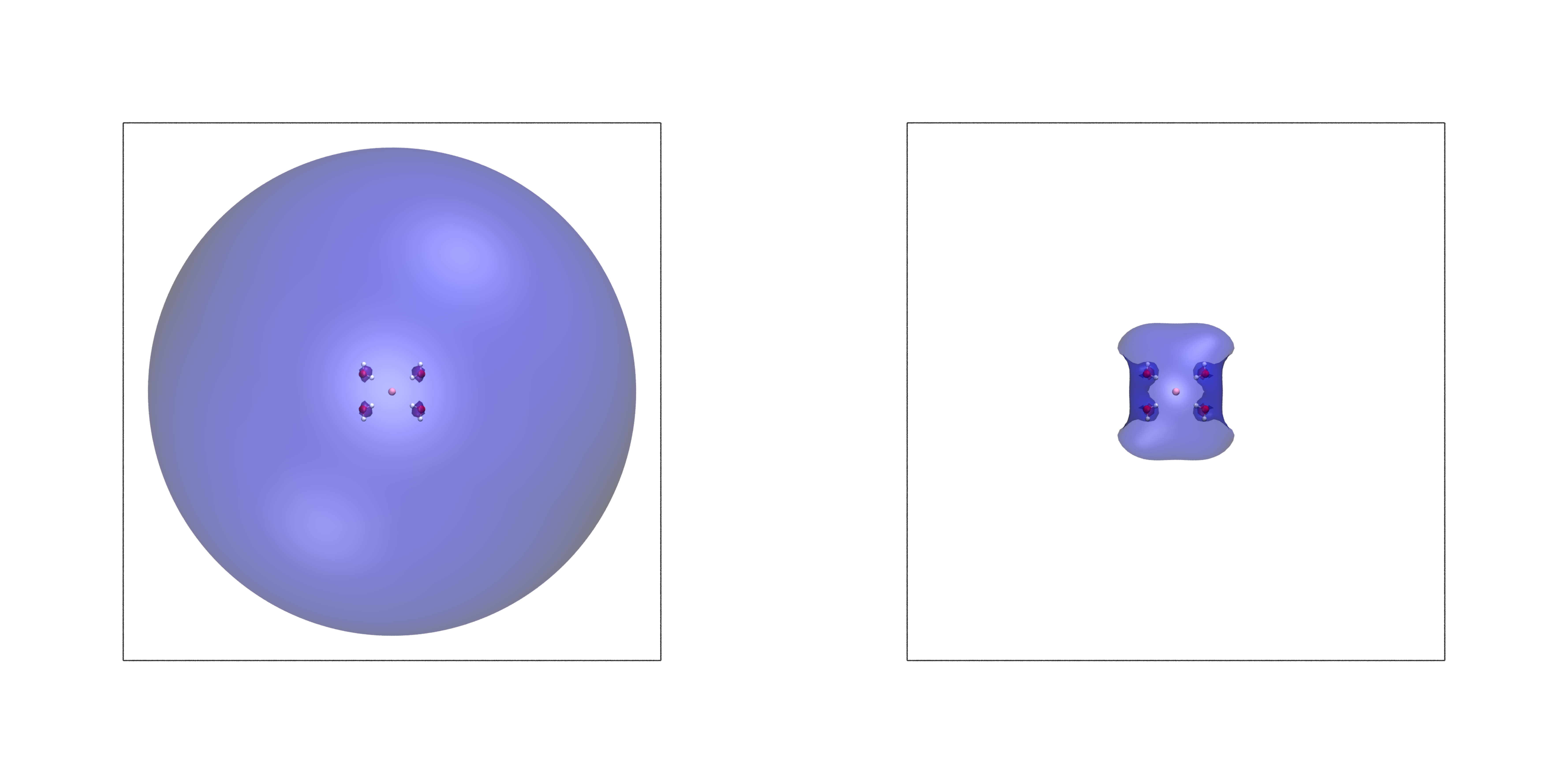}
  \caption{\label{Fig:anions} The singly occupied orbital from (left) HF and (right) B3LYP calculations on the anion of the (H$_2$O)$_4$ cluster model. These were carried out using the ccECP and the corresponding aug-cc-pVDZ basis set, augmented with a set of $7s7p$ diffuse functions centered at the origin. The bounding box is $100$ a.u. on each side, and the isosurface value is set to 0.0005 to enable comparison between the images. The highly diffuse orbital from the HF calculation is actually describing an approximation to a continuum function in the finite Gaussian basis rather than the orbital appropriate for the anion.}
\end{figure*}

For the QMC calculations, Slater-Jastrow trial wave functions that are products of a single Slater determinant comprised of HF or DFT orbitals and a Jastrow factor were employed.
All calculations were carried out using the ccECP pseudopotentials with the corresponding aug-cc-pVDZ type basis sets\cite{Bennett2017,Annaberdiyev2018} augmented with the same 7s7p set of diffuse functions as employed in ref~\onlinecite{Voora2017}.
The DMC calculations were carried out at three imaginary time steps (0.001 Ha$^{-1}$, 0.003 Ha$^{-1}$, 0.005 Ha$^{-1}$), and a linear extrapolation was performed to obtain the zero time step limit.
HF calculations with this basis set fail to bind the excess electron, and a plot (see Fig.~\ref{Fig:anions}) of the singly occupied orbital of the excess electron system reveals that it is very diffuse because it has collapsed onto the lowest ``continuum'' solution as described by the discrete basis set.
The QMC calculations were carried out with the QMCPACK code.

The Jastrow factors used in the trial wave functions included electron-electron, electron-nuclei, and electron-electron-nuclei terms.
Normally, one would optimize the parameters in the Jastrow factors separately for the neutral and for the anion.
However, this approach would not give meaningful results for an unbound anion, and, as a result, we adopted the strategy of using the Jastrow factor optimized for the neutral in the subsequent DMC calculations of the anion.
The DMC calculations using trial wave functions determined in this manner give an EBE of 183$\pm$10 meV, in good agreement with the previous EOM-EA-CCSDT result (174 meV).\cite{Voora2017}
This demonstrates that DMC calculations can recover from the use of a trial wave function for the anion that has collapsed onto a discretized continuum solution.

Even so, there remains the question of whether the EBE obtained from DMC calculations that use an unphysical (i.e., collapsed onto the continuum) trial wave function could incur an appreciable error due to an inadequate description of the nodal surface of the anion.
To address this question, we also carried out VMC and DMC calculations on the neutral and anion of the (H$_2$O)$_4$ cluster model using orbitals from B3LYP\cite{B3,LYP,VWN} calculations employing the same pseudopotential and basis sets as used for the HF calculations.
The anion is bound by 395 meV in the B3LYP calculations, and the singly occupied orbital of the excess electron system, while still diffuse, is localized much closer to the molecule than in the HF calculations.
A comparison of the charge distributions of the singly occupied orbitals from the HF and DFT calculations (see Fig.~\ref{Fig:anions})  shows that the DFT charge distribution is much less spatially extended.
Moreover, it more closely resembles the charge distribution of the relevant natural orbital from the EOM calculation of Ref.~\onlinecite{Voora2017}.
(The over-binding of the anion in the B3LYP calculations is likely due to the finite extent of the integration grid.)
The DMC calculations using trial wave functions derived from B3LYP orbitals give an EBE of 212$\pm$11 meV, $\sim$ 29 meV larger than the DMC result obtained using HF orbitals.
This increase indicates that employing a trial wave function with a more physical charge distribution for the singly occupied orbital of the anion does have an impact on the nodal surface for the exchange of the electrons.
The EOM-EA-CCSDT result obtained using the aug-cc-pVDZ+7s7p basis set is 38meV smaller than the DMC EBE obtained using the trial wave function employing B3LYP orbitals.
This suggests that the EBE from EOM calculations may not be fully converged in these large basis sets.
Overall, these results demonstrate that DMC is a viable approach for the characterization of NVCB anions.

\subsection{Excitation energies of localized defects}\label{sec:excited_defect}
For defects and interfaces, most ab-initio methods can only achieve qualitative agreement on the optical properties. 
We have recently studied emission energies of Mn$^{4+}$-doped solids using DMC, which is chosen as proof of principle
\cite{SaritasExcitation2018}. We show that our approach is applicable to similar systems, provided that the excitation is
sufficiently localized. In support of this work, Nexus scripts and new tutorials on excited state calculations were developed that
can be applied to any gapped system (Lab 5 in the QMCPACK manual).

Multivalent ionic defects, such as Mn$^{4+}$, can create multiple localized electronic states that are trapped within the band gap of wide gap materials. 
Thus, luminescent centers are created in the dopant sites through radiative recombination. 
Mn$^{4+}$ has $d^3$ electronic configuration all on the $t_{2g}$ orbitals. The ground state is in $t_{2g}^{\uparrow\uparrow\uparrow}$ ($^4$A$_{2g}$) configuration due to Hund's rules, but the excited state is found to be as $t_{2g}^{\uparrow\uparrow\downarrow}$ ($^2$E$_g$) \cite{Brik2013}. Therefore, the emission energy is simply defined as $E_{em}=E(^2$E$_g)-E(^4$A$_{2g})$. 

\begin{figure*}[t]
  \includegraphics[width=\textwidth]{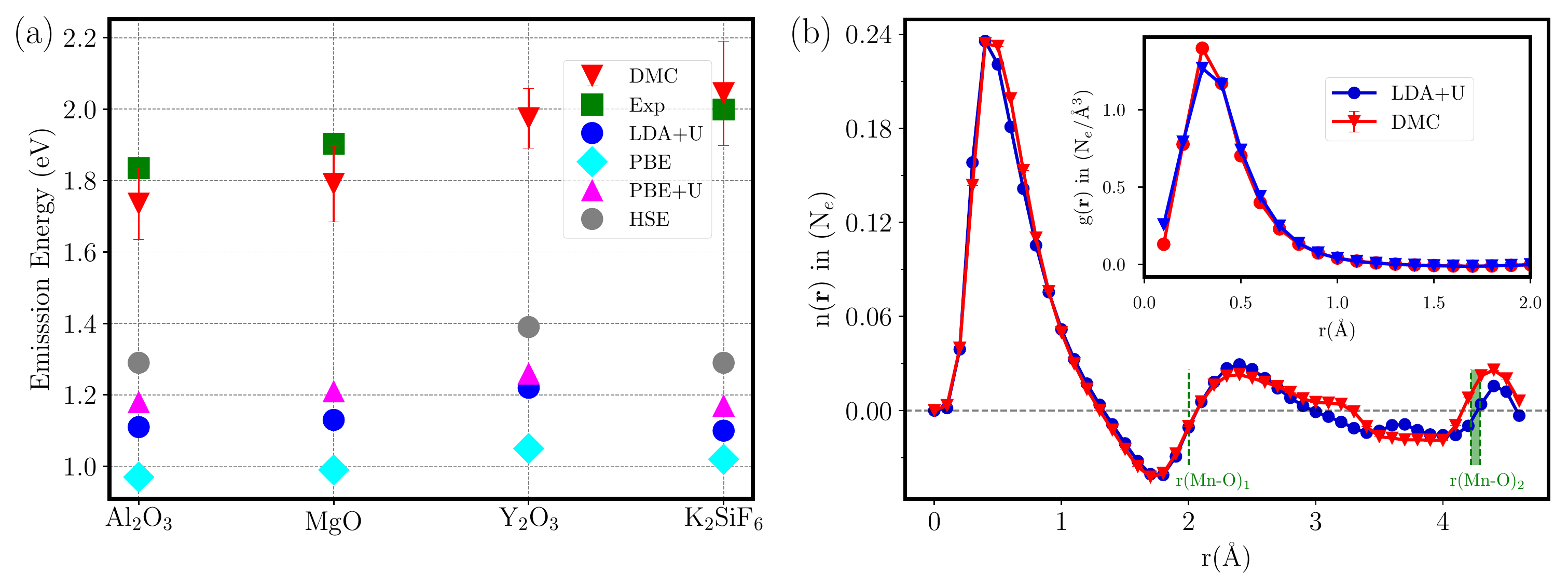}
  \caption{\label{Fig:phosphor} (a) Emission energies of Mn$^{4+}$ doped host compounds (on x-axis). (b) \replaced{Number of electrons per radial volume}{Radial density of the excited electron}, $n({\bf r})$, around the Mn atom. The inset shows the radial distribution function $g({\bf r})$ for the same density. \replaced{Number of electrons per radial volume and the radial distribution}{Radial density and the radial distribution} functions are related as $n({\bf r}) = g({\bf r})4{\pi}r^2dr$. Reproduced from Ref. ~\onlinecite{SaritasExcitation2018}.}
\end{figure*}

Fig. \ref{Fig:phosphor}(a) shows that DMC can reproduce experimental emission energies of Mn$^{4+}$ doped insulating host
compounds \cite{Geschwind1962, Dunphy1990, Brik2012a}. DFT+$U$ and hybrid-DFT, however, substantially underestimate. Relative
quantitative success of HSE with respect to PBE (or DFT+$U$) indicates that emission energies might be reproduced with a larger
portion of exact exchange. However, this would worsen the accuracy of the computed band gaps in the host compounds
\cite{SaritasExcitation2018}. In Fig. \ref{Fig:phosphor}(b), we show the spin flipped \replaced{electrons per radial volume}{electron density} $n({\bf r}) =
\rho_{ground}^{\uparrow}-\rho_{excited}^{\uparrow}$ which is spherically integrated around the Mn$^{4+}$ atom. $n({\bf r})$
approaches to zero with increasing radius indicating that excited electron density is strongly localized on the impurity atom. DMC
and LDA+$U$ $n({\bf r})$ densities are almost identical to each other despite the large difference in their emission energies
which underscore the difficulties that needed to be overcome for better DFT functionals \cite{Medvedev2017}.

\subsection{Calculation of the many-body properties: the momentum distribution}\label{sec:manybody}
As full-many body methods, QMC can be used to calculate many-body properties that can not be readily obtained from
single-particle or mean-field techniques. We have recently updated and optimized calculation of the momentum distribution. 

Experimentally the momentum distribution function (MDF) can be accessed, e.g., via angle-resolved photoemission spectroscopy and
scattering methods such as Compton scattering, positron annihilation, the $(e,2e)$ process, and high energy electron
scattering~\cite{Lindner1977,Williams1977,Cooper1985,Bell2001}. In general, the differential cross sections of the scattering can
be related to the momentum distribution. These experimental techniques are powerful probes for understanding subtle details in the
ground state properties of materials, which are manifested in the MDF. 

In normal Fermi liquids, the electron MDF has a discontinuity at the Fermi momentum $p_F$. In three-dimensional systems this
discontinuity defines the shape of the Fermi surface, which is also related to the screening properties of the
electrons~\cite{Dugdale2016}. The Fermi surface can be extracted from the $\mathbf{p}$-space MDF via back-folding~\cite{Lock1973}.
This leads to occupation density within the first Brillouin zone from which the Fermi surface topology can be
considered~\cite{Dugdale2006}. 

The magnitude of the discontinuity at the Fermi surface, however, quantifies the strength of a quasiparticle excitation and is
generally referred to as the renormalization factor~\cite{Huotari2010,Holzmann2011}. For strongly coupled systems the
renormalization factor tends to zero as the coupling strength increases, and thus, it provides an estimate for the strength of
electron correlations. Interestingly, the discontinuity at the Fermi momentum disappears for superfluids or superconducting
materials. For insulators the discontinuity is absent and the sharp drop is noticeably broadened, which also holds true for some
semi-metals~\cite{MomDist,Hiraoka2017}. Even small scale charge density oscillations lead to clear signatures in the
MDF~\cite{Kylaenpaeae2019}. Therefore, the momentum distribution function provides complementary and informative knowledge to
other characterizations of many-body systems. 

The MDF, $n(\mathbf{p})$, is obtained by taking the Fourier transform of the one-body density matrix:
\begin{align}
  n(\mathbf{p}) =\frac{N_e}{\Omega}\int dR d\mathbf{r}_1'
  ~e^{i(\mathbf{r}_1-\mathbf{r}_1')\cdot \mathbf{p}}
	\rho(\mathbf{r}_1,\ldots,\mathbf{r}_{N_e},\mathbf{r}_1',\ldots,\mathbf{r}_{N_e})
   =\frac{N_e}{\Omega}\int d\mathbf{s} ~e^{-i\mathbf{p}\cdot \mathbf{s}}n(\mathbf{s}),
  \notag
\end{align}
where $\Omega$ is the volume containing $N_e$ electrons, $R=\{\mathbf{r}_1,\ldots,\mathbf{r}_{N_e}\}$, $\mathbf{s}=\mathbf{r}_1'-\mathbf{r}_1$, and
\begin{align}
n(\mathbf{s}) = \int dR
	~\rho(\mathbf{r}_1,\ldots,\mathbf{r}_{N_e},\mathbf{r}_1+\mathbf{s},\ldots,\mathbf{r}_{N_e}).
\notag
\end{align}
In variational Monte Carlo this is expressed as
\begin{align}
n(\mathbf{s}) & = \int dR 
	~\Psi^*(\mathbf{r}_1,\ldots,\mathbf{r}_{N_e})\Psi(\mathbf{r}_1+\mathbf{s},\ldots,\mathbf{r}_{N_e})
 =\int dR~|\Psi(R)|^2 \frac{\Psi(R')}{\Psi(R)}
\notag,
\end{align}
where $R'=\{\mathbf{r}_1+\mathbf{s},\ldots,\mathbf{r}_{N_e}\}$. 
Thus, we get for the MDF:
\begin{align}
  n(\mathbf{p})& = \int dR~|\Psi(R)|^2
  \frac{N_e}{\Omega}\int d\mathbf{s} \frac{\Psi(R')}{\Psi(R)}e^{-i\mathbf{p}\cdot \mathbf{s}}.
  \notag
\end{align}
In practice, the Monte Carlo estimate for the MDF with $N_s$ samples is given by
\begin{align}
  \label{mom_dist_practice}
  n(\mathbf{p})& = \left\langle\frac{1}{\Omega N_s}\sum_{i=1}^{N_s}\sum_{j=1}^{N_e}\frac{\Psi(R+\mathbf{s}^i_j)}{\Psi(R)} e^{-i\mathbf{p}\cdot \mathbf{s}^i_j}\right\rangle_{|\Psi(R)|^2},
\end{align}
where $R$ includes the coordinates of all the electrons, and $\mathbf{s}^i_j$ is a displacement vector acting on the $j^{\rm th}$ electron of the $i^{\rm th}$ sample.
In diffusion Monte Carlo calculations the mixed distribution replaces $|\Psi(R)|^2$ , and additional measures must be taken to calculate or estimate the density matrix.
Notice that the momentum distribution normalizes to the number of electrons 
\begin{align}
\label{Ne}
\sum_{\mathbf{p}}n(\mathbf{p}) = N_e = \frac{\Omega}{{(2\pi)}^d}\int d\mathbf{p} ~n(\mathbf{p}),
\end{align}
in which $d$ refers to dimensionality. In Eq.~\eqref{Ne} a finite system and a system at the thermodynamic limit are described by
summation and integration, respectively. 

The MDF estimator is a one-body density matrix based estimator with very high computational cost resulting from the large number of
wavefunction evaluations required. A naive implementation can easily double the cost of a QMC calculation.
Thus, efficient algorithm and implementations are critical, and similar techniques can be used for related estimators.

In Eq.~\eqref{mom_dist_practice}, the computation of wavefunction ratios $\frac{\Psi(R+\mathbf{s}^i_j)}{\Psi(R)}$ and phase factor $e^{-i\mathbf{p}\cdot \mathbf{s}^i_j}$ are both expensive.
Direct $N_e N_s$ times of calculation is easy to implement but has a lot of repeated effort.
In the $\frac{\Psi(R+\mathbf{s}^i_j)}{\Psi(R)}$ term, the evaluation of single particle orbitals at $\mathbf{r}_j + \mathbf{s}^i_j$ dominates the cost.
In fact, its call count can be reduced from $N_s N_e$ to $N_s$ by making
\begin{align}
\mathbf{s}^i_j = \mathbf{r}'_i - \mathbf{r}_j \label{eq:separate_terms}
\end{align}
where $\mathbf{r}'_i$ is the electron coordinates of sample $i$. Although the leading cost of $\frac{\Psi(R+\mathbf{s}^i_j)}{\Psi(R)}$ is
optimized away, its remaining terms still scale as $O(N_s N^2_e)$ and the computational cost should be comparable to the non-local
pseudopotential calculation.

Unlike the wavefunction ratios which needs to be computed only once for all the $\mathbf{p}$, phase factors are computed for each
$\mathbf{p}$ and also take significant portion of time.
Similarly, for each $\mathbf{p}$, the number of evaluations can be reduced to $N_s + N_e$ times by separating indices $i$ and $j$
in two terms like Eq.~\eqref{eq:separate_terms}. The calculation of $e^{-i\mathbf{p}\cdot \mathbf{r}'_i}$ and
$e^{-i\mathbf{p}\cdot \mathbf{r}_j}$ can be efficiently vectorized using the single instruction multiple data (SIMD) unit in modern
processors.

By applying the above techniques and ensuring vectorization of all operations, the overhead for evaluating the MDF in a 48
atom cell VO$_2$ was reduced from additional 150\% to only 50\% cost increase compared to a DMC run without any estimators.

Within the so-called impulse approximation (IA) the Compton profile as well as the dynamical structure factor are proportional to the projection of $n(\mathbf{p})$ onto a scattering vector~\cite{MomDist,Zambelli2000}. In this case directional Compton profile in $z$-direction would be expressed as
\begin{align}
  J(q)=\frac{\Omega}{{(2\pi)}^3}\int\!\!\!\!\int n(p_x,p_y,p_z=q) dp_x dp_y.\notag
\end{align}
The IA is especially appropriate for X-ray Compton scattering from electronic systems~\cite{Cooper1985,MomDist}, and thus, it is capable of providing a unique perspective for understanding the electronic structure of materials; bulk properties, in particular.

\begin{figure*}[t]
  \includegraphics[width=\textwidth]{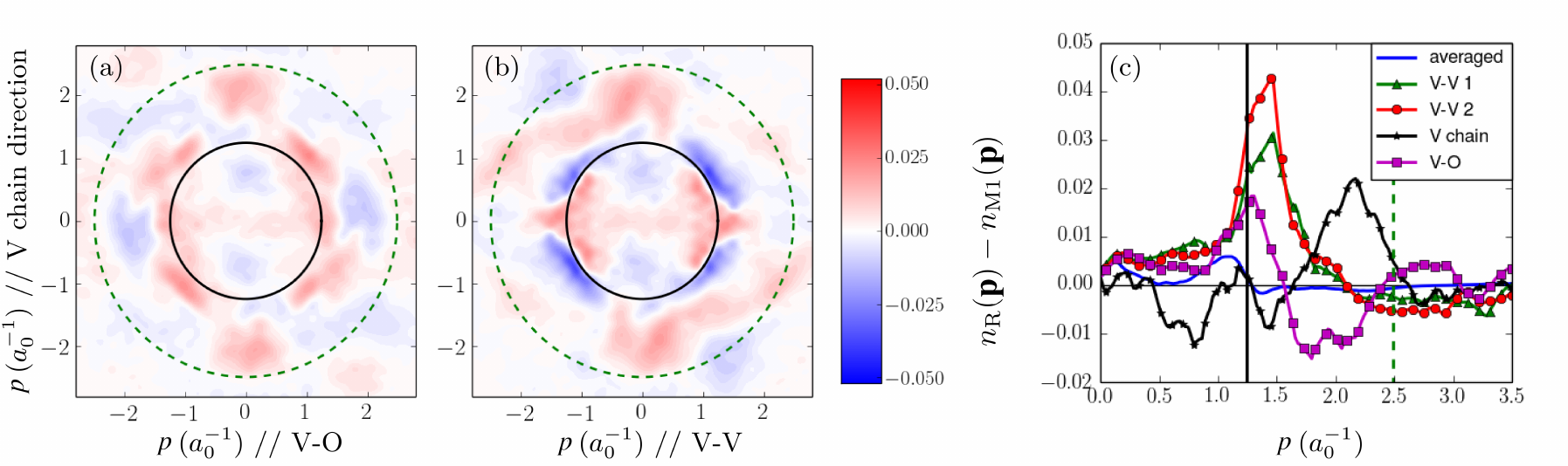}
  \caption{\label{Fig:np_diff}QMCPACK result for the difference in the momentum distribution of VO$_2$ across the metal-insulator phase transition based on Ref.~\onlinecite{Kylaenpaeae2019}. (a) and (b) show two planes of the 3D difference profile. In (c) the differences are given in four different directions and also for the angular averaged MDF. For more details, see Ref.~\onlinecite{Kylaenpaeae2019}.}
\end{figure*}

In Ref.~\onlinecite{Kylaenpaeae2019} QMCPACK was used in obtaining the MDFs and Compton profiles for VO$_2$ across its metal
insulator transition from metallic rutile (R) phase to insulating monoclinic (M1) phase. There the analysis of the MDF shows
signatures of the non-Fermi liquid character of the metallic phase of vanadium dioxide. Moreover, findings therein provide an
explanation for the experimentally observed anomalously low electronic thermal conductivity~\cite{Lee2017}, which manifests as
back scattering characteristics within the momentum distribution function. Fig.~\ref{Fig:np_diff} shows some examples of MDF
differences across the phase transition in two planes as well as for a few different directions~\cite{Kylaenpaeae2019}. 

\section{Summary}\label{sec:summary}

We have described recent enhancements to the open source QMCPACK package. Besides increases in capability for both real space and
auxiliary field Quantum Monte Carlo (QMC) methods, the surrounding ecosystem has also been improved. These enhancements include
the workflow system Nexus, which aims to reduce the complexity of performing research studies and the tens to hundreds of
individual calculations that might be entailed. A new set and open database of effective core potentials
has also been established at \url{https://pseudopotentiallibrary.org}, and we expect that these will be of interest for other
quantum chemical and many-body calculations due to their increased accuracy, including for stretched bonds and excited states. We
have also described how improvements in open software development have benefitted the project. Besides the activities described in
this article, we note that there is substantial
ongoing work to enhance the architecture of  QMCPACK for GPU accelerated machines and to obtain portable
performance from a single code base. Once the new design is proven on diverse GPUs it will be described in a future article.

Overall the applicability of QMC continues to expand, it is becoming easier to apply, and there are many
systems and phenomena where the higher accuracy and many-body nature of QMC is both warranted and can now be applied. We
hope that this article will help encourage these new applications.

\section*{Data availability}
The data that support the findings of this study are available from the corresponding author upon reasonable request.

\begin{acknowledgments}
Methodological development and scientific applications of QMCPACK are currently primarily  supported by the U.S. Department of
Energy, Office of Science, Basic Energy Sciences, Materials Sciences and Engineering Division, as part of the Computational
Materials Sciences Program and Center for Predictive Simulation of Functional Materials. Software developments focused on future
Exascale architectures are supported by the Exascale Computing Project (17-SC-20-SC), a collaborative effort of the U.S.
Department of Energy Office of Science and the National Nuclear Security Administration. S.U. and K.D.J. acknowledge the support
of NSF grant CHE-1762337. B.R. acknowledges previous support for the work described from the U.S. Department of Energy by Lawrence
Livermore National Laboratory under Contract DEAC52-07NA27344, 15-ERD-013 and NSF grant DMR-1726213. An award of computer time was
provided by the Innovative and Novel Computational Impact on Theory and Experiment (INCITE) program. This research used resources
of the Argonne Leadership Computing Facility, which is a DOE Office of Science User Facility supported under contract
DE-AC02-06CH11357. This research also used resources of the Oak Ridge Leadership Computing Facility, which is a DOE Office of
Science User Facility supported under Contract DE-AC05-00OR22725. Sandia National Laboratories is a multi-mission laboratory managed and operated by National Technology and Engineering Solutions of Sandia LLC, a wholly owned subsidiary of Honeywell International, Inc. for the U.S. Department of Energy’s National Nuclear Security Administration under Contract No. DE-NA0003525.
This paper describes objective technical results and analysis. Any subjective views or opinions that might be expressed in the paper do not necessarily represent the views of the U.S. Department of Energy or the United States Government.
\end{acknowledgments}

%\bibliography{specialqmc}

\begin{thebibliography}{153}%
\makeatletter
\providecommand \@ifxundefined [1]{%
 \@ifx{#1\undefined}
}%
\providecommand \@ifnum [1]{%
 \ifnum #1\expandafter \@firstoftwo
 \else \expandafter \@secondoftwo
 \fi
}%
\providecommand \@ifx [1]{%
 \ifx #1\expandafter \@firstoftwo
 \else \expandafter \@secondoftwo
 \fi
}%
\providecommand \natexlab [1]{#1}%
\providecommand \enquote  [1]{``#1''}%
\providecommand \bibnamefont  [1]{#1}%
\providecommand \bibfnamefont [1]{#1}%
\providecommand \citenamefont [1]{#1}%
\providecommand \href@noop [0]{\@secondoftwo}%
\providecommand \href [0]{\begingroup \@sanitize@url \@href}%
\providecommand \@href[1]{\@@startlink{#1}\@@href}%
\providecommand \@@href[1]{\endgroup#1\@@endlink}%
\providecommand \@sanitize@url [0]{\catcode `\\12\catcode `\$12\catcode
  `\&12\catcode `\#12\catcode `\^12\catcode `\_12\catcode `\%12\relax}%
\providecommand \@@startlink[1]{}%
\providecommand \@@endlink[0]{}%
\providecommand \url  [0]{\begingroup\@sanitize@url \@url }%
\providecommand \@url [1]{\endgroup\@href {#1}{\urlprefix }}%
\providecommand \urlprefix  [0]{URL }%
\providecommand \Eprint [0]{\href }%
\providecommand \doibase [0]{http://dx.doi.org/}%
\providecommand \selectlanguage [0]{\@gobble}%
\providecommand \bibinfo  [0]{\@secondoftwo}%
\providecommand \bibfield  [0]{\@secondoftwo}%
\providecommand \translation [1]{[#1]}%
\providecommand \BibitemOpen [0]{}%
\providecommand \bibitemStop [0]{}%
\providecommand \bibitemNoStop [0]{.\EOS\space}%
\providecommand \EOS [0]{\spacefactor3000\relax}%
\providecommand \BibitemShut  [1]{\csname bibitem#1\endcsname}%
\let\auto@bib@innerbib\@empty
%</preamble>
\bibitem [{\citenamefont {Becca}(2017)}]{BeccaQuantum2017}%
  \BibitemOpen
  \bibfield  {author} {\bibinfo {author} {\bibfnamefont {F.}~\bibnamefont
  {Becca}},\ }\href
  {https://www.ebook.de/de/product/29282633/federico_becca_quantum_monte_carlo_approaches_for_correlated_systems.html}
  {\emph {\bibinfo {title} {{Quantum Monte Carlo Approaches for Correlated
  Systems}}}}\ (\bibinfo  {publisher} {Cambridge University Press},\ \bibinfo
  {year} {2017})\BibitemShut {NoStop}%
\bibitem [{\citenamefont {Martin}, \citenamefont {Reining},\ and\ \citenamefont
  {Ceperley}(2016)}]{MartinInteracting2016}%
  \BibitemOpen
  \bibfield  {author} {\bibinfo {author} {\bibfnamefont {R.~M.}\ \bibnamefont
  {Martin}}, \bibinfo {author} {\bibfnamefont {L.}~\bibnamefont {Reining}}, \
  and\ \bibinfo {author} {\bibfnamefont {D.~M.}\ \bibnamefont {Ceperley}},\
  }\href
  {https://www.ebook.de/de/product/25518288/richard_m_martin_lucia_reining_david_m_ceperley_interacting_electrons.html}
  {\emph {\bibinfo {title} {{Interacting Electrons}}}}\ (\bibinfo  {publisher}
  {Cambridge University Press},\ \bibinfo {year} {2016})\BibitemShut {NoStop}%
\bibitem [{\citenamefont {Foulkes}\ \emph {et~al.}(2001)\citenamefont
  {Foulkes}, \citenamefont {Mitas}, \citenamefont {Needs},\ and\ \citenamefont
  {Rajagopal}}]{Foulkes2001}%
  \BibitemOpen
  \bibfield  {author} {\bibinfo {author} {\bibfnamefont {W.~M.~C.}\
  \bibnamefont {Foulkes}}, \bibinfo {author} {\bibfnamefont {L.}~\bibnamefont
  {Mitas}}, \bibinfo {author} {\bibfnamefont {R.~J.}\ \bibnamefont {Needs}}, \
  and\ \bibinfo {author} {\bibfnamefont {G.}~\bibnamefont {Rajagopal}},\
  }\bibfield  {title} {\enquote {\bibinfo {title} {{Quantum Monte Carlo
  simulations of solids}},}\ }\href {\doibase 10.1103/revmodphys.73.33}
  {\bibfield  {journal} {\bibinfo  {journal} {Reviews of Modern Physics}\
  }\textbf {\bibinfo {volume} {73}},\ \bibinfo {pages} {33--83} (\bibinfo
  {year} {2001})}\BibitemShut {NoStop}%
\bibitem [{\citenamefont {Kim}\ \emph {et~al.}(2018)\citenamefont {Kim},
  \citenamefont {Baczewski}, \citenamefont {Beaudet}, \citenamefont {Benali},
  \citenamefont {Bennett}, \citenamefont {Berrill}, \citenamefont {Blunt},
  \citenamefont {Borda}, \citenamefont {Casula}, \citenamefont {Ceperley},
  \citenamefont {Chiesa}, \citenamefont {Clark}, \citenamefont {Clay},
  \citenamefont {Delaney}, \citenamefont {Dewing}, \citenamefont {Esler},
  \citenamefont {Hao}, \citenamefont {Heinonen}, \citenamefont {Kent},
  \citenamefont {Krogel}, \citenamefont {Kyl{\"{a}}np{\"{a}}{\"{a}}},
  \citenamefont {Li}, \citenamefont {Lopez}, \citenamefont {Luo}, \citenamefont
  {Malone}, \citenamefont {Martin}, \citenamefont {Mathuriya}, \citenamefont
  {McMinis}, \citenamefont {Melton}, \citenamefont {Mitas}, \citenamefont
  {Morales}, \citenamefont {Neuscamman}, \citenamefont {Parker}, \citenamefont
  {Flores}, \citenamefont {Romero}, \citenamefont {Rubenstein}, \citenamefont
  {Shea}, \citenamefont {Shin}, \citenamefont {Shulenburger}, \citenamefont
  {Tillack}, \citenamefont {Townsend}, \citenamefont {Tubman}, \citenamefont
  {Goetz}, \citenamefont {Vincent}, \citenamefont {Yang}, \citenamefont {Yang},
  \citenamefont {Zhang},\ and\ \citenamefont {Zhao}}]{KimQMCPACKJCP2018}%
  \BibitemOpen
  \bibfield  {author} {\bibinfo {author} {\bibfnamefont {J.}~\bibnamefont
  {Kim}}, \bibinfo {author} {\bibfnamefont {A.~D.}\ \bibnamefont {Baczewski}},
  \bibinfo {author} {\bibfnamefont {T.~D.}\ \bibnamefont {Beaudet}}, \bibinfo
  {author} {\bibfnamefont {A.}~\bibnamefont {Benali}}, \bibinfo {author}
  {\bibfnamefont {M.~C.}\ \bibnamefont {Bennett}}, \bibinfo {author}
  {\bibfnamefont {M.~A.}\ \bibnamefont {Berrill}}, \bibinfo {author}
  {\bibfnamefont {N.~S.}\ \bibnamefont {Blunt}}, \bibinfo {author}
  {\bibfnamefont {E.~J.~L.}\ \bibnamefont {Borda}}, \bibinfo {author}
  {\bibfnamefont {M.}~\bibnamefont {Casula}}, \bibinfo {author} {\bibfnamefont
  {D.~M.}\ \bibnamefont {Ceperley}}, \bibinfo {author} {\bibfnamefont
  {S.}~\bibnamefont {Chiesa}}, \bibinfo {author} {\bibfnamefont {B.~K.}\
  \bibnamefont {Clark}}, \bibinfo {author} {\bibfnamefont {R.~C.}\ \bibnamefont
  {Clay}}, \bibinfo {author} {\bibfnamefont {K.~T.}\ \bibnamefont {Delaney}},
  \bibinfo {author} {\bibfnamefont {M.}~\bibnamefont {Dewing}}, \bibinfo
  {author} {\bibfnamefont {K.~P.}\ \bibnamefont {Esler}}, \bibinfo {author}
  {\bibfnamefont {H.}~\bibnamefont {Hao}}, \bibinfo {author} {\bibfnamefont
  {O.}~\bibnamefont {Heinonen}}, \bibinfo {author} {\bibfnamefont {P.~R.~C.}\
  \bibnamefont {Kent}}, \bibinfo {author} {\bibfnamefont {J.~T.}\ \bibnamefont
  {Krogel}}, \bibinfo {author} {\bibfnamefont {I.}~\bibnamefont
  {Kyl{\"{a}}np{\"{a}}{\"{a}}}}, \bibinfo {author} {\bibfnamefont {Y.~W.}\
  \bibnamefont {Li}}, \bibinfo {author} {\bibfnamefont {M.~G.}\ \bibnamefont
  {Lopez}}, \bibinfo {author} {\bibfnamefont {Y.}~\bibnamefont {Luo}}, \bibinfo
  {author} {\bibfnamefont {F.~D.}\ \bibnamefont {Malone}}, \bibinfo {author}
  {\bibfnamefont {R.~M.}\ \bibnamefont {Martin}}, \bibinfo {author}
  {\bibfnamefont {A.}~\bibnamefont {Mathuriya}}, \bibinfo {author}
  {\bibfnamefont {J.}~\bibnamefont {McMinis}}, \bibinfo {author} {\bibfnamefont
  {C.~A.}\ \bibnamefont {Melton}}, \bibinfo {author} {\bibfnamefont
  {L.}~\bibnamefont {Mitas}}, \bibinfo {author} {\bibfnamefont {M.~A.}\
  \bibnamefont {Morales}}, \bibinfo {author} {\bibfnamefont {E.}~\bibnamefont
  {Neuscamman}}, \bibinfo {author} {\bibfnamefont {W.~D.}\ \bibnamefont
  {Parker}}, \bibinfo {author} {\bibfnamefont {S.~D.~P.}\ \bibnamefont
  {Flores}}, \bibinfo {author} {\bibfnamefont {N.~A.}\ \bibnamefont {Romero}},
  \bibinfo {author} {\bibfnamefont {B.~M.}\ \bibnamefont {Rubenstein}},
  \bibinfo {author} {\bibfnamefont {J.~A.~R.}\ \bibnamefont {Shea}}, \bibinfo
  {author} {\bibfnamefont {H.}~\bibnamefont {Shin}}, \bibinfo {author}
  {\bibfnamefont {L.}~\bibnamefont {Shulenburger}}, \bibinfo {author}
  {\bibfnamefont {A.~F.}\ \bibnamefont {Tillack}}, \bibinfo {author}
  {\bibfnamefont {J.~P.}\ \bibnamefont {Townsend}}, \bibinfo {author}
  {\bibfnamefont {N.~M.}\ \bibnamefont {Tubman}}, \bibinfo {author}
  {\bibfnamefont {B.~V.~D.}\ \bibnamefont {Goetz}}, \bibinfo {author}
  {\bibfnamefont {J.~E.}\ \bibnamefont {Vincent}}, \bibinfo {author}
  {\bibfnamefont {D.~C.}\ \bibnamefont {Yang}}, \bibinfo {author}
  {\bibfnamefont {Y.}~\bibnamefont {Yang}}, \bibinfo {author} {\bibfnamefont
  {S.}~\bibnamefont {Zhang}}, \ and\ \bibinfo {author} {\bibfnamefont
  {L.}~\bibnamefont {Zhao}},\ }\bibfield  {title} {\enquote {\bibinfo {title}
  {{QMCPACK}: an open source ab initio quantum monte carlo package for the
  electronic structure of atoms, molecules and solids},}\ }\href {\doibase
  10.1088/1361-648x/aab9c3} {\bibfield  {journal} {\bibinfo  {journal} {Journal
  of Physics: Condensed Matter}\ }\textbf {\bibinfo {volume} {30}},\ \bibinfo
  {pages} {195901} (\bibinfo {year} {2018})}\BibitemShut {NoStop}%
\bibitem [{\citenamefont {Genovese}, \citenamefont {Shirakawa},\ and\
  \citenamefont {Sorella}(2019)}]{Genovesenature2019}%
  \BibitemOpen
  \bibfield  {author} {\bibinfo {author} {\bibfnamefont {C.}~\bibnamefont
  {Genovese}}, \bibinfo {author} {\bibfnamefont {T.}~\bibnamefont {Shirakawa}},
  \ and\ \bibinfo {author} {\bibfnamefont {S.}~\bibnamefont {Sorella}},\
  }\bibfield  {title} {\enquote {\bibinfo {title} {{The nature of the quadruple
  chemical bond in the dicarbon molecule}},}\ }\href@noop {} {\  (\bibinfo
  {year} {2019})},\ \Eprint
  {http://arxiv.org/abs/http://arxiv.org/abs/1911.09748v1}
  {arXiv:http://arxiv.org/abs/1911.09748v1} \BibitemShut {NoStop}%
\bibitem [{\citenamefont {Dupuy}\ and\ \citenamefont
  {Casula}(2018)}]{DupuyFate2018}%
  \BibitemOpen
  \bibfield  {author} {\bibinfo {author} {\bibfnamefont {N.}~\bibnamefont
  {Dupuy}}\ and\ \bibinfo {author} {\bibfnamefont {M.}~\bibnamefont {Casula}},\
  }\bibfield  {title} {\enquote {\bibinfo {title} {{Fate of the open-shell
  singlet ground state in the experimentally accessible acenes: A quantum Monte
  Carlo study}},}\ }\href {\doibase 10.1063/1.5016494} {\bibfield  {journal}
  {\bibinfo  {journal} {The Journal of Chemical Physics}\ }\textbf {\bibinfo
  {volume} {148}},\ \bibinfo {pages} {134112} (\bibinfo {year}
  {2018})}\BibitemShut {NoStop}%
\bibitem [{\citenamefont {Brandenburg}\ \emph {et~al.}(2019)\citenamefont
  {Brandenburg}, \citenamefont {Zen}, \citenamefont {Fitzner}, \citenamefont
  {Ramberger}, \citenamefont {Kresse}, \citenamefont {Tsatsoulis},
  \citenamefont {Gr{\"{u}}neis}, \citenamefont {Michaelides},\ and\
  \citenamefont {Alf{\`{e}}}}]{BrandenburgPhysisorption2019}%
  \BibitemOpen
  \bibfield  {author} {\bibinfo {author} {\bibfnamefont {J.~G.}\ \bibnamefont
  {Brandenburg}}, \bibinfo {author} {\bibfnamefont {A.}~\bibnamefont {Zen}},
  \bibinfo {author} {\bibfnamefont {M.}~\bibnamefont {Fitzner}}, \bibinfo
  {author} {\bibfnamefont {B.}~\bibnamefont {Ramberger}}, \bibinfo {author}
  {\bibfnamefont {G.}~\bibnamefont {Kresse}}, \bibinfo {author} {\bibfnamefont
  {T.}~\bibnamefont {Tsatsoulis}}, \bibinfo {author} {\bibfnamefont
  {A.}~\bibnamefont {Gr{\"{u}}neis}}, \bibinfo {author} {\bibfnamefont
  {A.}~\bibnamefont {Michaelides}}, \ and\ \bibinfo {author} {\bibfnamefont
  {D.}~\bibnamefont {Alf{\`{e}}}},\ }\bibfield  {title} {\enquote {\bibinfo
  {title} {{Physisorption of Water on Graphene: Subchemical Accuracy from
  Many-Body Electronic Structure Methods}},}\ }\href {\doibase
  10.1021/acs.jpclett.8b03679} {\bibfield  {journal} {\bibinfo  {journal} {The
  Journal of Physical Chemistry Letters}\ }\textbf {\bibinfo {volume} {10}},\
  \bibinfo {pages} {358--368} (\bibinfo {year} {2019})}\BibitemShut {NoStop}%
\bibitem [{\citenamefont {Shee}\ \emph {et~al.}(2019)\citenamefont {Shee},
  \citenamefont {Rudshteyn}, \citenamefont {Arthur}, \citenamefont {Zhang},
  \citenamefont {Reichman},\ and\ \citenamefont
  {Friesner}}]{SheeAchieving2019}%
  \BibitemOpen
  \bibfield  {author} {\bibinfo {author} {\bibfnamefont {J.}~\bibnamefont
  {Shee}}, \bibinfo {author} {\bibfnamefont {B.}~\bibnamefont {Rudshteyn}},
  \bibinfo {author} {\bibfnamefont {E.~J.}\ \bibnamefont {Arthur}}, \bibinfo
  {author} {\bibfnamefont {S.}~\bibnamefont {Zhang}}, \bibinfo {author}
  {\bibfnamefont {D.~R.}\ \bibnamefont {Reichman}}, \ and\ \bibinfo {author}
  {\bibfnamefont {R.~A.}\ \bibnamefont {Friesner}},\ }\bibfield  {title}
  {\enquote {\bibinfo {title} {{On Achieving High Accuracy in Quantum Chemical
  Calculations of 3d Transition Metal-Containing Systems: A Comparison of
  Auxiliary-Field Quantum Monte Carlo with Coupled Cluster, Density Functional
  Theory, and Experiment for Diatomic Molecules}},}\ }\href {\doibase
  10.1021/acs.jctc.9b00083} {\bibfield  {journal} {\bibinfo  {journal} {Journal
  of Chemical Theory and Computation}\ }\textbf {\bibinfo {volume} {15}},\
  \bibinfo {pages} {2346--2358} (\bibinfo {year} {2019})}\BibitemShut {NoStop}%
\bibitem [{\citenamefont {Qin}\ \emph {et~al.}(2020)\citenamefont {Qin},
  \citenamefont {Ichibha}, \citenamefont {Hongo},\ and\ \citenamefont
  {Maezono}}]{QinInconsistencies2020}%
  \BibitemOpen
  \bibfield  {author} {\bibinfo {author} {\bibfnamefont {K.~S.}\ \bibnamefont
  {Qin}}, \bibinfo {author} {\bibfnamefont {T.}~\bibnamefont {Ichibha}},
  \bibinfo {author} {\bibfnamefont {K.}~\bibnamefont {Hongo}}, \ and\ \bibinfo
  {author} {\bibfnamefont {R.}~\bibnamefont {Maezono}},\ }\bibfield  {title}
  {\enquote {\bibinfo {title} {{Inconsistencies in ab initio evaluations of
  non-additive contributions of {DNA} stacking energies}},}\ }\href {\doibase
  10.1016/j.chemphys.2019.110554} {\bibfield  {journal} {\bibinfo  {journal}
  {Chemical Physics}\ }\textbf {\bibinfo {volume} {529}},\ \bibinfo {pages}
  {110554} (\bibinfo {year} {2020})}\BibitemShut {NoStop}%
\bibitem [{\citenamefont {Yu}, \citenamefont {Wagner},\ and\ \citenamefont
  {Ertekin}(2017)}]{YuFixed2017}%
  \BibitemOpen
  \bibfield  {author} {\bibinfo {author} {\bibfnamefont {J.}~\bibnamefont
  {Yu}}, \bibinfo {author} {\bibfnamefont {L.~K.}\ \bibnamefont {Wagner}}, \
  and\ \bibinfo {author} {\bibfnamefont {E.}~\bibnamefont {Ertekin}},\
  }\bibfield  {title} {\enquote {\bibinfo {title} {{Fixed-node diffusion Monte
  Carlo description of nitrogen defects in zinc oxide}},}\ }\href {\doibase
  10.1103/physrevb.95.075209} {\bibfield  {journal} {\bibinfo  {journal}
  {Physical Review B}\ }\textbf {\bibinfo {volume} {95}},\ \bibinfo {pages}
  {075209} (\bibinfo {year} {2017})}\BibitemShut {NoStop}%
\bibitem [{\citenamefont {Saritas}\ \emph {et~al.}(2019)\citenamefont
  {Saritas}, \citenamefont {Ming}, \citenamefont {Du},\ and\ \citenamefont
  {Reboredo}}]{SaritasExcitation2018}%
  \BibitemOpen
  \bibfield  {author} {\bibinfo {author} {\bibfnamefont {K.}~\bibnamefont
  {Saritas}}, \bibinfo {author} {\bibfnamefont {W.}~\bibnamefont {Ming}},
  \bibinfo {author} {\bibfnamefont {M.-H.}\ \bibnamefont {Du}}, \ and\ \bibinfo
  {author} {\bibfnamefont {F.~A.}\ \bibnamefont {Reboredo}},\ }\bibfield
  {title} {\enquote {\bibinfo {title} {{Excitation Energies of Localized
  Correlated Defects via Quantum Monte Carlo: A Case Study of Mn$^{4+}$ Doped
  Phosphors}},}\ }\href {\doibase 10.1021/acs.jpclett.8b03015} {\bibfield
  {journal} {\bibinfo  {journal} {The Journal of Physical Chemistry Letters}\
  }\textbf {\bibinfo {volume} {10}},\ \bibinfo {pages} {67--74} (\bibinfo
  {year} {2019})}\BibitemShut {NoStop}%
\bibitem [{\citenamefont {Busemeyer}, \citenamefont {MacDougall},\ and\
  \citenamefont {Wagner}(2019)}]{BusemeyerPrediction2019}%
  \BibitemOpen
  \bibfield  {author} {\bibinfo {author} {\bibfnamefont {B.}~\bibnamefont
  {Busemeyer}}, \bibinfo {author} {\bibfnamefont {G.~J.}\ \bibnamefont
  {MacDougall}}, \ and\ \bibinfo {author} {\bibfnamefont {L.~K.}\ \bibnamefont
  {Wagner}},\ }\bibfield  {title} {\enquote {\bibinfo {title} {{Prediction for
  the singlet-triplet excitation energy for the spinel {MgTi}$_2$O$_4$ using
  first-principles diffusion Monte Carlo}},}\ }\href {\doibase
  10.1103/physrevb.99.081118} {\bibfield  {journal} {\bibinfo  {journal}
  {Physical Review B}\ }\textbf {\bibinfo {volume} {99}},\ \bibinfo {pages}
  {081118} (\bibinfo {year} {2019})}\BibitemShut {NoStop}%
\bibitem [{\citenamefont {Zen}\ \emph {et~al.}(2019)\citenamefont {Zen},
  \citenamefont {Brandenburg}, \citenamefont {Michaelides},\ and\ \citenamefont
  {Alf{\`{e}}}}]{Zennew2019}%
  \BibitemOpen
  \bibfield  {author} {\bibinfo {author} {\bibfnamefont {A.}~\bibnamefont
  {Zen}}, \bibinfo {author} {\bibfnamefont {J.~G.}\ \bibnamefont
  {Brandenburg}}, \bibinfo {author} {\bibfnamefont {A.}~\bibnamefont
  {Michaelides}}, \ and\ \bibinfo {author} {\bibfnamefont {D.}~\bibnamefont
  {Alf{\`{e}}}},\ }\bibfield  {title} {\enquote {\bibinfo {title} {{A new
  scheme for fixed node diffusion quantum Monte Carlo with pseudopotentials:
  Improving reproducibility and reducing the trial-wave-function bias}},}\
  }\href {\doibase 10.1063/1.5119729} {\bibfield  {journal} {\bibinfo
  {journal} {The Journal of Chemical Physics}\ }\textbf {\bibinfo {volume}
  {151}},\ \bibinfo {pages} {134105} (\bibinfo {year} {2019})}\BibitemShut
  {NoStop}%
\bibitem [{\citenamefont {Mussard}\ \emph {et~al.}(2018)\citenamefont
  {Mussard}, \citenamefont {Coccia}, \citenamefont {Assaraf}, \citenamefont
  {Otten}, \citenamefont {Umrigar},\ and\ \citenamefont
  {Toulouse}}]{MussardTime2018}%
  \BibitemOpen
  \bibfield  {author} {\bibinfo {author} {\bibfnamefont {B.}~\bibnamefont
  {Mussard}}, \bibinfo {author} {\bibfnamefont {E.}~\bibnamefont {Coccia}},
  \bibinfo {author} {\bibfnamefont {R.}~\bibnamefont {Assaraf}}, \bibinfo
  {author} {\bibfnamefont {M.}~\bibnamefont {Otten}}, \bibinfo {author}
  {\bibfnamefont {C.~J.}\ \bibnamefont {Umrigar}}, \ and\ \bibinfo {author}
  {\bibfnamefont {J.}~\bibnamefont {Toulouse}},\ }\bibfield  {title} {\enquote
  {\bibinfo {title} {{Time-Dependent Linear-Response Variational Monte
  Carlo}},}\ }in\ \href {\doibase 10.1016/bs.aiq.2017.05.005} {\emph {\bibinfo
  {booktitle} {Novel Electronic Structure Theory: General Innovations and
  Strongly Correlated Systems}}}\ (\bibinfo  {publisher} {Elsevier},\ \bibinfo
  {year} {2018})\ pp.\ \bibinfo {pages} {255--270}\BibitemShut {NoStop}%
\bibitem [{\citenamefont {Doblhoff-Dier}, \citenamefont {Kroes},\ and\
  \citenamefont {Libisch}(2018)}]{Doblhoff-DierDensity2018}%
  \BibitemOpen
  \bibfield  {author} {\bibinfo {author} {\bibfnamefont {K.}~\bibnamefont
  {Doblhoff-Dier}}, \bibinfo {author} {\bibfnamefont {G.-J.}\ \bibnamefont
  {Kroes}}, \ and\ \bibinfo {author} {\bibfnamefont {F.}~\bibnamefont
  {Libisch}},\ }\bibfield  {title} {\enquote {\bibinfo {title} {{Density
  functional embedding for periodic and nonperiodic diffusion Monte Carlo
  calculations}},}\ }\href {\doibase 10.1103/physrevb.98.085138} {\bibfield
  {journal} {\bibinfo  {journal} {Physical Review B}\ }\textbf {\bibinfo
  {volume} {98}},\ \bibinfo {pages} {085138} (\bibinfo {year}
  {2018})}\BibitemShut {NoStop}%
\bibitem [{\citenamefont {Dash}\ \emph {et~al.}(2019)\citenamefont {Dash},
  \citenamefont {Feldt}, \citenamefont {Moroni}, \citenamefont {Scemama},\ and\
  \citenamefont {Filippi}}]{DashExcited2019}%
  \BibitemOpen
  \bibfield  {author} {\bibinfo {author} {\bibfnamefont {M.}~\bibnamefont
  {Dash}}, \bibinfo {author} {\bibfnamefont {J.}~\bibnamefont {Feldt}},
  \bibinfo {author} {\bibfnamefont {S.}~\bibnamefont {Moroni}}, \bibinfo
  {author} {\bibfnamefont {A.}~\bibnamefont {Scemama}}, \ and\ \bibinfo
  {author} {\bibfnamefont {C.}~\bibnamefont {Filippi}},\ }\bibfield  {title}
  {\enquote {\bibinfo {title} {{Excited States with Selected Configuration
  Interaction-Quantum Monte Carlo: Chemically Accurate Excitation Energies and
  Geometries}},}\ }\href {\doibase 10.1021/acs.jctc.9b00476} {\bibfield
  {journal} {\bibinfo  {journal} {Journal of Chemical Theory and Computation}\
  }\textbf {\bibinfo {volume} {15}},\ \bibinfo {pages} {4896--4906} (\bibinfo
  {year} {2019})}\BibitemShut {NoStop}%
\bibitem [{\citenamefont {Azadi}\ and\ \citenamefont
  {Foulkes}(2019)}]{Azadiefficient2019}%
  \BibitemOpen
  \bibfield  {author} {\bibinfo {author} {\bibfnamefont {S.}~\bibnamefont
  {Azadi}}\ and\ \bibinfo {author} {\bibfnamefont {W.~M.~C.}\ \bibnamefont
  {Foulkes}},\ }\bibfield  {title} {\enquote {\bibinfo {title} {{Efficient
  method for grand-canonical twist averaging in quantum Monte Carlo
  calculations}},}\ }\href {\doibase 10.1103/physrevb.100.245142} {\bibfield
  {journal} {\bibinfo  {journal} {Physical Review B}\ }\textbf {\bibinfo
  {volume} {100}},\ \bibinfo {pages} {245142} (\bibinfo {year}
  {2019})}\BibitemShut {NoStop}%
\bibitem [{\citenamefont {Per}, \citenamefont {Fletcher},\ and\ \citenamefont
  {Cleland}(2019)}]{PerDensity2019}%
  \BibitemOpen
  \bibfield  {author} {\bibinfo {author} {\bibfnamefont {M.~C.}\ \bibnamefont
  {Per}}, \bibinfo {author} {\bibfnamefont {E.~K.}\ \bibnamefont {Fletcher}}, \
  and\ \bibinfo {author} {\bibfnamefont {D.~M.}\ \bibnamefont {Cleland}},\
  }\bibfield  {title} {\enquote {\bibinfo {title} {{Density functional orbitals
  in quantum Monte Carlo: The importance of accurate densities}},}\ }\href
  {\doibase 10.1063/1.5095158} {\bibfield  {journal} {\bibinfo  {journal} {The
  Journal of Chemical Physics}\ }\textbf {\bibinfo {volume} {150}},\ \bibinfo
  {pages} {184101} (\bibinfo {year} {2019})}\BibitemShut {NoStop}%
\bibitem [{\citenamefont {Motta}\ \emph {et~al.}(2017)\citenamefont {Motta},
  \citenamefont {Ceperley}, \citenamefont {Chan}, \citenamefont {Gomez},
  \citenamefont {Gull}, \citenamefont {Guo}, \citenamefont
  {Jim{\'{e}}nez-Hoyos}, \citenamefont {Lan}, \citenamefont {Li}, \citenamefont
  {Ma}, \citenamefont {Millis}, \citenamefont {Prokof'ev}, \citenamefont {Ray},
  \citenamefont {Scuseria}, \citenamefont {Sorella}, \citenamefont
  {Stoudenmire}, \citenamefont {Sun}, \citenamefont {Tupitsyn}, \citenamefont
  {White}, \citenamefont {Zgid},\ and\ \citenamefont {and}}]{MottaTowards2017}%
  \BibitemOpen
  \bibfield  {author} {\bibinfo {author} {\bibfnamefont {M.}~\bibnamefont
  {Motta}}, \bibinfo {author} {\bibfnamefont {D.~M.}\ \bibnamefont {Ceperley}},
  \bibinfo {author} {\bibfnamefont {G.~K.-L.}\ \bibnamefont {Chan}}, \bibinfo
  {author} {\bibfnamefont {J.~A.}\ \bibnamefont {Gomez}}, \bibinfo {author}
  {\bibfnamefont {E.}~\bibnamefont {Gull}}, \bibinfo {author} {\bibfnamefont
  {S.}~\bibnamefont {Guo}}, \bibinfo {author} {\bibfnamefont {C.~A.}\
  \bibnamefont {Jim{\'{e}}nez-Hoyos}}, \bibinfo {author} {\bibfnamefont
  {T.~N.}\ \bibnamefont {Lan}}, \bibinfo {author} {\bibfnamefont
  {J.}~\bibnamefont {Li}}, \bibinfo {author} {\bibfnamefont {F.}~\bibnamefont
  {Ma}}, \bibinfo {author} {\bibfnamefont {A.~J.}\ \bibnamefont {Millis}},
  \bibinfo {author} {\bibfnamefont {N.~V.}\ \bibnamefont {Prokof'ev}}, \bibinfo
  {author} {\bibfnamefont {U.}~\bibnamefont {Ray}}, \bibinfo {author}
  {\bibfnamefont {G.~E.}\ \bibnamefont {Scuseria}}, \bibinfo {author}
  {\bibfnamefont {S.}~\bibnamefont {Sorella}}, \bibinfo {author} {\bibfnamefont
  {E.~M.}\ \bibnamefont {Stoudenmire}}, \bibinfo {author} {\bibfnamefont
  {Q.}~\bibnamefont {Sun}}, \bibinfo {author} {\bibfnamefont {I.~S.}\
  \bibnamefont {Tupitsyn}}, \bibinfo {author} {\bibfnamefont {S.~R.}\
  \bibnamefont {White}}, \bibinfo {author} {\bibfnamefont {D.}~\bibnamefont
  {Zgid}}, \ and\ \bibinfo {author} {\bibfnamefont {S.~Z.}\ \bibnamefont
  {and}},\ }\bibfield  {title} {\enquote {\bibinfo {title} {{Towards the
  Solution of the Many-Electron Problem in Real Materials: Equation of State of
  the Hydrogen Chain with State-of-the-Art Many-Body Methods}},}\ }\href
  {\doibase 10.1103/physrevx.7.031059} {\bibfield  {journal} {\bibinfo
  {journal} {Physical Review X}\ }\textbf {\bibinfo {volume} {7}},\ \bibinfo
  {pages} {031059} (\bibinfo {year} {2017})}\BibitemShut {NoStop}%
\bibitem [{\citenamefont {Motta}\ and\ \citenamefont
  {Zhang}(2018)}]{Motta2019}%
  \BibitemOpen
  \bibfield  {author} {\bibinfo {author} {\bibfnamefont {M.}~\bibnamefont
  {Motta}}\ and\ \bibinfo {author} {\bibfnamefont {S.}~\bibnamefont {Zhang}},\
  }\bibfield  {title} {\enquote {\bibinfo {title} {{Ab initio computations of
  molecular systems by the auxiliary-field quantum Monte Carlo method}},}\
  }\href {\doibase 10.1002/wcms.1364} {\bibfield  {journal} {\bibinfo
  {journal} {WIREs Computational Molecular Science}\ }\textbf {\bibinfo
  {volume} {8}},\ \bibinfo {pages} {e1364} (\bibinfo {year}
  {2018})}\BibitemShut {NoStop}%
\bibitem [{\citenamefont {Alexander}\ \emph {et~al.}(2020)\citenamefont
  {Alexander}, \citenamefont {Almgren}, \citenamefont {Bell}, \citenamefont
  {Bhattacharjee}, \citenamefont {Chen}, \citenamefont {Colella}, \citenamefont
  {Daniel}, \citenamefont {DeSlippe}, \citenamefont {Diachin}, \citenamefont
  {Draeger}, \citenamefont {Dubey}, \citenamefont {Dunning}, \citenamefont
  {Evans}, \citenamefont {Foster}, \citenamefont {Francois}, \citenamefont
  {Germann}, \citenamefont {Gordon}, \citenamefont {Habib}, \citenamefont
  {Halappanavar}, \citenamefont {Hamilton}, \citenamefont {Hart}, \citenamefont
  {Huang}, \citenamefont {Hungerford}, \citenamefont {Kasen}, \citenamefont
  {Kent}, \citenamefont {Kolev}, \citenamefont {Kothe}, \citenamefont
  {Kronfeld}, \citenamefont {Luo}, \citenamefont {Mackenzie}, \citenamefont
  {McCallen}, \citenamefont {Messer}, \citenamefont {Mniszewski}, \citenamefont
  {Oehmen}, \citenamefont {Perazzo}, \citenamefont {Perez}, \citenamefont
  {Richards}, \citenamefont {Rider}, \citenamefont {Rieben}, \citenamefont
  {Roche}, \citenamefont {Siegel}, \citenamefont {Sprague}, \citenamefont
  {Steefel}, \citenamefont {Stevens}, \citenamefont {Syamlal}, \citenamefont
  {Taylor}, \citenamefont {Turner}, \citenamefont {Vay}, \citenamefont {Voter},
  \citenamefont {Windus},\ and\ \citenamefont {Yelick}}]{ExascaleSkin2020}%
  \BibitemOpen
  \bibfield  {author} {\bibinfo {author} {\bibfnamefont {F.}~\bibnamefont
  {Alexander}}, \bibinfo {author} {\bibfnamefont {A.}~\bibnamefont {Almgren}},
  \bibinfo {author} {\bibfnamefont {J.}~\bibnamefont {Bell}}, \bibinfo {author}
  {\bibfnamefont {A.}~\bibnamefont {Bhattacharjee}}, \bibinfo {author}
  {\bibfnamefont {J.}~\bibnamefont {Chen}}, \bibinfo {author} {\bibfnamefont
  {P.}~\bibnamefont {Colella}}, \bibinfo {author} {\bibfnamefont
  {D.}~\bibnamefont {Daniel}}, \bibinfo {author} {\bibfnamefont
  {J.}~\bibnamefont {DeSlippe}}, \bibinfo {author} {\bibfnamefont
  {L.}~\bibnamefont {Diachin}}, \bibinfo {author} {\bibfnamefont
  {E.}~\bibnamefont {Draeger}}, \bibinfo {author} {\bibfnamefont
  {A.}~\bibnamefont {Dubey}}, \bibinfo {author} {\bibfnamefont
  {T.}~\bibnamefont {Dunning}}, \bibinfo {author} {\bibfnamefont
  {T.}~\bibnamefont {Evans}}, \bibinfo {author} {\bibfnamefont
  {I.}~\bibnamefont {Foster}}, \bibinfo {author} {\bibfnamefont
  {M.}~\bibnamefont {Francois}}, \bibinfo {author} {\bibfnamefont
  {T.}~\bibnamefont {Germann}}, \bibinfo {author} {\bibfnamefont
  {M.}~\bibnamefont {Gordon}}, \bibinfo {author} {\bibfnamefont
  {S.}~\bibnamefont {Habib}}, \bibinfo {author} {\bibfnamefont
  {M.}~\bibnamefont {Halappanavar}}, \bibinfo {author} {\bibfnamefont
  {S.}~\bibnamefont {Hamilton}}, \bibinfo {author} {\bibfnamefont
  {W.}~\bibnamefont {Hart}}, \bibinfo {author} {\bibfnamefont {Z.~H.}\
  \bibnamefont {Huang}}, \bibinfo {author} {\bibfnamefont {A.}~\bibnamefont
  {Hungerford}}, \bibinfo {author} {\bibfnamefont {D.}~\bibnamefont {Kasen}},
  \bibinfo {author} {\bibfnamefont {P.~R.~C.}\ \bibnamefont {Kent}}, \bibinfo
  {author} {\bibfnamefont {T.}~\bibnamefont {Kolev}}, \bibinfo {author}
  {\bibfnamefont {D.~B.}\ \bibnamefont {Kothe}}, \bibinfo {author}
  {\bibfnamefont {A.}~\bibnamefont {Kronfeld}}, \bibinfo {author}
  {\bibfnamefont {Y.}~\bibnamefont {Luo}}, \bibinfo {author} {\bibfnamefont
  {P.}~\bibnamefont {Mackenzie}}, \bibinfo {author} {\bibfnamefont
  {D.}~\bibnamefont {McCallen}}, \bibinfo {author} {\bibfnamefont
  {B.}~\bibnamefont {Messer}}, \bibinfo {author} {\bibfnamefont
  {S.}~\bibnamefont {Mniszewski}}, \bibinfo {author} {\bibfnamefont
  {C.}~\bibnamefont {Oehmen}}, \bibinfo {author} {\bibfnamefont
  {A.}~\bibnamefont {Perazzo}}, \bibinfo {author} {\bibfnamefont
  {D.}~\bibnamefont {Perez}}, \bibinfo {author} {\bibfnamefont
  {D.}~\bibnamefont {Richards}}, \bibinfo {author} {\bibfnamefont {W.~J.}\
  \bibnamefont {Rider}}, \bibinfo {author} {\bibfnamefont {R.}~\bibnamefont
  {Rieben}}, \bibinfo {author} {\bibfnamefont {K.}~\bibnamefont {Roche}},
  \bibinfo {author} {\bibfnamefont {A.}~\bibnamefont {Siegel}}, \bibinfo
  {author} {\bibfnamefont {M.}~\bibnamefont {Sprague}}, \bibinfo {author}
  {\bibfnamefont {C.}~\bibnamefont {Steefel}}, \bibinfo {author} {\bibfnamefont
  {R.}~\bibnamefont {Stevens}}, \bibinfo {author} {\bibfnamefont
  {M.}~\bibnamefont {Syamlal}}, \bibinfo {author} {\bibfnamefont
  {M.}~\bibnamefont {Taylor}}, \bibinfo {author} {\bibfnamefont
  {J.}~\bibnamefont {Turner}}, \bibinfo {author} {\bibfnamefont {J.-L.}\
  \bibnamefont {Vay}}, \bibinfo {author} {\bibfnamefont {A.~F.}\ \bibnamefont
  {Voter}}, \bibinfo {author} {\bibfnamefont {T.~L.}\ \bibnamefont {Windus}}, \
  and\ \bibinfo {author} {\bibfnamefont {K.}~\bibnamefont {Yelick}},\
  }\bibfield  {title} {\enquote {\bibinfo {title} {{Exascale applications: skin
  in the game}},}\ }\href {\doibase 10.1098/rsta.2019.0056} {\bibfield
  {journal} {\bibinfo  {journal} {Philosophical Transactions of the Royal
  Society A: Mathematical, Physical and Engineering Sciences}\ }\textbf
  {\bibinfo {volume} {378}},\ \bibinfo {pages} {20190056} (\bibinfo {year}
  {2020})}\BibitemShut {NoStop}%
\bibitem [{\citenamefont {Gygi}(2008)}]{GygiArchitecture2008}%
  \BibitemOpen
  \bibfield  {author} {\bibinfo {author} {\bibfnamefont {F.}~\bibnamefont
  {Gygi}},\ }\bibfield  {title} {\enquote {\bibinfo {title} {{Architecture of
  Qbox: A scalable first-principles molecular dynamics code}},}\ }\href
  {\doibase 10.1147/rd.521.0137} {\bibfield  {journal} {\bibinfo  {journal}
  {{IBM} Journal of Research and Development}\ }\textbf {\bibinfo {volume}
  {52}},\ \bibinfo {pages} {137--144} (\bibinfo {year} {2008})}\BibitemShut
  {NoStop}%
\bibitem [{\citenamefont {Sun}\ \emph {et~al.}(2017)\citenamefont {Sun},
  \citenamefont {Berkelbach}, \citenamefont {Blunt}, \citenamefont {Booth},
  \citenamefont {Guo}, \citenamefont {Li}, \citenamefont {Liu}, \citenamefont
  {McClain}, \citenamefont {Sayfutyarova}, \citenamefont {Sharma},
  \citenamefont {Wouters},\ and\ \citenamefont {Chan}}]{PYSCF}%
  \BibitemOpen
  \bibfield  {author} {\bibinfo {author} {\bibfnamefont {Q.}~\bibnamefont
  {Sun}}, \bibinfo {author} {\bibfnamefont {T.~C.}\ \bibnamefont {Berkelbach}},
  \bibinfo {author} {\bibfnamefont {N.~S.}\ \bibnamefont {Blunt}}, \bibinfo
  {author} {\bibfnamefont {G.~H.}\ \bibnamefont {Booth}}, \bibinfo {author}
  {\bibfnamefont {S.}~\bibnamefont {Guo}}, \bibinfo {author} {\bibfnamefont
  {Z.}~\bibnamefont {Li}}, \bibinfo {author} {\bibfnamefont {J.}~\bibnamefont
  {Liu}}, \bibinfo {author} {\bibfnamefont {J.~D.}\ \bibnamefont {McClain}},
  \bibinfo {author} {\bibfnamefont {E.~R.}\ \bibnamefont {Sayfutyarova}},
  \bibinfo {author} {\bibfnamefont {S.}~\bibnamefont {Sharma}}, \bibinfo
  {author} {\bibfnamefont {S.}~\bibnamefont {Wouters}}, \ and\ \bibinfo
  {author} {\bibfnamefont {G.~K.~L.}\ \bibnamefont {Chan}},\ }\bibfield
  {title} {\enquote {\bibinfo {title} {{PySCF: the Python-based simulations of
  chemistry framework}},}\ }\href {\doibase 10.1002/wcms.1340} {\bibfield
  {journal} {\bibinfo  {journal} {WIREs Computational Molecular Science}\
  }\textbf {\bibinfo {volume} {8}},\ \bibinfo {pages} {e1340} (\bibinfo {year}
  {2017})}\BibitemShut {NoStop}%
\bibitem [{\citenamefont {Giannozzi}\ \emph {et~al.}(2009)\citenamefont
  {Giannozzi}, \citenamefont {Baroni}, \citenamefont {Bonini}, \citenamefont
  {Calandra}, \citenamefont {Car}, \citenamefont {Cavazzoni}, \citenamefont
  {Ceresoli}, \citenamefont {Chiarotti}, \citenamefont {Cococcioni},
  \citenamefont {Dabo}, \citenamefont {Corso}, \citenamefont {de~Gironcoli},
  \citenamefont {Fabris}, \citenamefont {Fratesi}, \citenamefont {Gebauer},
  \citenamefont {Gerstmann}, \citenamefont {Gougoussis}, \citenamefont
  {Kokalj}, \citenamefont {Lazzeri}, \citenamefont {Martin-Samos},
  \citenamefont {Marzari}, \citenamefont {Mauri}, \citenamefont {Mazzarello},
  \citenamefont {Paolini}, \citenamefont {Pasquarello}, \citenamefont
  {Paulatto}, \citenamefont {Sbraccia}, \citenamefont {Scandolo}, \citenamefont
  {Sclauzero}, \citenamefont {Seitsonen}, \citenamefont {Smogunov},
  \citenamefont {Umari},\ and\ \citenamefont {Wentzcovitch}}]{Gianozzi2009}%
  \BibitemOpen
  \bibfield  {author} {\bibinfo {author} {\bibfnamefont {P.}~\bibnamefont
  {Giannozzi}}, \bibinfo {author} {\bibfnamefont {S.}~\bibnamefont {Baroni}},
  \bibinfo {author} {\bibfnamefont {N.}~\bibnamefont {Bonini}}, \bibinfo
  {author} {\bibfnamefont {M.}~\bibnamefont {Calandra}}, \bibinfo {author}
  {\bibfnamefont {R.}~\bibnamefont {Car}}, \bibinfo {author} {\bibfnamefont
  {C.}~\bibnamefont {Cavazzoni}}, \bibinfo {author} {\bibfnamefont
  {D.}~\bibnamefont {Ceresoli}}, \bibinfo {author} {\bibfnamefont {G.~L.}\
  \bibnamefont {Chiarotti}}, \bibinfo {author} {\bibfnamefont {M.}~\bibnamefont
  {Cococcioni}}, \bibinfo {author} {\bibfnamefont {I.}~\bibnamefont {Dabo}},
  \bibinfo {author} {\bibfnamefont {A.~D.}\ \bibnamefont {Corso}}, \bibinfo
  {author} {\bibfnamefont {S.}~\bibnamefont {de~Gironcoli}}, \bibinfo {author}
  {\bibfnamefont {S.}~\bibnamefont {Fabris}}, \bibinfo {author} {\bibfnamefont
  {G.}~\bibnamefont {Fratesi}}, \bibinfo {author} {\bibfnamefont
  {R.}~\bibnamefont {Gebauer}}, \bibinfo {author} {\bibfnamefont
  {U.}~\bibnamefont {Gerstmann}}, \bibinfo {author} {\bibfnamefont
  {C.}~\bibnamefont {Gougoussis}}, \bibinfo {author} {\bibfnamefont
  {A.}~\bibnamefont {Kokalj}}, \bibinfo {author} {\bibfnamefont
  {M.}~\bibnamefont {Lazzeri}}, \bibinfo {author} {\bibfnamefont
  {L.}~\bibnamefont {Martin-Samos}}, \bibinfo {author} {\bibfnamefont
  {N.}~\bibnamefont {Marzari}}, \bibinfo {author} {\bibfnamefont
  {F.}~\bibnamefont {Mauri}}, \bibinfo {author} {\bibfnamefont
  {R.}~\bibnamefont {Mazzarello}}, \bibinfo {author} {\bibfnamefont
  {S.}~\bibnamefont {Paolini}}, \bibinfo {author} {\bibfnamefont
  {A.}~\bibnamefont {Pasquarello}}, \bibinfo {author} {\bibfnamefont
  {L.}~\bibnamefont {Paulatto}}, \bibinfo {author} {\bibfnamefont
  {C.}~\bibnamefont {Sbraccia}}, \bibinfo {author} {\bibfnamefont
  {S.}~\bibnamefont {Scandolo}}, \bibinfo {author} {\bibfnamefont
  {G.}~\bibnamefont {Sclauzero}}, \bibinfo {author} {\bibfnamefont {A.~P.}\
  \bibnamefont {Seitsonen}}, \bibinfo {author} {\bibfnamefont {A.}~\bibnamefont
  {Smogunov}}, \bibinfo {author} {\bibfnamefont {P.}~\bibnamefont {Umari}}, \
  and\ \bibinfo {author} {\bibfnamefont {R.~M.}\ \bibnamefont {Wentzcovitch}},\
  }\bibfield  {title} {\enquote {\bibinfo {title} {{QUANTUM ESPRESSO: a modular
  and open-source software project for quantum simulations of materials}},}\
  }\href {http://stacks.iop.org/0953-8984/21/i=39/a=395502} {\bibfield
  {journal} {\bibinfo  {journal} {Journal of Physics: Condensed Matter}\
  }\textbf {\bibinfo {volume} {21}},\ \bibinfo {pages} {395502} (\bibinfo
  {year} {2009})}\BibitemShut {NoStop}%
\bibitem [{\citenamefont {Garniron}\ \emph {et~al.}(2019)\citenamefont
  {Garniron}, \citenamefont {Applencourt}, \citenamefont {Gasperich},
  \citenamefont {Benali}, \citenamefont {Fert{\'{e}}}, \citenamefont {Paquier},
  \citenamefont {Pradines}, \citenamefont {Assaraf}, \citenamefont {Reinhardt},
  \citenamefont {Toulouse}, \citenamefont {Barbaresco}, \citenamefont {Renon},
  \citenamefont {David}, \citenamefont {Malrieu}, \citenamefont {V{\'{e}}ril},
  \citenamefont {Caffarel}, \citenamefont {Loos}, \citenamefont {Giner},\ and\
  \citenamefont {Scemama}}]{QP2019}%
  \BibitemOpen
  \bibfield  {author} {\bibinfo {author} {\bibfnamefont {Y.}~\bibnamefont
  {Garniron}}, \bibinfo {author} {\bibfnamefont {T.}~\bibnamefont
  {Applencourt}}, \bibinfo {author} {\bibfnamefont {K.}~\bibnamefont
  {Gasperich}}, \bibinfo {author} {\bibfnamefont {A.}~\bibnamefont {Benali}},
  \bibinfo {author} {\bibfnamefont {A.}~\bibnamefont {Fert{\'{e}}}}, \bibinfo
  {author} {\bibfnamefont {J.}~\bibnamefont {Paquier}}, \bibinfo {author}
  {\bibfnamefont {B.}~\bibnamefont {Pradines}}, \bibinfo {author}
  {\bibfnamefont {R.}~\bibnamefont {Assaraf}}, \bibinfo {author} {\bibfnamefont
  {P.}~\bibnamefont {Reinhardt}}, \bibinfo {author} {\bibfnamefont
  {J.}~\bibnamefont {Toulouse}}, \bibinfo {author} {\bibfnamefont
  {P.}~\bibnamefont {Barbaresco}}, \bibinfo {author} {\bibfnamefont
  {N.}~\bibnamefont {Renon}}, \bibinfo {author} {\bibfnamefont
  {G.}~\bibnamefont {David}}, \bibinfo {author} {\bibfnamefont {J.-P.}\
  \bibnamefont {Malrieu}}, \bibinfo {author} {\bibfnamefont {M.}~\bibnamefont
  {V{\'{e}}ril}}, \bibinfo {author} {\bibfnamefont {M.}~\bibnamefont
  {Caffarel}}, \bibinfo {author} {\bibfnamefont {P.-F.}\ \bibnamefont {Loos}},
  \bibinfo {author} {\bibfnamefont {E.}~\bibnamefont {Giner}}, \ and\ \bibinfo
  {author} {\bibfnamefont {A.}~\bibnamefont {Scemama}},\ }\bibfield  {title}
  {\enquote {\bibinfo {title} {{Quantum Package 2.0: An Open-Source
  Determinant-Driven Suite of Programs}},}\ }\href {\doibase
  10.1021/acs.jctc.9b00176} {\bibfield  {journal} {\bibinfo  {journal} {Journal
  of Chemical Theory and Computation}\ }\textbf {\bibinfo {volume} {15}},\
  \bibinfo {pages} {3591--3609} (\bibinfo {year} {2019})}\BibitemShut {NoStop}%
\bibitem [{\citenamefont {Schmidt}\ \emph {et~al.}(1993)\citenamefont
  {Schmidt}, \citenamefont {Baldridge}, \citenamefont {Boatz}, \citenamefont
  {Elbert}, \citenamefont {Gordon}, \citenamefont {Jensen}, \citenamefont
  {Koseki}, \citenamefont {Matsunaga}, \citenamefont {Nguyen}, \citenamefont
  {Su}, \citenamefont {Windus}, \citenamefont {Dupuis},\ and\ \citenamefont
  {Montgomery}}]{SchmidtGeneral1993}%
  \BibitemOpen
  \bibfield  {author} {\bibinfo {author} {\bibfnamefont {M.~W.}\ \bibnamefont
  {Schmidt}}, \bibinfo {author} {\bibfnamefont {K.~K.}\ \bibnamefont
  {Baldridge}}, \bibinfo {author} {\bibfnamefont {J.~A.}\ \bibnamefont
  {Boatz}}, \bibinfo {author} {\bibfnamefont {S.~T.}\ \bibnamefont {Elbert}},
  \bibinfo {author} {\bibfnamefont {M.~S.}\ \bibnamefont {Gordon}}, \bibinfo
  {author} {\bibfnamefont {J.~H.}\ \bibnamefont {Jensen}}, \bibinfo {author}
  {\bibfnamefont {S.}~\bibnamefont {Koseki}}, \bibinfo {author} {\bibfnamefont
  {N.}~\bibnamefont {Matsunaga}}, \bibinfo {author} {\bibfnamefont {K.~A.}\
  \bibnamefont {Nguyen}}, \bibinfo {author} {\bibfnamefont {S.}~\bibnamefont
  {Su}}, \bibinfo {author} {\bibfnamefont {T.~L.}\ \bibnamefont {Windus}},
  \bibinfo {author} {\bibfnamefont {M.}~\bibnamefont {Dupuis}}, \ and\ \bibinfo
  {author} {\bibfnamefont {J.~A.}\ \bibnamefont {Montgomery}},\ }\bibfield
  {title} {\enquote {\bibinfo {title} {{General atomic and molecular electronic
  structure system}},}\ }\href {\doibase 10.1002/jcc.540141112} {\bibfield
  {journal} {\bibinfo  {journal} {Journal of Computational Chemistry}\ }\textbf
  {\bibinfo {volume} {14}},\ \bibinfo {pages} {1347--1363} (\bibinfo {year}
  {1993})}\BibitemShut {NoStop}%
\bibitem [{\citenamefont {Valiev}\ \emph {et~al.}(2010)\citenamefont {Valiev},
  \citenamefont {Bylaska}, \citenamefont {Govind}, \citenamefont {Kowalski},
  \citenamefont {Straatsma}, \citenamefont {Dam}, \citenamefont {Wang},
  \citenamefont {Nieplocha}, \citenamefont {Apra}, \citenamefont {Windus},\
  and\ \citenamefont {de~Jong}}]{ValievNWChem2010}%
  \BibitemOpen
  \bibfield  {author} {\bibinfo {author} {\bibfnamefont {M.}~\bibnamefont
  {Valiev}}, \bibinfo {author} {\bibfnamefont {E.}~\bibnamefont {Bylaska}},
  \bibinfo {author} {\bibfnamefont {N.}~\bibnamefont {Govind}}, \bibinfo
  {author} {\bibfnamefont {K.}~\bibnamefont {Kowalski}}, \bibinfo {author}
  {\bibfnamefont {T.}~\bibnamefont {Straatsma}}, \bibinfo {author}
  {\bibfnamefont {H.~V.}\ \bibnamefont {Dam}}, \bibinfo {author} {\bibfnamefont
  {D.}~\bibnamefont {Wang}}, \bibinfo {author} {\bibfnamefont {J.}~\bibnamefont
  {Nieplocha}}, \bibinfo {author} {\bibfnamefont {E.}~\bibnamefont {Apra}},
  \bibinfo {author} {\bibfnamefont {T.}~\bibnamefont {Windus}}, \ and\ \bibinfo
  {author} {\bibfnamefont {W.}~\bibnamefont {de~Jong}},\ }\bibfield  {title}
  {\enquote {\bibinfo {title} {{NWChem}: A comprehensive and scalable
  open-source solution for large scale molecular simulations},}\ }\href
  {\doibase 10.1016/j.cpc.2010.04.018} {\bibfield  {journal} {\bibinfo
  {journal} {Computer Physics Communications}\ }\textbf {\bibinfo {volume}
  {181}},\ \bibinfo {pages} {1477--1489} (\bibinfo {year} {2010})}\BibitemShut
  {NoStop}%
\bibitem [{\citenamefont {Ortiz}, \citenamefont {Eriksson},\ and\ \citenamefont
  {Klintenberg}(2009)}]{Ortiz2009}%
  \BibitemOpen
  \bibfield  {author} {\bibinfo {author} {\bibfnamefont {C.}~\bibnamefont
  {Ortiz}}, \bibinfo {author} {\bibfnamefont {O.}~\bibnamefont {Eriksson}}, \
  and\ \bibinfo {author} {\bibfnamefont {M.}~\bibnamefont {Klintenberg}},\
  }\bibfield  {title} {\enquote {\bibinfo {title} {{Data mining and accelerated
  electronic structure theory as a tool in the search for new functional
  materials}},}\ }\href {\doibase
  http://dx.doi.org/10.1016/j.commatsci.2008.07.016} {\bibfield  {journal}
  {\bibinfo  {journal} {Computational Materials Science}\ }\textbf {\bibinfo
  {volume} {44}},\ \bibinfo {pages} {1042--1049} (\bibinfo {year}
  {2009})}\BibitemShut {NoStop}%
\bibitem [{\citenamefont {Hachmann}\ \emph {et~al.}(2011)\citenamefont
  {Hachmann}, \citenamefont {Olivares-Amaya}, \citenamefont {Atahan-Evrenk},
  \citenamefont {Amador-Bedolla}, \citenamefont {S\'{a}nchez-Carrera},
  \citenamefont {Gold-Parker}, \citenamefont {Vogt}, \citenamefont {Brockway},\
  and\ \citenamefont {Aspuru-Guzik}}]{Hachmann2011}%
  \BibitemOpen
  \bibfield  {author} {\bibinfo {author} {\bibfnamefont {J.}~\bibnamefont
  {Hachmann}}, \bibinfo {author} {\bibfnamefont {R.}~\bibnamefont
  {Olivares-Amaya}}, \bibinfo {author} {\bibfnamefont {S.}~\bibnamefont
  {Atahan-Evrenk}}, \bibinfo {author} {\bibfnamefont {C.}~\bibnamefont
  {Amador-Bedolla}}, \bibinfo {author} {\bibfnamefont {R.~S.}\ \bibnamefont
  {S\'{a}nchez-Carrera}}, \bibinfo {author} {\bibfnamefont {A.}~\bibnamefont
  {Gold-Parker}}, \bibinfo {author} {\bibfnamefont {L.}~\bibnamefont {Vogt}},
  \bibinfo {author} {\bibfnamefont {A.~M.}\ \bibnamefont {Brockway}}, \ and\
  \bibinfo {author} {\bibfnamefont {A.}~\bibnamefont {Aspuru-Guzik}},\
  }\bibfield  {title} {\enquote {\bibinfo {title} {{The Harvard Clean Energy
  Project: Large-Scale Computational Screening and Design of Organic
  Photovoltaics on the World Community Grid}},}\ }\href {\doibase
  10.1021/jz200866s} {\bibfield  {journal} {\bibinfo  {journal} {The Journal of
  Physical Chemistry Letters}\ }\textbf {\bibinfo {volume} {2}},\ \bibinfo
  {pages} {2241--2251} (\bibinfo {year} {2011})},\ \Eprint
  {http://arxiv.org/abs/http://dx.doi.org/10.1021/jz200866s}
  {http://dx.doi.org/10.1021/jz200866s} \BibitemShut {NoStop}%
\bibitem [{\citenamefont {Curtarolo}\ \emph {et~al.}(2012)\citenamefont
  {Curtarolo}, \citenamefont {Setyawan}, \citenamefont {Hart}, \citenamefont
  {Jahnatek}, \citenamefont {Chepulskii}, \citenamefont {Taylor}, \citenamefont
  {Wang}, \citenamefont {Xue}, \citenamefont {Yang}, \citenamefont {Levy},
  \citenamefont {Mehl}, \citenamefont {Stokes}, \citenamefont {Demchenko},\
  and\ \citenamefont {Morgan}}]{Curtarolo2012}%
  \BibitemOpen
  \bibfield  {author} {\bibinfo {author} {\bibfnamefont {S.}~\bibnamefont
  {Curtarolo}}, \bibinfo {author} {\bibfnamefont {W.}~\bibnamefont {Setyawan}},
  \bibinfo {author} {\bibfnamefont {G.~L.}\ \bibnamefont {Hart}}, \bibinfo
  {author} {\bibfnamefont {M.}~\bibnamefont {Jahnatek}}, \bibinfo {author}
  {\bibfnamefont {R.~V.}\ \bibnamefont {Chepulskii}}, \bibinfo {author}
  {\bibfnamefont {R.~H.}\ \bibnamefont {Taylor}}, \bibinfo {author}
  {\bibfnamefont {S.}~\bibnamefont {Wang}}, \bibinfo {author} {\bibfnamefont
  {J.}~\bibnamefont {Xue}}, \bibinfo {author} {\bibfnamefont {K.}~\bibnamefont
  {Yang}}, \bibinfo {author} {\bibfnamefont {O.}~\bibnamefont {Levy}}, \bibinfo
  {author} {\bibfnamefont {M.~J.}\ \bibnamefont {Mehl}}, \bibinfo {author}
  {\bibfnamefont {H.~T.}\ \bibnamefont {Stokes}}, \bibinfo {author}
  {\bibfnamefont {D.~O.}\ \bibnamefont {Demchenko}}, \ and\ \bibinfo {author}
  {\bibfnamefont {D.}~\bibnamefont {Morgan}},\ }\bibfield  {title} {\enquote
  {\bibinfo {title} {{AFLOW: An automatic framework for high-throughput
  materials discovery}},}\ }\href {\doibase
  http://dx.doi.org/10.1016/j.commatsci.2012.02.005} {\bibfield  {journal}
  {\bibinfo  {journal} {Computational Materials Science}\ }\textbf {\bibinfo
  {volume} {58}},\ \bibinfo {pages} {218--226} (\bibinfo {year}
  {2012})}\BibitemShut {NoStop}%
\bibitem [{\citenamefont {Jain}\ \emph {et~al.}(2013)\citenamefont {Jain},
  \citenamefont {Ong}, \citenamefont {Hautier}, \citenamefont {Chen},
  \citenamefont {Richards}, \citenamefont {Dacek}, \citenamefont {Cholia},
  \citenamefont {Gunter}, \citenamefont {Skinner}, \citenamefont {Ceder},\ and\
  \citenamefont {Persson}}]{Jain2013}%
  \BibitemOpen
  \bibfield  {author} {\bibinfo {author} {\bibfnamefont {A.}~\bibnamefont
  {Jain}}, \bibinfo {author} {\bibfnamefont {S.~P.}\ \bibnamefont {Ong}},
  \bibinfo {author} {\bibfnamefont {G.}~\bibnamefont {Hautier}}, \bibinfo
  {author} {\bibfnamefont {W.}~\bibnamefont {Chen}}, \bibinfo {author}
  {\bibfnamefont {W.~D.}\ \bibnamefont {Richards}}, \bibinfo {author}
  {\bibfnamefont {S.}~\bibnamefont {Dacek}}, \bibinfo {author} {\bibfnamefont
  {S.}~\bibnamefont {Cholia}}, \bibinfo {author} {\bibfnamefont
  {D.}~\bibnamefont {Gunter}}, \bibinfo {author} {\bibfnamefont
  {D.}~\bibnamefont {Skinner}}, \bibinfo {author} {\bibfnamefont
  {G.}~\bibnamefont {Ceder}}, \ and\ \bibinfo {author} {\bibfnamefont {K.~A.}\
  \bibnamefont {Persson}},\ }\bibfield  {title} {\enquote {\bibinfo {title}
  {{Commentary: The Materials Project: A materials genome approach to
  accelerating materials innovation}},}\ }\href {\doibase
  http://dx.doi.org/10.1063/1.4812323} {\bibfield  {journal} {\bibinfo
  {journal} {APL Materials}\ }\textbf {\bibinfo {volume} {1}},\ \bibinfo {eid}
  {011002} (\bibinfo {year} {2013})}\BibitemShut {NoStop}%
\bibitem [{\citenamefont {Konkov}\ and\ \citenamefont
  {Peverati}(2019)}]{KonkovQMC2019}%
  \BibitemOpen
  \bibfield  {author} {\bibinfo {author} {\bibfnamefont {V.}~\bibnamefont
  {Konkov}}\ and\ \bibinfo {author} {\bibfnamefont {R.}~\bibnamefont
  {Peverati}},\ }\bibfield  {title} {\enquote {\bibinfo {title} {{QMC}-{SW}: A
  simple workflow for quantum monte carlo calculations in chemistry},}\ }\href
  {\doibase 10.1016/j.softx.2018.11.001} {\bibfield  {journal} {\bibinfo
  {journal} {{SoftwareX}}\ }\textbf {\bibinfo {volume} {9}},\ \bibinfo {pages}
  {7--14} (\bibinfo {year} {2019})}\BibitemShut {NoStop}%
\bibitem [{\citenamefont {Saritas}\ \emph {et~al.}(2017)\citenamefont
  {Saritas}, \citenamefont {Mueller}, \citenamefont {Wagner},\ and\
  \citenamefont {Grossman}}]{Saritas2017}%
  \BibitemOpen
  \bibfield  {author} {\bibinfo {author} {\bibfnamefont {K.}~\bibnamefont
  {Saritas}}, \bibinfo {author} {\bibfnamefont {T.}~\bibnamefont {Mueller}},
  \bibinfo {author} {\bibfnamefont {L.}~\bibnamefont {Wagner}}, \ and\ \bibinfo
  {author} {\bibfnamefont {J.~C.}\ \bibnamefont {Grossman}},\ }\bibfield
  {title} {\enquote {\bibinfo {title} {{Investigation of a Quantum Monte Carlo
  Protocol To Achieve High Accuracy and High-Throughput Materials Formation
  Energies}},}\ }\href {\doibase 10.1021/acs.jctc.6b01179} {\bibfield
  {journal} {\bibinfo  {journal} {Journal of Chemical Theory and Computation}\
  }\textbf {\bibinfo {volume} {13}},\ \bibinfo {pages} {1943--1951} (\bibinfo
  {year} {2017})}\BibitemShut {NoStop}%
\bibitem [{\citenamefont {Krogel}(2016)}]{Krogel2016Nexus}%
  \BibitemOpen
  \bibfield  {author} {\bibinfo {author} {\bibfnamefont {J.~T.}\ \bibnamefont
  {Krogel}},\ }\bibfield  {title} {\enquote {\bibinfo {title} {{Nexus: A
  modular workflow management system for quantum simulation codes}},}\ }\href
  {\doibase http://dx.doi.org/10.1016/j.cpc.2015.08.012} {\bibfield  {journal}
  {\bibinfo  {journal} {Computer Physics Communications}\ }\textbf {\bibinfo
  {volume} {198}},\ \bibinfo {pages} {154--168} (\bibinfo {year}
  {2016})}\BibitemShut {NoStop}%
\bibitem [{\citenamefont {Ceperley}(1986)}]{Ceperley1986}%
  \BibitemOpen
  \bibfield  {author} {\bibinfo {author} {\bibfnamefont {D.~M.}\ \bibnamefont
  {Ceperley}},\ }\bibfield  {title} {\enquote {\bibinfo {title} {{The
  statistical error of green's function Monte Carlo}},}\ }\href {\doibase
  10.1007/BF02628307} {\bibfield  {journal} {\bibinfo  {journal} {Journal of
  Statistical Physics}\ }\textbf {\bibinfo {volume} {43}},\ \bibinfo {pages}
  {815--826} (\bibinfo {year} {1986})}\BibitemShut {NoStop}%
\bibitem [{\citenamefont {Hammond}, \citenamefont {Reynolds},\ and\
  \citenamefont {Lester}(1987)}]{Hammond1987}%
  \BibitemOpen
  \bibfield  {author} {\bibinfo {author} {\bibfnamefont {B.~L.}\ \bibnamefont
  {Hammond}}, \bibinfo {author} {\bibfnamefont {P.~J.}\ \bibnamefont
  {Reynolds}}, \ and\ \bibinfo {author} {\bibfnamefont {W.~A.}\ \bibnamefont
  {Lester}},\ }\bibfield  {title} {\enquote {\bibinfo {title} {{Valence quantum
  Monte Carlo with ab initio effective core potentials}},}\ }\href {\doibase
  10.1063/1.453345} {\bibfield  {journal} {\bibinfo  {journal} {The Journal of
  Chemical Physics}\ }\textbf {\bibinfo {volume} {87}},\ \bibinfo {pages}
  {1130--1136} (\bibinfo {year} {1987})}\BibitemShut {NoStop}%
\bibitem [{\citenamefont {Bennett}\ \emph {et~al.}(2017)\citenamefont
  {Bennett}, \citenamefont {Melton}, \citenamefont {Annaberdiyev},
  \citenamefont {Wang}, \citenamefont {Shulenburger},\ and\ \citenamefont
  {Mitas}}]{Bennett2017}%
  \BibitemOpen
  \bibfield  {author} {\bibinfo {author} {\bibfnamefont {M.~C.}\ \bibnamefont
  {Bennett}}, \bibinfo {author} {\bibfnamefont {C.~A.}\ \bibnamefont {Melton}},
  \bibinfo {author} {\bibfnamefont {A.}~\bibnamefont {Annaberdiyev}}, \bibinfo
  {author} {\bibfnamefont {G.}~\bibnamefont {Wang}}, \bibinfo {author}
  {\bibfnamefont {L.}~\bibnamefont {Shulenburger}}, \ and\ \bibinfo {author}
  {\bibfnamefont {L.}~\bibnamefont {Mitas}},\ }\bibfield  {title} {\enquote
  {\bibinfo {title} {{A new generation of effective core potentials for
  correlated calculations}},}\ }\href {\doibase 10.1063/1.4995643} {\bibfield
  {journal} {\bibinfo  {journal} {The Journal of Chemical Physics}\ }\textbf
  {\bibinfo {volume} {147}},\ \bibinfo {pages} {224106} (\bibinfo {year}
  {2017})}\BibitemShut {NoStop}%
\bibitem [{\citenamefont {Bennett}\ \emph {et~al.}(2018)\citenamefont
  {Bennett}, \citenamefont {Wang}, \citenamefont {Annaberdiyev}, \citenamefont
  {Melton}, \citenamefont {Shulenburger},\ and\ \citenamefont
  {Mitas}}]{Bennett2018}%
  \BibitemOpen
  \bibfield  {author} {\bibinfo {author} {\bibfnamefont {M.~C.}\ \bibnamefont
  {Bennett}}, \bibinfo {author} {\bibfnamefont {G.}~\bibnamefont {Wang}},
  \bibinfo {author} {\bibfnamefont {A.}~\bibnamefont {Annaberdiyev}}, \bibinfo
  {author} {\bibfnamefont {C.~A.}\ \bibnamefont {Melton}}, \bibinfo {author}
  {\bibfnamefont {L.}~\bibnamefont {Shulenburger}}, \ and\ \bibinfo {author}
  {\bibfnamefont {L.}~\bibnamefont {Mitas}},\ }\bibfield  {title} {\enquote
  {\bibinfo {title} {{A new generation of effective core potentials from
  correlated calculations: 2\textsuperscript{nd} row elements}},}\ }\href
  {\doibase 10.1063/1.5038135} {\bibfield  {journal} {\bibinfo  {journal} {The
  Journal of Chemical Physics}\ }\textbf {\bibinfo {volume} {149}},\ \bibinfo
  {pages} {104108} (\bibinfo {year} {2018})}\BibitemShut {NoStop}%
\bibitem [{\citenamefont {Annaberdiyev}\ \emph {et~al.}(2018)\citenamefont
  {Annaberdiyev}, \citenamefont {Wang}, \citenamefont {Melton}, \citenamefont
  {Chandler~Bennett}, \citenamefont {Shulenburger},\ and\ \citenamefont
  {Mitas}}]{Annaberdiyev2018}%
  \BibitemOpen
  \bibfield  {author} {\bibinfo {author} {\bibfnamefont {A.}~\bibnamefont
  {Annaberdiyev}}, \bibinfo {author} {\bibfnamefont {G.}~\bibnamefont {Wang}},
  \bibinfo {author} {\bibfnamefont {C.~A.}\ \bibnamefont {Melton}}, \bibinfo
  {author} {\bibfnamefont {M.}~\bibnamefont {Chandler~Bennett}}, \bibinfo
  {author} {\bibfnamefont {L.}~\bibnamefont {Shulenburger}}, \ and\ \bibinfo
  {author} {\bibfnamefont {L.}~\bibnamefont {Mitas}},\ }\bibfield  {title}
  {\enquote {\bibinfo {title} {{A new generation of effective core potentials
  from correlated calculations: 3d transition metal series}},}\ }\href
  {\doibase 10.1063/1.5040472} {\bibfield  {journal} {\bibinfo  {journal} {The
  Journal of Chemical Physics}\ }\textbf {\bibinfo {volume} {149}},\ \bibinfo
  {pages} {134108} (\bibinfo {year} {2018})}\BibitemShut {NoStop}%
\bibitem [{\citenamefont {Wang}\ \emph {et~al.}(2019)\citenamefont {Wang},
  \citenamefont {Annaberdiyev}, \citenamefont {Melton}, \citenamefont
  {Bennett}, \citenamefont {Shulenburger},\ and\ \citenamefont
  {Mitas}}]{Wang2019}%
  \BibitemOpen
  \bibfield  {author} {\bibinfo {author} {\bibfnamefont {G.}~\bibnamefont
  {Wang}}, \bibinfo {author} {\bibfnamefont {A.}~\bibnamefont {Annaberdiyev}},
  \bibinfo {author} {\bibfnamefont {C.~A.}\ \bibnamefont {Melton}}, \bibinfo
  {author} {\bibfnamefont {M.~C.}\ \bibnamefont {Bennett}}, \bibinfo {author}
  {\bibfnamefont {L.}~\bibnamefont {Shulenburger}}, \ and\ \bibinfo {author}
  {\bibfnamefont {L.}~\bibnamefont {Mitas}},\ }\bibfield  {title} {\enquote
  {\bibinfo {title} {{A new generation of effective core potentials from
  correlated calculations: 4s and 4p main group elements and first row
  additions}},}\ }\href {\doibase 10.1063/1.5121006} {\bibfield  {journal}
  {\bibinfo  {journal} {The Journal of Chemical Physics}\ }\textbf {\bibinfo
  {volume} {151}},\ \bibinfo {pages} {144110} (\bibinfo {year}
  {2019})}\BibitemShut {NoStop}%
\bibitem [{\citenamefont {Burkatzki}, \citenamefont {Filippi},\ and\
  \citenamefont {Dolg}(2008)}]{BFD_2008}%
  \BibitemOpen
  \bibfield  {author} {\bibinfo {author} {\bibfnamefont {M.}~\bibnamefont
  {Burkatzki}}, \bibinfo {author} {\bibfnamefont {C.}~\bibnamefont {Filippi}},
  \ and\ \bibinfo {author} {\bibfnamefont {M.}~\bibnamefont {Dolg}},\
  }\bibfield  {title} {\enquote {\bibinfo {title} {{Energy-consistent
  small-core pseudopotentials for 3d-transition metals adapted to quantum
  {Monte} {Carlo} calculations}},}\ }\href {\doibase 10.1063/1.2987872}
  {\bibfield  {journal} {\bibinfo  {journal} {The Journal of Chemical Physics}\
  }\textbf {\bibinfo {volume} {129}},\ \bibinfo {pages} {164115} (\bibinfo
  {year} {2008})}\BibitemShut {NoStop}%
\bibitem [{\citenamefont {Dolg}\ \emph {et~al.}(1987)\citenamefont {Dolg},
  \citenamefont {Wedig}, \citenamefont {Stoll},\ and\ \citenamefont
  {Preuss}}]{STU_1987}%
  \BibitemOpen
  \bibfield  {author} {\bibinfo {author} {\bibfnamefont {M.}~\bibnamefont
  {Dolg}}, \bibinfo {author} {\bibfnamefont {U.}~\bibnamefont {Wedig}},
  \bibinfo {author} {\bibfnamefont {H.}~\bibnamefont {Stoll}}, \ and\ \bibinfo
  {author} {\bibfnamefont {H.}~\bibnamefont {Preuss}},\ }\bibfield  {title}
  {\enquote {\bibinfo {title} {{Energy-adjusted ab initio pseudopotentials for
  the first row transition elements}},}\ }\href {\doibase 10.1063/1.452288}
  {\bibfield  {journal} {\bibinfo  {journal} {The Journal of Chemical Physics}\
  }\textbf {\bibinfo {volume} {86}},\ \bibinfo {pages} {866--872} (\bibinfo
  {year} {1987})}\BibitemShut {NoStop}%
\bibitem [{\citenamefont {Trail}\ and\ \citenamefont {Needs}(2017)}]{TN_2017}%
  \BibitemOpen
  \bibfield  {author} {\bibinfo {author} {\bibfnamefont {J.~R.}\ \bibnamefont
  {Trail}}\ and\ \bibinfo {author} {\bibfnamefont {R.~J.}\ \bibnamefont
  {Needs}},\ }\bibfield  {title} {\enquote {\bibinfo {title} {{Shape and energy
  consistent pseudopotentials for correlated electron systems}},}\ }\href
  {\doibase 10.1063/1.4984046} {\bibfield  {journal} {\bibinfo  {journal} {The
  Journal of Chemical Physics}\ }\textbf {\bibinfo {volume} {146}},\ \bibinfo
  {pages} {204107} (\bibinfo {year} {2017})}\BibitemShut {NoStop}%
\bibitem [{\citenamefont {Fernandez~Pacios}\ and\ \citenamefont
  {Christiansen}(1985)}]{CRENBL_1985}%
  \BibitemOpen
  \bibfield  {author} {\bibinfo {author} {\bibfnamefont {L.}~\bibnamefont
  {Fernandez~Pacios}}\ and\ \bibinfo {author} {\bibfnamefont {P.~A.}\
  \bibnamefont {Christiansen}},\ }\bibfield  {title} {\enquote {\bibinfo
  {title} {Ab initio relativistic effective potentials with spin-orbit
  operators. {I}. {Li} through {Ar}},}\ }\href {\doibase 10.1063/1.448263}
  {\bibfield  {journal} {\bibinfo  {journal} {The Journal of Chemical Physics}\
  }\textbf {\bibinfo {volume} {82}},\ \bibinfo {pages} {2664--2671} (\bibinfo
  {year} {1985})}\BibitemShut {NoStop}%
\bibitem [{\citenamefont {Stevens}, \citenamefont {Basch},\ and\ \citenamefont
  {Krauss}(1984)}]{SBKJC_1984}%
  \BibitemOpen
  \bibfield  {author} {\bibinfo {author} {\bibfnamefont {W.~J.}\ \bibnamefont
  {Stevens}}, \bibinfo {author} {\bibfnamefont {H.}~\bibnamefont {Basch}}, \
  and\ \bibinfo {author} {\bibfnamefont {M.}~\bibnamefont {Krauss}},\
  }\bibfield  {title} {\enquote {\bibinfo {title} {{Compact effective
  potentials and efficient shared-exponent basis sets for the first- and
  second-row atoms}},}\ }\href {\doibase 10.1063/1.447604} {\bibfield
  {journal} {\bibinfo  {journal} {The Journal of Chemical Physics}\ }\textbf
  {\bibinfo {volume} {81}},\ \bibinfo {pages} {6026--6033} (\bibinfo {year}
  {1984})}\BibitemShut {NoStop}%
\bibitem [{\citenamefont {Annaberdiyev}\ \emph {et~al.}(2020)\citenamefont
  {Annaberdiyev}, \citenamefont {Melton}, \citenamefont {Bennett},
  \citenamefont {Wang},\ and\ \citenamefont
  {Mitas}}]{annaberdiyev_accurate_2019}%
  \BibitemOpen
  \bibfield  {author} {\bibinfo {author} {\bibfnamefont {A.}~\bibnamefont
  {Annaberdiyev}}, \bibinfo {author} {\bibfnamefont {C.~A.}\ \bibnamefont
  {Melton}}, \bibinfo {author} {\bibfnamefont {M.~C.}\ \bibnamefont {Bennett}},
  \bibinfo {author} {\bibfnamefont {G.}~\bibnamefont {Wang}}, \ and\ \bibinfo
  {author} {\bibfnamefont {L.}~\bibnamefont {Mitas}},\ }\bibfield  {title}
  {\enquote {\bibinfo {title} {{Accurate Atomic Correlation and Total Energies
  for Correlation Consistent Effective Core Potentials}},}\ }\href {\doibase
  10.1021/acs.jctc.9b00962} {\bibfield  {journal} {\bibinfo  {journal} {Journal
  of Chemical Theory and Computation}\ }\textbf {\bibinfo {volume} {16}},\
  \bibinfo {pages} {1482} (\bibinfo {year} {2020})}\BibitemShut {NoStop}%
\bibitem [{\citenamefont {Casula}(2006)}]{CasulaLocality2006}%
  \BibitemOpen
  \bibfield  {author} {\bibinfo {author} {\bibfnamefont {M.}~\bibnamefont
  {Casula}},\ }\bibfield  {title} {\enquote {\bibinfo {title} {Beyond the
  locality approximation in the standard diffusion monte carlo method},}\
  }\href {\doibase 10.1103/physrevb.74.161102} {\bibfield  {journal} {\bibinfo
  {journal} {Physical Review B}\ }\textbf {\bibinfo {volume} {74}} (\bibinfo
  {year} {2006}),\ 10.1103/physrevb.74.161102}\BibitemShut {NoStop}%
\bibitem [{\citenamefont {Dunning}, \citenamefont {Peterson},\ and\
  \citenamefont {Woon}(2002)}]{dunning_basis_2002}%
  \BibitemOpen
  \bibfield  {author} {\bibinfo {author} {\bibfnamefont {T.~H.}\ \bibnamefont
  {Dunning}}, \bibinfo {author} {\bibfnamefont {K.~A.}\ \bibnamefont
  {Peterson}}, \ and\ \bibinfo {author} {\bibfnamefont {D.~E.}\ \bibnamefont
  {Woon}},\ }\bibfield  {title} {\enquote {\bibinfo {title} {Basis {Sets}:
  {Correlation} {Consistent} {Sets}},}\ }in\ \href {\doibase
  10.1002/0470845015.cca053} {\emph {\bibinfo {booktitle} {Encyclopedia of
  {Computational} {Chemistry}}}}\ (\bibinfo  {publisher} {American Cancer
  Society},\ \bibinfo {year} {2002})\BibitemShut {NoStop}%
\bibitem [{\citenamefont {Kleinman}\ and\ \citenamefont
  {Bylander}(1982)}]{Kleinman1982}%
  \BibitemOpen
  \bibfield  {author} {\bibinfo {author} {\bibfnamefont {L.}~\bibnamefont
  {Kleinman}}\ and\ \bibinfo {author} {\bibfnamefont {D.~M.}\ \bibnamefont
  {Bylander}},\ }\bibfield  {title} {\enquote {\bibinfo {title} {{Efficacious
  Form for Model Pseudopotentials}},}\ }\href {\doibase
  10.1103/PhysRevLett.48.1425} {\bibfield  {journal} {\bibinfo  {journal}
  {Physical Review Letters}\ }\textbf {\bibinfo {volume} {48}},\ \bibinfo
  {pages} {1425--1428} (\bibinfo {year} {1982})}\BibitemShut {NoStop}%
\bibitem [{\citenamefont {Wang}, \citenamefont {Zhou},\ and\ \citenamefont
  {Wang}(2019)}]{WangTingWang2019}%
  \BibitemOpen
  \bibfield  {author} {\bibinfo {author} {\bibfnamefont {T.}~\bibnamefont
  {Wang}}, \bibinfo {author} {\bibfnamefont {X.}~\bibnamefont {Zhou}}, \ and\
  \bibinfo {author} {\bibfnamefont {F.}~\bibnamefont {Wang}},\ }\bibfield
  {title} {\enquote {\bibinfo {title} {{Performance of the Diffusion Quantum
  Monte Carlo Method with a Single-Slater-Jastrow Trial Wavefunction Using
  Natural Orbitals and Density Functional Theory Orbitals on Atomization
  Energies of the Gaussian-2 Set}},}\ }\href {\doibase
  10.1021/acs.jpca.9b01933} {\bibfield  {journal} {\bibinfo  {journal} {The
  Journal of Physical Chemistry A}\ }\textbf {\bibinfo {volume} {123}},\
  \bibinfo {pages} {3809--3817} (\bibinfo {year} {2019})}\BibitemShut {NoStop}%
\bibitem [{\citenamefont {Zhou}\ \emph {et~al.}(2019)\citenamefont {Zhou},
  \citenamefont {Zhao}, \citenamefont {Wang},\ and\ \citenamefont
  {Wang}}]{zhou_diffusion_2019}%
  \BibitemOpen
  \bibfield  {author} {\bibinfo {author} {\bibfnamefont {X.}~\bibnamefont
  {Zhou}}, \bibinfo {author} {\bibfnamefont {H.}~\bibnamefont {Zhao}}, \bibinfo
  {author} {\bibfnamefont {T.}~\bibnamefont {Wang}}, \ and\ \bibinfo {author}
  {\bibfnamefont {F.}~\bibnamefont {Wang}},\ }\bibfield  {title} {\enquote
  {\bibinfo {title} {{Diffusion quantum {Monte} {Carlo} calculations with a
  recent generation of effective core potentials for ionization potentials and
  electron affinities}},}\ }\href {\doibase 10.1103/PhysRevA.100.062502}
  {\bibfield  {journal} {\bibinfo  {journal} {Physical Review A}\ }\textbf
  {\bibinfo {volume} {100}},\ \bibinfo {pages} {062502} (\bibinfo {year}
  {2019})}\BibitemShut {NoStop}%
\bibitem [{\citenamefont {Zhang}\ and\ \citenamefont
  {Krakauer}(2003)}]{Zhang_phaseless}%
  \BibitemOpen
  \bibfield  {author} {\bibinfo {author} {\bibfnamefont {S.}~\bibnamefont
  {Zhang}}\ and\ \bibinfo {author} {\bibfnamefont {H.}~\bibnamefont
  {Krakauer}},\ }\bibfield  {title} {\enquote {\bibinfo {title} {{Quantum Monte
  Carlo Method using Phase-Free Random Walks with Slater Determinants}},}\
  }\href {\doibase 10.1103/PhysRevLett.90.136401} {\bibfield  {journal}
  {\bibinfo  {journal} {Physical Review Letters}\ }\textbf {\bibinfo {volume}
  {90}},\ \bibinfo {pages} {136401} (\bibinfo {year} {2003})}\BibitemShut
  {NoStop}%
\bibitem [{\citenamefont {Landinez~Borda}, \citenamefont {Gomez},\ and\
  \citenamefont {Morales}(2019)}]{borda_nomsd}%
  \BibitemOpen
  \bibfield  {author} {\bibinfo {author} {\bibfnamefont {E.~J.}\ \bibnamefont
  {Landinez~Borda}}, \bibinfo {author} {\bibfnamefont {J.}~\bibnamefont
  {Gomez}}, \ and\ \bibinfo {author} {\bibfnamefont {M.~A.}\ \bibnamefont
  {Morales}},\ }\bibfield  {title} {\enquote {\bibinfo {title} {{Non-orthogonal
  multi-Slater determinant expansions in auxiliary field quantum Monte
  Carlo}},}\ }\href {\doibase 10.1063/1.5049143} {\bibfield  {journal}
  {\bibinfo  {journal} {The Journal of Chemical Physics}\ }\textbf {\bibinfo
  {volume} {150}},\ \bibinfo {pages} {074105} (\bibinfo {year}
  {2019})}\BibitemShut {NoStop}%
\bibitem [{\citenamefont {Lee}, \citenamefont {Malone},\ and\ \citenamefont
  {Morales}(2020)}]{LeeSymmetryBreaking2020}%
  \BibitemOpen
  \bibfield  {author} {\bibinfo {author} {\bibfnamefont {J.}~\bibnamefont
  {Lee}}, \bibinfo {author} {\bibfnamefont {F.~D.}\ \bibnamefont {Malone}}, \
  and\ \bibinfo {author} {\bibfnamefont {M.~A.}\ \bibnamefont {Morales}},\
  }\bibfield  {title} {\enquote {\bibinfo {title} {Utilizing essential symmetry
  breaking in auxiliary-field quantum monte carlo: Application to the spin gaps
  of the c$_{36}$ fullerene and an iron porphyrin model complex},}\ }\href@noop
  {} {\bibfield  {journal} {\bibinfo  {journal} {arXiv preprint
  arXiv:2001.05109}\ } (\bibinfo {year} {2020})}\BibitemShut {NoStop}%
\bibitem [{\citenamefont {Zhang}, \citenamefont {Malone},\ and\ \citenamefont
  {Morales}(2018)}]{zhang_nio}%
  \BibitemOpen
  \bibfield  {author} {\bibinfo {author} {\bibfnamefont {S.}~\bibnamefont
  {Zhang}}, \bibinfo {author} {\bibfnamefont {F.~D.}\ \bibnamefont {Malone}}, \
  and\ \bibinfo {author} {\bibfnamefont {M.~A.}\ \bibnamefont {Morales}},\
  }\bibfield  {title} {\enquote {\bibinfo {title} {{Auxiliary-field quantum
  Monte Carlo calculations of the structural properties of nickel oxide}},}\
  }\href {\doibase 10.1063/1.5040900} {\bibfield  {journal} {\bibinfo
  {journal} {The Journal of Chemical Physics}\ }\textbf {\bibinfo {volume}
  {149}},\ \bibinfo {pages} {164102} (\bibinfo {year} {2018})}\BibitemShut
  {NoStop}%
\bibitem [{\citenamefont {Malone}, \citenamefont {Zhang},\ and\ \citenamefont
  {Morales}(2019)}]{malone_isdf}%
  \BibitemOpen
  \bibfield  {author} {\bibinfo {author} {\bibfnamefont {F.~D.}\ \bibnamefont
  {Malone}}, \bibinfo {author} {\bibfnamefont {S.}~\bibnamefont {Zhang}}, \
  and\ \bibinfo {author} {\bibfnamefont {M.~A.}\ \bibnamefont {Morales}},\
  }\bibfield  {title} {\enquote {\bibinfo {title} {{Overcoming the Memory
  Bottleneck in Auxiliary Field Quantum Monte Carlo Simulations with
  Interpolative Separable Density Fitting}},}\ }\href {\doibase
  10.1021/acs.jctc.8b00944} {\bibfield  {journal} {\bibinfo  {journal} {J.
  Chem. Theory. Comput.}\ }\textbf {\bibinfo {volume} {15}},\ \bibinfo {pages}
  {256} (\bibinfo {year} {2019})}\BibitemShut {NoStop}%
\bibitem [{\citenamefont {Lee}, \citenamefont {Malone},\ and\ \citenamefont
  {Morales}(2019)}]{lee_2019_UEG}%
  \BibitemOpen
  \bibfield  {author} {\bibinfo {author} {\bibfnamefont {J.}~\bibnamefont
  {Lee}}, \bibinfo {author} {\bibfnamefont {F.~D.}\ \bibnamefont {Malone}}, \
  and\ \bibinfo {author} {\bibfnamefont {M.~A.}\ \bibnamefont {Morales}},\
  }\bibfield  {title} {\enquote {\bibinfo {title} {{An auxiliary-Field quantum
  Monte Carlo perspective on the ground state of the dense uniform electron
  gas: An investigation with Hartree-Fock trial wavefunctions}},}\ }\href
  {https://doi.org/10.1063/1.5109572} {\bibfield  {journal} {\bibinfo
  {journal} {The Journal of Chemical Physics}\ }\textbf {\bibinfo {volume}
  {151}},\ \bibinfo {pages} {064122} (\bibinfo {year} {2019})}\BibitemShut
  {NoStop}%
\bibitem [{\citenamefont {Hubbard}(1959)}]{hubbard_strat}%
  \BibitemOpen
  \bibfield  {author} {\bibinfo {author} {\bibfnamefont {J.}~\bibnamefont
  {Hubbard}},\ }\bibfield  {title} {\enquote {\bibinfo {title} {{Calculation of
  Partition Functions}},}\ }\href
  {https://link.aps.org/doi/10.1103/PhysRevLett.3.77} {\bibfield  {journal}
  {\bibinfo  {journal} {Physical Review Letters}\ }\textbf {\bibinfo {volume}
  {3}},\ \bibinfo {pages} {77} (\bibinfo {year} {1959})}\BibitemShut {NoStop}%
\bibitem [{\citenamefont {Zhang}, \citenamefont {Carlson},\ and\ \citenamefont
  {Gubernatis}(1997)}]{zhang_cpmc}%
  \BibitemOpen
  \bibfield  {author} {\bibinfo {author} {\bibfnamefont {S.}~\bibnamefont
  {Zhang}}, \bibinfo {author} {\bibfnamefont {J.}~\bibnamefont {Carlson}}, \
  and\ \bibinfo {author} {\bibfnamefont {J.~E.}\ \bibnamefont {Gubernatis}},\
  }\bibfield  {title} {\enquote {\bibinfo {title} {{Constrained path Monte
  Carlo method for fermion ground states}},}\ }\href {\doibase
  10.1103/PhysRevB.55.7464} {\bibfield  {journal} {\bibinfo  {journal}
  {Physical Review B}\ }\textbf {\bibinfo {volume} {55}},\ \bibinfo {pages}
  {7464} (\bibinfo {year} {1997})}\BibitemShut {NoStop}%
\bibitem [{\citenamefont {Beebe}\ and\ \citenamefont
  {Linderberg}(1977)}]{modified_chol_1}%
  \BibitemOpen
  \bibfield  {author} {\bibinfo {author} {\bibfnamefont {N.~H.~F.}\
  \bibnamefont {Beebe}}\ and\ \bibinfo {author} {\bibfnamefont
  {J.}~\bibnamefont {Linderberg}},\ }\bibfield  {title} {\enquote {\bibinfo
  {title} {{Simplifications in the generation and transformation of
  two{-}electron integrals in molecular calculations}},}\ }\href
  {https://onlinelibrary.wiley.com/doi/abs/10.1002/qua.560120408} {\bibfield
  {journal} {\bibinfo  {journal} {Int. J. Quantum Chem.}\ }\textbf {\bibinfo
  {volume} {12}},\ \bibinfo {pages} {683} (\bibinfo {year} {1977})}\BibitemShut
  {NoStop}%
\bibitem [{\citenamefont {Koch}, \citenamefont {de~Mer{\'a}s},\ and\
  \citenamefont {Pedersen}(2003)}]{modified_chol_2}%
  \BibitemOpen
  \bibfield  {author} {\bibinfo {author} {\bibfnamefont {H.}~\bibnamefont
  {Koch}}, \bibinfo {author} {\bibfnamefont {A.~S.}\ \bibnamefont
  {de~Mer{\'a}s}}, \ and\ \bibinfo {author} {\bibfnamefont {T.~B.}\
  \bibnamefont {Pedersen}},\ }\bibfield  {title} {\enquote {\bibinfo {title}
  {{Reduced scaling in electronic structure calculations using Cholesky
  decompositions}},}\ }\href {https://doi.org/10.1063/1.1578621} {\bibfield
  {journal} {\bibinfo  {journal} {The Journal of Chemical Physics}\ }\textbf
  {\bibinfo {volume} {118}},\ \bibinfo {pages} {9481} (\bibinfo {year}
  {2003})}\BibitemShut {NoStop}%
\bibitem [{\citenamefont {Aquilante}\ \emph {et~al.}(2009)\citenamefont
  {Aquilante}, \citenamefont {De~Vico}, \citenamefont {Ferr{\'e}},
  \citenamefont {Ghigo}, \citenamefont {Malmqvist}, \citenamefont
  {Neogr{\'a}dy}, \citenamefont {Pedersen}, \citenamefont {Pito{\v n}{\'a}k},
  \citenamefont {Reiher}, \citenamefont {Roos}, \citenamefont
  {Serrano{-}Andr{\'e}s}, \citenamefont {Urban}, \citenamefont {Veryazov},\
  and\ \citenamefont {Lindh}}]{modified_chol_3}%
  \BibitemOpen
  \bibfield  {author} {\bibinfo {author} {\bibfnamefont {F.}~\bibnamefont
  {Aquilante}}, \bibinfo {author} {\bibfnamefont {L.}~\bibnamefont {De~Vico}},
  \bibinfo {author} {\bibfnamefont {N.}~\bibnamefont {Ferr{\'e}}}, \bibinfo
  {author} {\bibfnamefont {G.}~\bibnamefont {Ghigo}}, \bibinfo {author}
  {\bibfnamefont {P.-{\aa}.}\ \bibnamefont {Malmqvist}}, \bibinfo {author}
  {\bibfnamefont {P.}~\bibnamefont {Neogr{\'a}dy}}, \bibinfo {author}
  {\bibfnamefont {T.~B.}\ \bibnamefont {Pedersen}}, \bibinfo {author}
  {\bibfnamefont {M.}~\bibnamefont {Pito{\v n}{\'a}k}}, \bibinfo {author}
  {\bibfnamefont {M.}~\bibnamefont {Reiher}}, \bibinfo {author} {\bibfnamefont
  {B.~O.}\ \bibnamefont {Roos}}, \bibinfo {author} {\bibfnamefont
  {L.}~\bibnamefont {Serrano{-}Andr{\'e}s}}, \bibinfo {author} {\bibfnamefont
  {M.}~\bibnamefont {Urban}}, \bibinfo {author} {\bibfnamefont
  {V.}~\bibnamefont {Veryazov}}, \ and\ \bibinfo {author} {\bibfnamefont
  {R.}~\bibnamefont {Lindh}},\ }\bibfield  {title} {\enquote {\bibinfo {title}
  {{MOLCAS 7: The Next Generation}},}\ }\href
  {https://onlinelibrary.wiley.com/doi/abs/10.1002/jcc.21318} {\bibfield
  {journal} {\bibinfo  {journal} {J. Comput. Chem.}\ }\textbf {\bibinfo
  {volume} {31}},\ \bibinfo {pages} {224} (\bibinfo {year} {2009})}\BibitemShut
  {NoStop}%
\bibitem [{\citenamefont {Purwanto}\ \emph {et~al.}(2011)\citenamefont
  {Purwanto}, \citenamefont {Krakauer}, \citenamefont {Virgus},\ and\
  \citenamefont {Zhang}}]{purwanto_cholesky}%
  \BibitemOpen
  \bibfield  {author} {\bibinfo {author} {\bibfnamefont {W.}~\bibnamefont
  {Purwanto}}, \bibinfo {author} {\bibfnamefont {H.}~\bibnamefont {Krakauer}},
  \bibinfo {author} {\bibfnamefont {Y.}~\bibnamefont {Virgus}}, \ and\ \bibinfo
  {author} {\bibfnamefont {S.}~\bibnamefont {Zhang}},\ }\bibfield  {title}
  {\enquote {\bibinfo {title} {{Assessing weak hydrogen binding on Ca+ centers:
  An accurate many-body study with large basis sets}},}\ }\href {\doibase
  10.1063/1.3654002} {\bibfield  {journal} {\bibinfo  {journal} {The Journal of
  Chemical Physics}\ }\textbf {\bibinfo {volume} {135}},\ \bibinfo {pages}
  {164105} (\bibinfo {year} {2011})}\BibitemShut {NoStop}%
\bibitem [{\citenamefont {Purwanto}, \citenamefont {Zhang},\ and\ \citenamefont
  {Krakauer}(2013)}]{purwanto_downfolding_jctc}%
  \BibitemOpen
  \bibfield  {author} {\bibinfo {author} {\bibfnamefont {W.}~\bibnamefont
  {Purwanto}}, \bibinfo {author} {\bibfnamefont {S.}~\bibnamefont {Zhang}}, \
  and\ \bibinfo {author} {\bibfnamefont {H.}~\bibnamefont {Krakauer}},\
  }\bibfield  {title} {\enquote {\bibinfo {title} {{Frozen-Orbital and
  Downfolding Calculations with Auxiliary-Field Quantum Monte Carlo}},}\ }\href
  {https://doi.org/10.1021/ct4006486} {\bibfield  {journal} {\bibinfo
  {journal} {Journal of Chemical Theory and Computation}\ }\textbf {\bibinfo
  {volume} {9}},\ \bibinfo {pages} {4825--4833} (\bibinfo {year}
  {2013})}\BibitemShut {NoStop}%
\bibitem [{\citenamefont {Hohenstein}, \citenamefont {Parrish},\ and\
  \citenamefont {Mart{\'\i}nez}(2012)}]{thc_1}%
  \BibitemOpen
  \bibfield  {author} {\bibinfo {author} {\bibfnamefont {E.~G.}\ \bibnamefont
  {Hohenstein}}, \bibinfo {author} {\bibfnamefont {R.~M.}\ \bibnamefont
  {Parrish}}, \ and\ \bibinfo {author} {\bibfnamefont {T.~J.}\ \bibnamefont
  {Mart{\'\i}nez}},\ }\bibfield  {title} {\enquote {\bibinfo {title} {{Tensor
  hypercontraction density fitting. I. Quartic scaling second- and third-order
  M{\o}ller-Plesset perturbation theory}},}\ }\href {\doibase
  10.1063/1.4732310} {\bibfield  {journal} {\bibinfo  {journal} {The Journal of
  Chemical Physics}\ }\textbf {\bibinfo {volume} {137}},\ \bibinfo {pages}
  {044103} (\bibinfo {year} {2012})}\BibitemShut {NoStop}%
\bibitem [{\citenamefont {Parrish}\ \emph {et~al.}(2012)\citenamefont
  {Parrish}, \citenamefont {Hohenstein}, \citenamefont {Mart{\'\i}nez},\ and\
  \citenamefont {Sherrill}}]{thc_2}%
  \BibitemOpen
  \bibfield  {author} {\bibinfo {author} {\bibfnamefont {R.~M.}\ \bibnamefont
  {Parrish}}, \bibinfo {author} {\bibfnamefont {E.~G.}\ \bibnamefont
  {Hohenstein}}, \bibinfo {author} {\bibfnamefont {T.~J.}\ \bibnamefont
  {Mart{\'\i}nez}}, \ and\ \bibinfo {author} {\bibfnamefont {C.~D.}\
  \bibnamefont {Sherrill}},\ }\bibfield  {title} {\enquote {\bibinfo {title}
  {{Tensor hypercontraction. II. Least-squares renormalization}},}\ }\href
  {\doibase 10.1063/1.4768233} {\bibfield  {journal} {\bibinfo  {journal} {The
  Journal of Chemical Physics}\ }\textbf {\bibinfo {volume} {137}},\ \bibinfo
  {pages} {224106} (\bibinfo {year} {2012})}\BibitemShut {NoStop}%
\bibitem [{\citenamefont {Hohenstein}\ \emph {et~al.}(2012)\citenamefont
  {Hohenstein}, \citenamefont {Parrish}, \citenamefont {Sherrill},\ and\
  \citenamefont {Mart{\'\i}nez}}]{thc_3}%
  \BibitemOpen
  \bibfield  {author} {\bibinfo {author} {\bibfnamefont {E.~G.}\ \bibnamefont
  {Hohenstein}}, \bibinfo {author} {\bibfnamefont {R.~M.}\ \bibnamefont
  {Parrish}}, \bibinfo {author} {\bibfnamefont {C.~D.}\ \bibnamefont
  {Sherrill}}, \ and\ \bibinfo {author} {\bibfnamefont {T.~J.}\ \bibnamefont
  {Mart{\'\i}nez}},\ }\bibfield  {title} {\enquote {\bibinfo {title}
  {{Communication: Tensor hypercontraction. III. Least-squares tensor
  hypercontraction for the determination of correlated wavefunctions}},}\
  }\href {\doibase 10.1063/1.4768241} {\bibfield  {journal} {\bibinfo
  {journal} {The Journal of Chemical Physics}\ }\textbf {\bibinfo {volume}
  {137}},\ \bibinfo {pages} {221101} (\bibinfo {year} {2012})}\BibitemShut
  {NoStop}%
\bibitem [{\citenamefont {Lu}\ and\ \citenamefont {Ying}(2015)}]{Ying_ISDF}%
  \BibitemOpen
  \bibfield  {author} {\bibinfo {author} {\bibfnamefont {J.}~\bibnamefont
  {Lu}}\ and\ \bibinfo {author} {\bibfnamefont {L.}~\bibnamefont {Ying}},\
  }\bibfield  {title} {\enquote {\bibinfo {title} {{Compression of the electron
  repulsion integral tensor in tensor hypercontraction format with cubic
  scaling cost}},}\ }\href {\doibase https://doi.org/10.1016/j.jcp.2015.09.014}
  {\bibfield  {journal} {\bibinfo  {journal} {J. Comput. Phys.}\ }\textbf
  {\bibinfo {volume} {302}},\ \bibinfo {pages} {329} (\bibinfo {year}
  {2015})}\BibitemShut {NoStop}%
\bibitem [{\citenamefont {Hu}, \citenamefont {Lin},\ and\ \citenamefont
  {Yang}(2017)}]{ISDF_LinLin}%
  \BibitemOpen
  \bibfield  {author} {\bibinfo {author} {\bibfnamefont {W.}~\bibnamefont
  {Hu}}, \bibinfo {author} {\bibfnamefont {L.}~\bibnamefont {Lin}}, \ and\
  \bibinfo {author} {\bibfnamefont {C.}~\bibnamefont {Yang}},\ }\bibfield
  {title} {\enquote {\bibinfo {title} {{Interpolative Separable Density Fitting
  Decomposition for Accelerating Hybrid Density Functional Calculations with
  Applications to Defects in Silicon}},}\ }\href {\doibase
  10.1021/acs.jctc.7b00807} {\bibfield  {journal} {\bibinfo  {journal} {Journal
  of Chemical Theory and Computation}\ }\textbf {\bibinfo {volume} {13}},\
  \bibinfo {pages} {5420} (\bibinfo {year} {2017})}\BibitemShut {NoStop}%
\bibitem [{\citenamefont {Dong}, \citenamefont {Hu},\ and\ \citenamefont
  {Lin}(2018)}]{ISDF_CVT}%
  \BibitemOpen
  \bibfield  {author} {\bibinfo {author} {\bibfnamefont {K.}~\bibnamefont
  {Dong}}, \bibinfo {author} {\bibfnamefont {W.}~\bibnamefont {Hu}}, \ and\
  \bibinfo {author} {\bibfnamefont {L.}~\bibnamefont {Lin}},\ }\bibfield
  {title} {\enquote {\bibinfo {title} {{Interpolative Separable Density Fitting
  through Centroidal Voronoi Tessellation with Applications to Hybrid
  Functional Electronic Structure Calculations}},}\ }\href {\doibase
  10.1021/acs.jctc.7b01113} {\bibfield  {journal} {\bibinfo  {journal} {Journal
  of Chemical Theory and Computation}\ }\textbf {\bibinfo {volume} {14}},\
  \bibinfo {pages} {1311} (\bibinfo {year} {2018})}\BibitemShut {NoStop}%
\bibitem [{\citenamefont {Lee}, \citenamefont {Lin},\ and\ \citenamefont
  {Head-Gordon}(2019)}]{lee2019systematically}%
  \BibitemOpen
  \bibfield  {author} {\bibinfo {author} {\bibfnamefont {J.}~\bibnamefont
  {Lee}}, \bibinfo {author} {\bibfnamefont {L.}~\bibnamefont {Lin}}, \ and\
  \bibinfo {author} {\bibfnamefont {M.}~\bibnamefont {Head-Gordon}},\
  }\bibfield  {title} {\enquote {\bibinfo {title} {{Systematically Improvable
  Tensor Hypercontraction: Interpolative Separable Density-Fitting for
  Molecules Applied to Exact Exchange, Second- and Third-Order
  M{\o}ller{\textendash}Plesset Perturbation Theory}},}\ }\href {\doibase
  10.1021/acs.jctc.9b00820} {\bibfield  {journal} {\bibinfo  {journal} {Journal
  of Chemical Theory and Computation}\ }\textbf {\bibinfo {volume} {16}},\
  \bibinfo {pages} {243--263} (\bibinfo {year} {2019})}\BibitemShut {NoStop}%
\bibitem [{\citenamefont {Motta}, \citenamefont {Zhang},\ and\ \citenamefont
  {Chan}(2019)}]{motta_kpoint}%
  \BibitemOpen
  \bibfield  {author} {\bibinfo {author} {\bibfnamefont {M.}~\bibnamefont
  {Motta}}, \bibinfo {author} {\bibfnamefont {S.}~\bibnamefont {Zhang}}, \ and\
  \bibinfo {author} {\bibfnamefont {G.~K.-L.}\ \bibnamefont {Chan}},\
  }\bibfield  {title} {\enquote {\bibinfo {title} {{Hamiltonian symmetries in
  auxiliary-field quantum Monte Carlo calculations for electronic
  structure}},}\ }\href {\doibase 10.1103/PhysRevB.100.045127} {\bibfield
  {journal} {\bibinfo  {journal} {Physical Review B}\ }\textbf {\bibinfo
  {volume} {100}},\ \bibinfo {pages} {045127} (\bibinfo {year}
  {2019})}\BibitemShut {NoStop}%
\bibitem [{\citenamefont {Suewattana}\ \emph {et~al.}(2007)\citenamefont
  {Suewattana}, \citenamefont {Purwanto}, \citenamefont {Zhang}, \citenamefont
  {Krakauer},\ and\ \citenamefont {Walter}}]{suewattana_pw_afqmc}%
  \BibitemOpen
  \bibfield  {author} {\bibinfo {author} {\bibfnamefont {M.}~\bibnamefont
  {Suewattana}}, \bibinfo {author} {\bibfnamefont {W.}~\bibnamefont
  {Purwanto}}, \bibinfo {author} {\bibfnamefont {S.}~\bibnamefont {Zhang}},
  \bibinfo {author} {\bibfnamefont {H.}~\bibnamefont {Krakauer}}, \ and\
  \bibinfo {author} {\bibfnamefont {E.~J.}\ \bibnamefont {Walter}},\ }\bibfield
   {title} {\enquote {\bibinfo {title} {{Phaseless auxiliary-field quantum
  Monte Carlo calculations with plane waves and pseudopotentials: Applications
  to atoms and molecules}},}\ }\href {\doibase 10.1103/PhysRevB.75.245123}
  {\bibfield  {journal} {\bibinfo  {journal} {Physical Review B}\ }\textbf
  {\bibinfo {volume} {75}},\ \bibinfo {pages} {245123} (\bibinfo {year}
  {2007})}\BibitemShut {NoStop}%
\bibitem [{\citenamefont {Ma}, \citenamefont {Zhang},\ and\ \citenamefont
  {Krakauer}(2017)}]{ma_multiple_proj}%
  \BibitemOpen
  \bibfield  {author} {\bibinfo {author} {\bibfnamefont {F.}~\bibnamefont
  {Ma}}, \bibinfo {author} {\bibfnamefont {S.}~\bibnamefont {Zhang}}, \ and\
  \bibinfo {author} {\bibfnamefont {H.}~\bibnamefont {Krakauer}},\ }\bibfield
  {title} {\enquote {\bibinfo {title} {{Auxiliary-field quantum Monte Carlo
  calculations with multiple-projector pseudopotentials}},}\ }\href {\doibase
  10.1103/PhysRevB.95.165103} {\bibfield  {journal} {\bibinfo  {journal}
  {Physical Review B}\ }\textbf {\bibinfo {volume} {95}},\ \bibinfo {pages}
  {165103} (\bibinfo {year} {2017})}\BibitemShut {NoStop}%
\bibitem [{\citenamefont {Sharma}\ \emph {et~al.}(2017)\citenamefont {Sharma},
  \citenamefont {Holmes}, \citenamefont {Jeanmairet}, \citenamefont {Alavi},\
  and\ \citenamefont {Umrigar}}]{sharma_dice_1}%
  \BibitemOpen
  \bibfield  {author} {\bibinfo {author} {\bibfnamefont {S.}~\bibnamefont
  {Sharma}}, \bibinfo {author} {\bibfnamefont {A.~A.}\ \bibnamefont {Holmes}},
  \bibinfo {author} {\bibfnamefont {G.}~\bibnamefont {Jeanmairet}}, \bibinfo
  {author} {\bibfnamefont {A.}~\bibnamefont {Alavi}}, \ and\ \bibinfo {author}
  {\bibfnamefont {C.~J.}\ \bibnamefont {Umrigar}},\ }\bibfield  {title}
  {\enquote {\bibinfo {title} {{Semistochastic Heat-Bath Configuration
  Interaction Method: Selected Configuration Interaction with Semistochastic
  Perturbation Theory}},}\ }\href {https://doi.org/10.1021/acs.jctc.6b01028}
  {\bibfield  {journal} {\bibinfo  {journal} {Journal of Chemical Theory and
  Computation}\ }\textbf {\bibinfo {volume} {13}},\ \bibinfo {pages}
  {1595--1604} (\bibinfo {year} {2017})}\BibitemShut {NoStop}%
\bibitem [{\citenamefont {Holmes}, \citenamefont {Tubman},\ and\ \citenamefont
  {Umrigar}(2016)}]{holmes_dice_2}%
  \BibitemOpen
  \bibfield  {author} {\bibinfo {author} {\bibfnamefont {A.~A.}\ \bibnamefont
  {Holmes}}, \bibinfo {author} {\bibfnamefont {N.~M.}\ \bibnamefont {Tubman}},
  \ and\ \bibinfo {author} {\bibfnamefont {C.~J.}\ \bibnamefont {Umrigar}},\
  }\bibfield  {title} {\enquote {\bibinfo {title} {{Heat-Bath Configuration
  Interaction: An Efficient Selected Configuration Interaction Algorithm
  Inspired by Heat-Bath Sampling}},}\ }\href {\doibase
  10.1021/acs.jctc.6b00407} {\bibfield  {journal} {\bibinfo  {journal} {Journal
  of Chemical Theory and Computation}\ }\textbf {\bibinfo {volume} {12}},\
  \bibinfo {pages} {3674--3680} (\bibinfo {year} {2016})}\BibitemShut {NoStop}%
\bibitem [{\citenamefont {Jim{\'e}nez-Hoyos}\ \emph {et~al.}(2012)\citenamefont
  {Jim{\'e}nez-Hoyos}, \citenamefont {Henderson}, \citenamefont {Tsuchimochi},\
  and\ \citenamefont {Scuseria}}]{Jimenez_phf}%
  \BibitemOpen
  \bibfield  {author} {\bibinfo {author} {\bibfnamefont {C.~A.}\ \bibnamefont
  {Jim{\'e}nez-Hoyos}}, \bibinfo {author} {\bibfnamefont {T.~M.}\ \bibnamefont
  {Henderson}}, \bibinfo {author} {\bibfnamefont {T.}~\bibnamefont
  {Tsuchimochi}}, \ and\ \bibinfo {author} {\bibfnamefont {G.~E.}\ \bibnamefont
  {Scuseria}},\ }\bibfield  {title} {\enquote {\bibinfo {title} {{Projected
  Hartree--Fock theory}},}\ }\href {\doibase 10.1063/1.4705280} {\bibfield
  {journal} {\bibinfo  {journal} {The Journal of Chemical Physics}\ }\textbf
  {\bibinfo {volume} {136}},\ \bibinfo {pages} {164109} (\bibinfo {year}
  {2012})}\BibitemShut {NoStop}%
\bibitem [{\citenamefont {Jim{\'e}nez-Hoyos}, \citenamefont
  {Rodr{\'\i}guez-Guzm{\'a}n},\ and\ \citenamefont
  {Scuseria}(2013)}]{scuseria_phf}%
  \BibitemOpen
  \bibfield  {author} {\bibinfo {author} {\bibfnamefont {C.~A.}\ \bibnamefont
  {Jim{\'e}nez-Hoyos}}, \bibinfo {author} {\bibfnamefont {R.}~\bibnamefont
  {Rodr{\'\i}guez-Guzm{\'a}n}}, \ and\ \bibinfo {author} {\bibfnamefont
  {G.~E.}\ \bibnamefont {Scuseria}},\ }\bibfield  {title} {\enquote {\bibinfo
  {title} {{Multi-component symmetry-projected approach for molecular ground
  state correlations}},}\ }\href {\doibase 10.1063/1.4832476} {\bibfield
  {journal} {\bibinfo  {journal} {The Journal of Chemical Physics}\ }\textbf
  {\bibinfo {volume} {139}},\ \bibinfo {pages} {204102} (\bibinfo {year}
  {2013})}\BibitemShut {NoStop}%
\bibitem [{\citenamefont {Schutski}, \citenamefont {Jim{\'e}nez-Hoyos},\ and\
  \citenamefont {Scuseria}(2014)}]{scuseria_phf_grad}%
  \BibitemOpen
  \bibfield  {author} {\bibinfo {author} {\bibfnamefont {R.}~\bibnamefont
  {Schutski}}, \bibinfo {author} {\bibfnamefont {C.~A.}\ \bibnamefont
  {Jim{\'e}nez-Hoyos}}, \ and\ \bibinfo {author} {\bibfnamefont {G.~E.}\
  \bibnamefont {Scuseria}},\ }\bibfield  {title} {\enquote {\bibinfo {title}
  {{Analytic energy gradient for the projected Hartree--Fock method}},}\ }\href
  {\doibase 10.1063/1.4876490} {\bibfield  {journal} {\bibinfo  {journal} {The
  Journal of Chemical Physics}\ }\textbf {\bibinfo {volume} {140}},\ \bibinfo
  {pages} {204101} (\bibinfo {year} {2014})}\BibitemShut {NoStop}%
\bibitem [{\citenamefont {Solomonik}\ \emph {et~al.}(2014)\citenamefont
  {Solomonik}, \citenamefont {Matthews}, \citenamefont {Hammond}, \citenamefont
  {Stanton},\ and\ \citenamefont {Demmel}}]{SolomonikAquarius2014}%
  \BibitemOpen
  \bibfield  {author} {\bibinfo {author} {\bibfnamefont {E.}~\bibnamefont
  {Solomonik}}, \bibinfo {author} {\bibfnamefont {D.}~\bibnamefont {Matthews}},
  \bibinfo {author} {\bibfnamefont {J.~R.}\ \bibnamefont {Hammond}}, \bibinfo
  {author} {\bibfnamefont {J.~F.}\ \bibnamefont {Stanton}}, \ and\ \bibinfo
  {author} {\bibfnamefont {J.}~\bibnamefont {Demmel}},\ }\bibfield  {title}
  {\enquote {\bibinfo {title} {{A massively parallel tensor contraction
  framework for coupled-cluster computations}},}\ }\href {\doibase
  10.1016/j.jpdc.2014.06.002} {\bibfield  {journal} {\bibinfo  {journal}
  {Journal of Parallel and Distributed Computing}\ }\textbf {\bibinfo {volume}
  {74}},\ \bibinfo {pages} {3176--3190} (\bibinfo {year} {2014})}\BibitemShut
  {NoStop}%
\bibitem [{\citenamefont {Purwanto}\ and\ \citenamefont
  {Zhang}(2004)}]{purwanto_back_prop}%
  \BibitemOpen
  \bibfield  {author} {\bibinfo {author} {\bibfnamefont {W.}~\bibnamefont
  {Purwanto}}\ and\ \bibinfo {author} {\bibfnamefont {S.}~\bibnamefont
  {Zhang}},\ }\bibfield  {title} {\enquote {\bibinfo {title} {{Quantum Monte
  Carlo method for the ground state of many-boson systems}},}\ }\href {\doibase
  10.1103/PhysRevE.70.056702} {\bibfield  {journal} {\bibinfo  {journal}
  {Physical Review E}\ }\textbf {\bibinfo {volume} {70}},\ \bibinfo {pages}
  {056702} (\bibinfo {year} {2004})}\BibitemShut {NoStop}%
\bibitem [{\citenamefont {Motta}\ and\ \citenamefont
  {Zhang}(2017)}]{motta_back_prop}%
  \BibitemOpen
  \bibfield  {author} {\bibinfo {author} {\bibfnamefont {M.}~\bibnamefont
  {Motta}}\ and\ \bibinfo {author} {\bibfnamefont {S.}~\bibnamefont {Zhang}},\
  }\bibfield  {title} {\enquote {\bibinfo {title} {{Computation of Ground-State
  Properties in Molecular Systems: Back-Propagation with Auxiliary-Field
  Quantum Monte Carlo}},}\ }\href {\doibase 10.1021/acs.jctc.7b00730}
  {\bibfield  {journal} {\bibinfo  {journal} {Journal of Chemical Theory and
  Computation}\ }\textbf {\bibinfo {volume} {13}},\ \bibinfo {pages} {5367}
  (\bibinfo {year} {2017})}\BibitemShut {NoStop}%
\bibitem [{\citenamefont {Lee}\ \emph {et~al.}(2017{\natexlab{a}})\citenamefont
  {Lee}, \citenamefont {Small}, \citenamefont {Epifanovsky},\ and\
  \citenamefont {Head-Gordon}}]{lee_ccvb}%
  \BibitemOpen
  \bibfield  {author} {\bibinfo {author} {\bibfnamefont {J.}~\bibnamefont
  {Lee}}, \bibinfo {author} {\bibfnamefont {D.~W.}\ \bibnamefont {Small}},
  \bibinfo {author} {\bibfnamefont {E.}~\bibnamefont {Epifanovsky}}, \ and\
  \bibinfo {author} {\bibfnamefont {M.}~\bibnamefont {Head-Gordon}},\
  }\bibfield  {title} {\enquote {\bibinfo {title} {{Coupled-Cluster
  Valence-Bond Singles and Doubles for Strongly Correlated Systems:
  Block-Tensor Based Implementation and Application to Oligoacenes}},}\ }\href
  {\doibase 10.1021/acs.jctc.6b01092} {\bibfield  {journal} {\bibinfo
  {journal} {Journal of Chemical Theory and Computation}\ }\textbf {\bibinfo
  {volume} {13}},\ \bibinfo {pages} {602--615} (\bibinfo {year}
  {2017}{\natexlab{a}})}\BibitemShut {NoStop}%
\bibitem [{\citenamefont {Zhang}(1999)}]{zhang_ftafqmc_99}%
  \BibitemOpen
  \bibfield  {author} {\bibinfo {author} {\bibfnamefont {S.}~\bibnamefont
  {Zhang}},\ }\bibfield  {title} {\enquote {\bibinfo {title}
  {{Finite-Temperature Monte Carlo Calculations for Systems with Fermions}},}\
  }\href {\doibase 10.1103/PhysRevLett.83.2777} {\bibfield  {journal} {\bibinfo
   {journal} {Physical Review Letters}\ }\textbf {\bibinfo {volume} {83}},\
  \bibinfo {pages} {2777--2780} (\bibinfo {year} {1999})}\BibitemShut {NoStop}%
\bibitem [{\citenamefont {Rubenstein}, \citenamefont {Zhang},\ and\
  \citenamefont {Reichman}(2012)}]{rubenstein_ft}%
  \BibitemOpen
  \bibfield  {author} {\bibinfo {author} {\bibfnamefont {B.~M.}\ \bibnamefont
  {Rubenstein}}, \bibinfo {author} {\bibfnamefont {S.}~\bibnamefont {Zhang}}, \
  and\ \bibinfo {author} {\bibfnamefont {D.~R.}\ \bibnamefont {Reichman}},\
  }\bibfield  {title} {\enquote {\bibinfo {title} {{Finite-temperature
  auxiliary-field quantum Monte Carlo technique for Bose-Fermi mixtures}},}\
  }\href {\doibase 10.1103/PhysRevA.86.053606} {\bibfield  {journal} {\bibinfo
  {journal} {Physical Review A}\ }\textbf {\bibinfo {volume} {86}},\ \bibinfo
  {pages} {053606} (\bibinfo {year} {2012})}\BibitemShut {NoStop}%
\bibitem [{\citenamefont {He}\ \emph {et~al.}(2019)\citenamefont {He},
  \citenamefont {Qin}, \citenamefont {Shi}, \citenamefont {Lu},\ and\
  \citenamefont {Zhang}}]{he_sconsis_ft}%
  \BibitemOpen
  \bibfield  {author} {\bibinfo {author} {\bibfnamefont {Y.-Y.}\ \bibnamefont
  {He}}, \bibinfo {author} {\bibfnamefont {M.}~\bibnamefont {Qin}}, \bibinfo
  {author} {\bibfnamefont {H.}~\bibnamefont {Shi}}, \bibinfo {author}
  {\bibfnamefont {Z.-Y.}\ \bibnamefont {Lu}}, \ and\ \bibinfo {author}
  {\bibfnamefont {S.}~\bibnamefont {Zhang}},\ }\bibfield  {title} {\enquote
  {\bibinfo {title} {{Finite-temperature auxiliary-field quantum Monte Carlo:
  Self-consistent constraint and systematic approach to low temperatures}},}\
  }\href {\doibase 10.1103/PhysRevB.99.045108} {\bibfield  {journal} {\bibinfo
  {journal} {Physical Review B}\ }\textbf {\bibinfo {volume} {99}},\ \bibinfo
  {pages} {045108} (\bibinfo {year} {2019})}\BibitemShut {NoStop}%
\bibitem [{\citenamefont {Liu}, \citenamefont {Cho},\ and\ \citenamefont
  {Rubenstein}(2018)}]{liu_ab_2018}%
  \BibitemOpen
  \bibfield  {author} {\bibinfo {author} {\bibfnamefont {Y.}~\bibnamefont
  {Liu}}, \bibinfo {author} {\bibfnamefont {M.}~\bibnamefont {Cho}}, \ and\
  \bibinfo {author} {\bibfnamefont {B.}~\bibnamefont {Rubenstein}},\ }\bibfield
   {title} {\enquote {\bibinfo {title} {{Ab Initio Finite Temperature Auxiliary
  Field Quantum Monte Carlo}},}\ }\href {\doibase 10.1021/acs.jctc.8b00569}
  {\bibfield  {journal} {\bibinfo  {journal} {Journal of Chemical Theory and
  Computation}\ }\textbf {\bibinfo {volume} {14}},\ \bibinfo {pages}
  {4722--4732} (\bibinfo {year} {2018})}\BibitemShut {NoStop}%
\bibitem [{\citenamefont {He}, \citenamefont {Shi},\ and\ \citenamefont
  {Zhang}(2019)}]{he_continuum_prl}%
  \BibitemOpen
  \bibfield  {author} {\bibinfo {author} {\bibfnamefont {Y.-Y.}\ \bibnamefont
  {He}}, \bibinfo {author} {\bibfnamefont {H.}~\bibnamefont {Shi}}, \ and\
  \bibinfo {author} {\bibfnamefont {S.}~\bibnamefont {Zhang}},\ }\bibfield
  {title} {\enquote {\bibinfo {title} {{Reaching the Continuum Limit in
  Finite-Temperature Ab Initio Field-Theory Computations in Many-Fermion
  Systems}},}\ }\href {\doibase 10.1103/PhysRevLett.123.136402} {\bibfield
  {journal} {\bibinfo  {journal} {Physical Review Letters}\ }\textbf {\bibinfo
  {volume} {123}},\ \bibinfo {pages} {136402} (\bibinfo {year}
  {2019})}\BibitemShut {NoStop}%
\bibitem [{\citenamefont {Dubeck{\'{y}}}(2017)}]{DubeckyNoncovalent2017}%
  \BibitemOpen
  \bibfield  {author} {\bibinfo {author} {\bibfnamefont {M.}~\bibnamefont
  {Dubeck{\'{y}}}},\ }\bibfield  {title} {\enquote {\bibinfo {title}
  {{Noncovalent Interactions by Fixed-Node Diffusion Monte Carlo: Convergence
  of Nodes and Energy Differences vs Gaussian Basis-Set Size}},}\ }\href
  {\doibase 10.1021/acs.jctc.7b00537} {\bibfield  {journal} {\bibinfo
  {journal} {Journal of Chemical Theory and Computation}\ }\textbf {\bibinfo
  {volume} {13}},\ \bibinfo {pages} {3626--3635} (\bibinfo {year}
  {2017})}\BibitemShut {NoStop}%
\bibitem [{\citenamefont {Toulouse}\ and\ \citenamefont
  {Umrigar}(2008)}]{ToulouseFull2008}%
  \BibitemOpen
  \bibfield  {author} {\bibinfo {author} {\bibfnamefont {J.}~\bibnamefont
  {Toulouse}}\ and\ \bibinfo {author} {\bibfnamefont {C.~J.}\ \bibnamefont
  {Umrigar}},\ }\bibfield  {title} {\enquote {\bibinfo {title} {{Full
  optimization of Jastrow{\textendash}Slater wave functions with application to
  the first-row atoms and homonuclear diatomic molecules}},}\ }\href {\doibase
  10.1063/1.2908237} {\bibfield  {journal} {\bibinfo  {journal} {The Journal of
  Chemical Physics}\ }\textbf {\bibinfo {volume} {128}},\ \bibinfo {pages}
  {174101} (\bibinfo {year} {2008})}\BibitemShut {NoStop}%
\bibitem [{\citenamefont {Assaraf}, \citenamefont {Moroni},\ and\ \citenamefont
  {Filippi}(2017)}]{AssarafOptimizing2017}%
  \BibitemOpen
  \bibfield  {author} {\bibinfo {author} {\bibfnamefont {R.}~\bibnamefont
  {Assaraf}}, \bibinfo {author} {\bibfnamefont {S.}~\bibnamefont {Moroni}}, \
  and\ \bibinfo {author} {\bibfnamefont {C.}~\bibnamefont {Filippi}},\
  }\bibfield  {title} {\enquote {\bibinfo {title} {Optimizing the energy with
  quantum monte carlo: A lower numerical scaling for jastrow{\textendash}slater
  expansions},}\ }\href {\doibase 10.1021/acs.jctc.7b00648} {\bibfield
  {journal} {\bibinfo  {journal} {Journal of Chemical Theory and Computation}\
  }\textbf {\bibinfo {volume} {13}},\ \bibinfo {pages} {5273--5281} (\bibinfo
  {year} {2017})}\BibitemShut {NoStop}%
\bibitem [{\citenamefont {Taddei}\ \emph {et~al.}(2015)\citenamefont {Taddei},
  \citenamefont {Ruggeri}, \citenamefont {Moroni},\ and\ \citenamefont
  {Holzmann}}]{TaddeiIterative2015}%
  \BibitemOpen
  \bibfield  {author} {\bibinfo {author} {\bibfnamefont {M.}~\bibnamefont
  {Taddei}}, \bibinfo {author} {\bibfnamefont {M.}~\bibnamefont {Ruggeri}},
  \bibinfo {author} {\bibfnamefont {S.}~\bibnamefont {Moroni}}, \ and\ \bibinfo
  {author} {\bibfnamefont {M.}~\bibnamefont {Holzmann}},\ }\bibfield  {title}
  {\enquote {\bibinfo {title} {Iterative backflow renormalization procedure for
  many-body ground-state wave functions of strongly interacting normal fermi
  liquids},}\ }\href {\doibase 10.1103/physrevb.91.115106} {\bibfield
  {journal} {\bibinfo  {journal} {Physical Review B}\ }\textbf {\bibinfo
  {volume} {91}} (\bibinfo {year} {2015}),\
  10.1103/physrevb.91.115106}\BibitemShut {NoStop}%
\bibitem [{\citenamefont {Holzmann}\ and\ \citenamefont
  {Moroni}(2019)}]{HolzmannOrbital2019}%
  \BibitemOpen
  \bibfield  {author} {\bibinfo {author} {\bibfnamefont {M.}~\bibnamefont
  {Holzmann}}\ and\ \bibinfo {author} {\bibfnamefont {S.}~\bibnamefont
  {Moroni}},\ }\bibfield  {title} {\enquote {\bibinfo {title}
  {Orbital-dependent backflow wave functions for real-space quantum monte
  carlo},}\ }\href {\doibase 10.1103/physrevb.99.085121} {\bibfield  {journal}
  {\bibinfo  {journal} {Physical Review B}\ }\textbf {\bibinfo {volume} {99}}
  (\bibinfo {year} {2019}),\ 10.1103/physrevb.99.085121}\BibitemShut {NoStop}%
\bibitem [{\citenamefont {Casula}, \citenamefont {Attaccalite},\ and\
  \citenamefont {Sorella}(2004)}]{CasulaCorrelated2004}%
  \BibitemOpen
  \bibfield  {author} {\bibinfo {author} {\bibfnamefont {M.}~\bibnamefont
  {Casula}}, \bibinfo {author} {\bibfnamefont {C.}~\bibnamefont {Attaccalite}},
  \ and\ \bibinfo {author} {\bibfnamefont {S.}~\bibnamefont {Sorella}},\
  }\bibfield  {title} {\enquote {\bibinfo {title} {Correlated geminal wave
  function for molecules:{\hspace{0.6em}}an efficient resonating valence bond
  approach},}\ }\href {\doibase 10.1063/1.1794632} {\bibfield  {journal}
  {\bibinfo  {journal} {The Journal of Chemical Physics}\ }\textbf {\bibinfo
  {volume} {121}},\ \bibinfo {pages} {7110--7126} (\bibinfo {year}
  {2004})}\BibitemShut {NoStop}%
\bibitem [{\citenamefont {Nukala}\ and\ \citenamefont
  {Kent}(2009)}]{NukalaFast2009}%
  \BibitemOpen
  \bibfield  {author} {\bibinfo {author} {\bibfnamefont {P.~K. V.~V.}\
  \bibnamefont {Nukala}}\ and\ \bibinfo {author} {\bibfnamefont {P.~R.~C.}\
  \bibnamefont {Kent}},\ }\bibfield  {title} {\enquote {\bibinfo {title} {{A
  fast and efficient algorithm for Slater determinant updates in quantum Monte
  Carlo simulations}},}\ }\href {\doibase 10.1063/1.3142703} {\bibfield
  {journal} {\bibinfo  {journal} {The Journal of Chemical Physics}\ }\textbf
  {\bibinfo {volume} {130}},\ \bibinfo {pages} {204105} (\bibinfo {year}
  {2009})}\BibitemShut {NoStop}%
\bibitem [{\citenamefont {Clark}\ \emph {et~al.}(2011)\citenamefont {Clark},
  \citenamefont {Morales}, \citenamefont {McMinis}, \citenamefont {Kim},\ and\
  \citenamefont {Scuseria}}]{Scuseria2011}%
  \BibitemOpen
  \bibfield  {author} {\bibinfo {author} {\bibfnamefont {B.~K.}\ \bibnamefont
  {Clark}}, \bibinfo {author} {\bibfnamefont {M.~A.}\ \bibnamefont {Morales}},
  \bibinfo {author} {\bibfnamefont {J.}~\bibnamefont {McMinis}}, \bibinfo
  {author} {\bibfnamefont {J.}~\bibnamefont {Kim}}, \ and\ \bibinfo {author}
  {\bibfnamefont {G.~E.}\ \bibnamefont {Scuseria}},\ }\bibfield  {title}
  {\enquote {\bibinfo {title} {{Computing the energy of a water molecule using
  multideterminants: A simple, efficient algorithm}},}\ }\href@noop {}
  {\bibfield  {journal} {\bibinfo  {journal} {The Journal of Chemical Physics}\
  }\textbf {\bibinfo {volume} {135}},\ \bibinfo {pages} {244105} (\bibinfo
  {year} {2011})}\BibitemShut {NoStop}%
\bibitem [{\citenamefont {Morales}\ \emph {et~al.}(2012)\citenamefont
  {Morales}, \citenamefont {McMinis}, \citenamefont {Clark}, \citenamefont
  {Kim},\ and\ \citenamefont {Scuseria}}]{Scuseria2012}%
  \BibitemOpen
  \bibfield  {author} {\bibinfo {author} {\bibfnamefont {M.~A.}\ \bibnamefont
  {Morales}}, \bibinfo {author} {\bibfnamefont {J.}~\bibnamefont {McMinis}},
  \bibinfo {author} {\bibfnamefont {B.~K.}\ \bibnamefont {Clark}}, \bibinfo
  {author} {\bibfnamefont {J.}~\bibnamefont {Kim}}, \ and\ \bibinfo {author}
  {\bibfnamefont {G.~E.}\ \bibnamefont {Scuseria}},\ }\bibfield  {title}
  {\enquote {\bibinfo {title} {{Multideterminant Wave Functions in Quantum
  Monte Carlo}},}\ }\href@noop {} {\bibfield  {journal} {\bibinfo  {journal}
  {Journal of Chemical Theory and Computation}\ }\textbf {\bibinfo {volume}
  {8}},\ \bibinfo {pages} {2181--2188} (\bibinfo {year} {2012})}\BibitemShut
  {NoStop}%
\bibitem [{\citenamefont {Huron}, \citenamefont {Malrieu},\ and\ \citenamefont
  {Rancurel}(1973)}]{Huron1973}%
  \BibitemOpen
  \bibfield  {author} {\bibinfo {author} {\bibfnamefont {B.}~\bibnamefont
  {Huron}}, \bibinfo {author} {\bibfnamefont {J.~P.}\ \bibnamefont {Malrieu}},
  \ and\ \bibinfo {author} {\bibfnamefont {P.}~\bibnamefont {Rancurel}},\
  }\bibfield  {title} {\enquote {\bibinfo {title} {{Iterative perturbation
  calculations of ground and excited state energies from multiconfigurational
  zeroth-order wavefunctions}},}\ }\href {\doibase 10.1063/1.1679199}
  {\bibfield  {journal} {\bibinfo  {journal} {The Journal of Chemical Physics}\
  }\textbf {\bibinfo {volume} {58}},\ \bibinfo {pages} {5745--5759} (\bibinfo
  {year} {1973})}\BibitemShut {NoStop}%
\bibitem [{\citenamefont {Scemama}\ \emph {et~al.}(2019)\citenamefont
  {Scemama}, \citenamefont {Caffarel}, \citenamefont {Benali}, \citenamefont
  {Jacquemin},\ and\ \citenamefont {Loos}}]{SCEMAMA2019}%
  \BibitemOpen
  \bibfield  {author} {\bibinfo {author} {\bibfnamefont {A.}~\bibnamefont
  {Scemama}}, \bibinfo {author} {\bibfnamefont {M.}~\bibnamefont {Caffarel}},
  \bibinfo {author} {\bibfnamefont {A.}~\bibnamefont {Benali}}, \bibinfo
  {author} {\bibfnamefont {D.}~\bibnamefont {Jacquemin}}, \ and\ \bibinfo
  {author} {\bibfnamefont {P.-F.}\ \bibnamefont {Loos}},\ }\bibfield  {title}
  {\enquote {\bibinfo {title} {{Influence of pseudopotentials on excitation
  energies from selected configuration interaction and diffusion Monte
  Carlo}},}\ }\href {\doibase 0.1016/j.rechem.2019.100002} {\bibfield
  {journal} {\bibinfo  {journal} {Results in Chemistry}\ }\textbf {\bibinfo
  {volume} {1}},\ \bibinfo {pages} {100002} (\bibinfo {year}
  {2019})}\BibitemShut {NoStop}%
\bibitem [{\citenamefont {Scemama}\ \emph {et~al.}(2018)\citenamefont
  {Scemama}, \citenamefont {Benali}, \citenamefont {Jacquemin}, \citenamefont
  {Caffarel},\ and\ \citenamefont {Loos}}]{Scemama2018}%
  \BibitemOpen
  \bibfield  {author} {\bibinfo {author} {\bibfnamefont {A.}~\bibnamefont
  {Scemama}}, \bibinfo {author} {\bibfnamefont {A.}~\bibnamefont {Benali}},
  \bibinfo {author} {\bibfnamefont {D.}~\bibnamefont {Jacquemin}}, \bibinfo
  {author} {\bibfnamefont {M.}~\bibnamefont {Caffarel}}, \ and\ \bibinfo
  {author} {\bibfnamefont {P.-F.}\ \bibnamefont {Loos}},\ }\bibfield  {title}
  {\enquote {\bibinfo {title} {{Excitation energies from diffusion Monte Carlo
  using selected configuration interaction nodes}},}\ }\href {\doibase
  10.1063/1.5041327} {\bibfield  {journal} {\bibinfo  {journal} {The Journal of
  Chemical Physics}\ }\textbf {\bibinfo {volume} {149}},\ \bibinfo {pages}
  {034108} (\bibinfo {year} {2018})}\BibitemShut {NoStop}%
\bibitem [{\citenamefont {Chien}\ \emph {et~al.}(2018)\citenamefont {Chien},
  \citenamefont {Holmes}, \citenamefont {Otten}, \citenamefont {Umrigar},
  \citenamefont {Sharma},\ and\ \citenamefont {Zimmerman}}]{Chien2018}%
  \BibitemOpen
  \bibfield  {author} {\bibinfo {author} {\bibfnamefont {A.~D.}\ \bibnamefont
  {Chien}}, \bibinfo {author} {\bibfnamefont {A.~A.}\ \bibnamefont {Holmes}},
  \bibinfo {author} {\bibfnamefont {M.}~\bibnamefont {Otten}}, \bibinfo
  {author} {\bibfnamefont {C.~J.}\ \bibnamefont {Umrigar}}, \bibinfo {author}
  {\bibfnamefont {S.}~\bibnamefont {Sharma}}, \ and\ \bibinfo {author}
  {\bibfnamefont {P.~M.}\ \bibnamefont {Zimmerman}},\ }\bibfield  {title}
  {\enquote {\bibinfo {title} {{Excited States of Methylene, Polyenes, and
  Ozone from Heat-Bath Configuration Interaction}},}\ }\href {\doibase
  10.1021/acs.jpca.8b01554} {\bibfield  {journal} {\bibinfo  {journal} {The
  Journal of Physical Chemistry A}\ }\textbf {\bibinfo {volume} {122}},\
  \bibinfo {pages} {2714--2722} (\bibinfo {year} {2018})}\BibitemShut {NoStop}%
\bibitem [{\citenamefont {Li}\ \emph {et~al.}(2018)\citenamefont {Li},
  \citenamefont {Otten}, \citenamefont {Holmes}, \citenamefont {Sharma},\ and\
  \citenamefont {Umrigar}}]{Li2018}%
  \BibitemOpen
  \bibfield  {author} {\bibinfo {author} {\bibfnamefont {J.}~\bibnamefont
  {Li}}, \bibinfo {author} {\bibfnamefont {M.}~\bibnamefont {Otten}}, \bibinfo
  {author} {\bibfnamefont {A.~A.}\ \bibnamefont {Holmes}}, \bibinfo {author}
  {\bibfnamefont {S.}~\bibnamefont {Sharma}}, \ and\ \bibinfo {author}
  {\bibfnamefont {C.~J.}\ \bibnamefont {Umrigar}},\ }\bibfield  {title}
  {\enquote {\bibinfo {title} {{Fast semistochastic heat-bath configuration
  interaction}},}\ }\href {\doibase 10.1063/1.5055390} {\bibfield  {journal}
  {\bibinfo  {journal} {The Journal of Chemical Physics}\ }\textbf {\bibinfo
  {volume} {149}},\ \bibinfo {pages} {214110} (\bibinfo {year}
  {2018})}\BibitemShut {NoStop}%
\bibitem [{\citenamefont {Caffarel}\ \emph {et~al.}(2016)\citenamefont
  {Caffarel}, \citenamefont {Applencourt}, \citenamefont {Giner},\ and\
  \citenamefont {Scemama}}]{Caffarel2016}%
  \BibitemOpen
  \bibfield  {author} {\bibinfo {author} {\bibfnamefont {M.}~\bibnamefont
  {Caffarel}}, \bibinfo {author} {\bibfnamefont {T.}~\bibnamefont
  {Applencourt}}, \bibinfo {author} {\bibfnamefont {E.}~\bibnamefont {Giner}},
  \ and\ \bibinfo {author} {\bibfnamefont {A.}~\bibnamefont {Scemama}},\
  }\bibfield  {title} {\enquote {\bibinfo {title} {{Communication: Toward an
  improved control of the fixed-node error in quantum Monte Carlo: The case of
  the water molecule}},}\ }\href {\doibase 10.1063/1.4947093} {\bibfield
  {journal} {\bibinfo  {journal} {The Journal of Chemical Physics}\ }\textbf
  {\bibinfo {volume} {144}},\ \bibinfo {pages} {151103} (\bibinfo {year}
  {2016})}\BibitemShut {NoStop}%
\bibitem [{\citenamefont {Scemama}\ \emph {et~al.}(2014)\citenamefont
  {Scemama}, \citenamefont {Applencourt}, \citenamefont {Giner},\ and\
  \citenamefont {Caffarel}}]{Scemama2014}%
  \BibitemOpen
  \bibfield  {author} {\bibinfo {author} {\bibfnamefont {A.}~\bibnamefont
  {Scemama}}, \bibinfo {author} {\bibfnamefont {T.}~\bibnamefont
  {Applencourt}}, \bibinfo {author} {\bibfnamefont {E.}~\bibnamefont {Giner}},
  \ and\ \bibinfo {author} {\bibfnamefont {M.}~\bibnamefont {Caffarel}},\
  }\bibfield  {title} {\enquote {\bibinfo {title} {{Accurate nonrelativistic
  ground-state energies of 3d transition metal atoms}},}\ }\href {\doibase
  10.1063/1.4903985} {\bibfield  {journal} {\bibinfo  {journal} {The Journal of
  Chemical Physics}\ }\textbf {\bibinfo {volume} {141}},\ \bibinfo {pages}
  {244110} (\bibinfo {year} {2014})}\BibitemShut {NoStop}%
\bibitem [{\citenamefont {Giner}, \citenamefont {Scemama},\ and\ \citenamefont
  {Caffarel}(2013)}]{Giner2013}%
  \BibitemOpen
  \bibfield  {author} {\bibinfo {author} {\bibfnamefont {E.}~\bibnamefont
  {Giner}}, \bibinfo {author} {\bibfnamefont {A.}~\bibnamefont {Scemama}}, \
  and\ \bibinfo {author} {\bibfnamefont {M.}~\bibnamefont {Caffarel}},\
  }\bibfield  {title} {\enquote {\bibinfo {title} {{Using perturbatively
  selected configuration interaction in quantum Monte Carlo calculations}},}\
  }\href {\doibase 10.1139/cjc-2013-0017} {\bibfield  {journal} {\bibinfo
  {journal} {Canadian Journal of Chemistry}\ }\textbf {\bibinfo {volume}
  {91}},\ \bibinfo {pages} {879--885} (\bibinfo {year} {2013})}\BibitemShut
  {NoStop}%
\bibitem [{\citenamefont {Berkowitz}, \citenamefont {Tasman},\ and\
  \citenamefont {Chupka}(1962)}]{Berkowitz1962}%
  \BibitemOpen
  \bibfield  {author} {\bibinfo {author} {\bibfnamefont {J.}~\bibnamefont
  {Berkowitz}}, \bibinfo {author} {\bibfnamefont {H.~A.}\ \bibnamefont
  {Tasman}}, \ and\ \bibinfo {author} {\bibfnamefont {W.~A.}\ \bibnamefont
  {Chupka}},\ }\bibfield  {title} {\enquote {\bibinfo {title} {{Double-Oven
  Experiments with Lithium Halide Vapors}},}\ }\href {\doibase
  10.1063/1.1732848} {\bibfield  {journal} {\bibinfo  {journal} {The Journal of
  Chemical Physics}\ }\textbf {\bibinfo {volume} {36}},\ \bibinfo {pages}
  {2170--2179} (\bibinfo {year} {1962})}\BibitemShut {NoStop}%
\bibitem [{\citenamefont {Aigueperse}\ \emph {et~al.}(2000)\citenamefont
  {Aigueperse}, \citenamefont {Mollard}, \citenamefont {Devilliers},
  \citenamefont {Chemla}, \citenamefont {Faron}, \citenamefont {Romano},\ and\
  \citenamefont {Cuer}}]{Aigueperse2000}%
  \BibitemOpen
  \bibfield  {author} {\bibinfo {author} {\bibfnamefont {J.}~\bibnamefont
  {Aigueperse}}, \bibinfo {author} {\bibfnamefont {P.}~\bibnamefont {Mollard}},
  \bibinfo {author} {\bibfnamefont {D.}~\bibnamefont {Devilliers}}, \bibinfo
  {author} {\bibfnamefont {M.}~\bibnamefont {Chemla}}, \bibinfo {author}
  {\bibfnamefont {R.}~\bibnamefont {Faron}}, \bibinfo {author} {\bibfnamefont
  {R.}~\bibnamefont {Romano}}, \ and\ \bibinfo {author} {\bibfnamefont {J.~P.}\
  \bibnamefont {Cuer}},\ }\enquote {\bibinfo {title} {{Fluorine Compounds,
  Inorganic}},}\ in\ \href {\doibase 10.1002/14356007.a11_307} {\emph {\bibinfo
  {booktitle} {Ullmann's Encyclopedia of Industrial Chemistry}}}\ (\bibinfo
  {publisher} {American Cancer Society},\ \bibinfo {year} {2000})\BibitemShut
  {NoStop}%
\bibitem [{\citenamefont {Zhao}\ and\ \citenamefont
  {Neuscamman}(2016)}]{Zhao2016}%
  \BibitemOpen
  \bibfield  {author} {\bibinfo {author} {\bibfnamefont {L.}~\bibnamefont
  {Zhao}}\ and\ \bibinfo {author} {\bibfnamefont {E.}~\bibnamefont
  {Neuscamman}},\ }\bibfield  {title} {\enquote {\bibinfo {title} {{An
  Efficient Variational Principle for the Direct Optimization of Excited
  states}},}\ }\href@noop {} {\bibfield  {journal} {\bibinfo  {journal}
  {Journal of Chemical Theory and Computation}\ }\textbf {\bibinfo {volume}
  {12}},\ \bibinfo {pages} {3436--3440} (\bibinfo {year} {2016})}\BibitemShut
  {NoStop}%
\bibitem [{\citenamefont {Shea}\ and\ \citenamefont
  {Neuscamman}(2017)}]{Shea2017}%
  \BibitemOpen
  \bibfield  {author} {\bibinfo {author} {\bibfnamefont {J.~A.~R.}\
  \bibnamefont {Shea}}\ and\ \bibinfo {author} {\bibfnamefont {E.}~\bibnamefont
  {Neuscamman}},\ }\bibfield  {title} {\enquote {\bibinfo {title} {{Size
  Consistent Excited States via Algorithmic Transformations between Variational
  Principles}},}\ }\href@noop {} {\bibfield  {journal} {\bibinfo  {journal}
  {Journal of Chemical Theory and Computation}\ }\textbf {\bibinfo {volume}
  {13}},\ \bibinfo {pages} {6078--6088} (\bibinfo {year} {2017})}\BibitemShut
  {NoStop}%
\bibitem [{\citenamefont {Robinson}, \citenamefont {Flores},\ and\
  \citenamefont {Neuscamman}(2017)}]{Robinson2017}%
  \BibitemOpen
  \bibfield  {author} {\bibinfo {author} {\bibfnamefont {P.~J.}\ \bibnamefont
  {Robinson}}, \bibinfo {author} {\bibfnamefont {S.~D.~P.}\ \bibnamefont
  {Flores}}, \ and\ \bibinfo {author} {\bibfnamefont {E.}~\bibnamefont
  {Neuscamman}},\ }\bibfield  {title} {\enquote {\bibinfo {title} {{Excitation
  variance matching with limited configuration interaction expansions in
  variational Monte Carlo}},}\ }\href@noop {} {\bibfield  {journal} {\bibinfo
  {journal} {The Journal of Chemical Physics}\ }\textbf {\bibinfo {volume}
  {147}},\ \bibinfo {pages} {164114} (\bibinfo {year} {2017})}\BibitemShut
  {NoStop}%
\bibitem [{\citenamefont {Onida}, \citenamefont {Reining},\ and\ \citenamefont
  {Rubio}(2002)}]{Rubio2002}%
  \BibitemOpen
  \bibfield  {author} {\bibinfo {author} {\bibfnamefont {G.}~\bibnamefont
  {Onida}}, \bibinfo {author} {\bibfnamefont {L.}~\bibnamefont {Reining}}, \
  and\ \bibinfo {author} {\bibfnamefont {A.}~\bibnamefont {Rubio}},\ }\bibfield
   {title} {\enquote {\bibinfo {title} {{Electronic excitations:
  density-functional versus many-body Green{\textquoteright}s-function
  approaches}},}\ }\href@noop {} {\bibfield  {journal} {\bibinfo  {journal}
  {Rev. Mod. Phys.}\ }\textbf {\bibinfo {volume} {74}},\ \bibinfo {pages}
  {601--659} (\bibinfo {year} {2002})}\BibitemShut {NoStop}%
\bibitem [{\citenamefont {Zhao}\ and\ \citenamefont
  {Neuscamman}(2019)}]{Zhao2019}%
  \BibitemOpen
  \bibfield  {author} {\bibinfo {author} {\bibfnamefont {L.}~\bibnamefont
  {Zhao}}\ and\ \bibinfo {author} {\bibfnamefont {E.}~\bibnamefont
  {Neuscamman}},\ }\bibfield  {title} {\enquote {\bibinfo {title} {{Variational
  Excitations in Real Solids: Optical Gaps and Insights into Many-Body
  Perturbation Theory}},}\ }\href@noop {} {\bibfield  {journal} {\bibinfo
  {journal} {Physical Review Letters}\ }\textbf {\bibinfo {volume} {123}},\
  \bibinfo {pages} {036402} (\bibinfo {year} {2019})}\BibitemShut {NoStop}%
\bibitem [{\citenamefont {Fuchs}\ \emph {et~al.}(2007)\citenamefont {Fuchs},
  \citenamefont {Furthm{\"u}ller}, \citenamefont {Bechstedt}, \citenamefont
  {Shishkin},\ and\ \citenamefont {Kresse}}]{Fuchs2007}%
  \BibitemOpen
  \bibfield  {author} {\bibinfo {author} {\bibfnamefont {F.}~\bibnamefont
  {Fuchs}}, \bibinfo {author} {\bibfnamefont {J.}~\bibnamefont
  {Furthm{\"u}ller}}, \bibinfo {author} {\bibfnamefont {F.}~\bibnamefont
  {Bechstedt}}, \bibinfo {author} {\bibfnamefont {M.}~\bibnamefont {Shishkin}},
  \ and\ \bibinfo {author} {\bibfnamefont {G.}~\bibnamefont {Kresse}},\
  }\bibfield  {title} {\enquote {\bibinfo {title} {{Quasiparticle band
  structure based on a generalized Kohn-Sham scheme}},}\ }\href@noop {}
  {\bibfield  {journal} {\bibinfo  {journal} {Physical Review B}\ }\textbf
  {\bibinfo {volume} {76}},\ \bibinfo {pages} {115109} (\bibinfo {year}
  {2007})}\BibitemShut {NoStop}%
\bibitem [{\citenamefont {Tsoi}\ \emph {et~al.}(2006)\citenamefont {Tsoi},
  \citenamefont {Lu}, \citenamefont {Ramdas}, \citenamefont {Alawadhi},
  \citenamefont {Grimsditch}, \citenamefont {Cardona},\ and\ \citenamefont
  {Lauck}}]{Lauck2006}%
  \BibitemOpen
  \bibfield  {author} {\bibinfo {author} {\bibfnamefont {S.}~\bibnamefont
  {Tsoi}}, \bibinfo {author} {\bibfnamefont {X.}~\bibnamefont {Lu}}, \bibinfo
  {author} {\bibfnamefont {A.~K.}\ \bibnamefont {Ramdas}}, \bibinfo {author}
  {\bibfnamefont {H.}~\bibnamefont {Alawadhi}}, \bibinfo {author}
  {\bibfnamefont {M.}~\bibnamefont {Grimsditch}}, \bibinfo {author}
  {\bibfnamefont {M.}~\bibnamefont {Cardona}}, \ and\ \bibinfo {author}
  {\bibfnamefont {R.}~\bibnamefont {Lauck}},\ }\bibfield  {title} {\enquote
  {\bibinfo {title} {{Isotopic-mass dependence of the A, B, and C excitonic
  band gaps in $\mathrm{ZnO}$ at low temperatures}},}\ }\href@noop {}
  {\bibfield  {journal} {\bibinfo  {journal} {Physical Review B}\ }\textbf
  {\bibinfo {volume} {74}},\ \bibinfo {pages} {165203} (\bibinfo {year}
  {2006})}\BibitemShut {NoStop}%
\bibitem [{\citenamefont {Shih}\ \emph {et~al.}(2010)\citenamefont {Shih},
  \citenamefont {Xue}, \citenamefont {Zhang}, \citenamefont {Cohen},\ and\
  \citenamefont {Louie}}]{Louie2010}%
  \BibitemOpen
  \bibfield  {author} {\bibinfo {author} {\bibfnamefont {B.-C.}\ \bibnamefont
  {Shih}}, \bibinfo {author} {\bibfnamefont {Y.}~\bibnamefont {Xue}}, \bibinfo
  {author} {\bibfnamefont {P.}~\bibnamefont {Zhang}}, \bibinfo {author}
  {\bibfnamefont {M.~L.}\ \bibnamefont {Cohen}}, \ and\ \bibinfo {author}
  {\bibfnamefont {S.~G.}\ \bibnamefont {Louie}},\ }\bibfield  {title} {\enquote
  {\bibinfo {title} {{Quasiparticle Band Gap of ZnO: High Accuracy from the
  Conventional G$^0$W$^0$ Approach}},}\ }\href@noop {} {\bibfield  {journal}
  {\bibinfo  {journal} {Physical Review Letters}\ }\textbf {\bibinfo {volume}
  {105}},\ \bibinfo {pages} {146401} (\bibinfo {year} {2010})}\BibitemShut
  {NoStop}%
\bibitem [{\citenamefont {Fermi}\ and\ \citenamefont
  {Teller}(1947)}]{Fermi1947}%
  \BibitemOpen
  \bibfield  {author} {\bibinfo {author} {\bibfnamefont {E.}~\bibnamefont
  {Fermi}}\ and\ \bibinfo {author} {\bibfnamefont {E.}~\bibnamefont {Teller}},\
  }\bibfield  {title} {\enquote {\bibinfo {title} {{The Capture of Negative
  Mesotrons in Matter}},}\ }\href {\doibase 10.1103/PhysRev.72.399} {\bibfield
  {journal} {\bibinfo  {journal} {Physical Review}\ }\textbf {\bibinfo {volume}
  {72}},\ \bibinfo {pages} {399--408} (\bibinfo {year} {1947})}\BibitemShut
  {NoStop}%
\bibitem [{\citenamefont {Wallis}, \citenamefont {Herman},\ and\ \citenamefont
  {Milnes}(1960)}]{Wallis1960}%
  \BibitemOpen
  \bibfield  {author} {\bibinfo {author} {\bibfnamefont {R.~F.}\ \bibnamefont
  {Wallis}}, \bibinfo {author} {\bibfnamefont {R.}~\bibnamefont {Herman}}, \
  and\ \bibinfo {author} {\bibfnamefont {H.~W.}\ \bibnamefont {Milnes}},\
  }\bibfield  {title} {\enquote {\bibinfo {title} {{Energy levels of an
  electron in the field of a finite dipole}},}\ }\href {\doibase
  https://doi.org/10.1016/0022-2852(60)90065-5} {\bibfield  {journal} {\bibinfo
   {journal} {Journal of Molecular Spectroscopy}\ }\textbf {\bibinfo {volume}
  {4}},\ \bibinfo {pages} {51--74} (\bibinfo {year} {1960})}\BibitemShut
  {NoStop}%
\bibitem [{\citenamefont {Crawford}(1967)}]{Crawford1967}%
  \BibitemOpen
  \bibfield  {author} {\bibinfo {author} {\bibfnamefont {O.~H.}\ \bibnamefont
  {Crawford}},\ }\bibfield  {title} {\enquote {\bibinfo {title} {{Bound states
  of a charged particle in a dipole field}},}\ }\href {\doibase
  10.1088/0370-1328/91/2/303} {\bibfield  {journal} {\bibinfo  {journal}
  {Proceedings of the Physical Society}\ }\textbf {\bibinfo {volume} {91}},\
  \bibinfo {pages} {279--284} (\bibinfo {year} {1967})}\BibitemShut {NoStop}%
\bibitem [{\citenamefont {Turner}, \citenamefont {Anderson},\ and\
  \citenamefont {Fox}(1968)}]{Turner1968}%
  \BibitemOpen
  \bibfield  {author} {\bibinfo {author} {\bibfnamefont {J.~E.}\ \bibnamefont
  {Turner}}, \bibinfo {author} {\bibfnamefont {V.~E.}\ \bibnamefont
  {Anderson}}, \ and\ \bibinfo {author} {\bibfnamefont {K.}~\bibnamefont
  {Fox}},\ }\bibfield  {title} {\enquote {\bibinfo {title} {{Ground-State
  Energy Eigenvalues and Eigenfunctions for an Electron in an Electric-Dipole
  Field}},}\ }\href {\doibase 10.1103/PhysRev.174.81} {\bibfield  {journal}
  {\bibinfo  {journal} {Physical Review}\ }\textbf {\bibinfo {volume} {174}},\
  \bibinfo {pages} {81--89} (\bibinfo {year} {1968})}\BibitemShut {NoStop}%
\bibitem [{\citenamefont {Simons}\ and\ \citenamefont
  {Jordan}(1987)}]{Simons1987}%
  \BibitemOpen
  \bibfield  {author} {\bibinfo {author} {\bibfnamefont {J.}~\bibnamefont
  {Simons}}\ and\ \bibinfo {author} {\bibfnamefont {K.~D.}\ \bibnamefont
  {Jordan}},\ }\bibfield  {title} {\enquote {\bibinfo {title} {{Ab initio
  electronic structure of anions}},}\ }\href {\doibase 10.1021/cr00079a004}
  {\bibfield  {journal} {\bibinfo  {journal} {Chemical Reviews}\ }\textbf
  {\bibinfo {volume} {87}},\ \bibinfo {pages} {535--555} (\bibinfo {year}
  {1987})},\ \Eprint {http://arxiv.org/abs/https://doi.org/10.1021/cr00079a004}
  {https://doi.org/10.1021/cr00079a004} \BibitemShut {NoStop}%
\bibitem [{\citenamefont {Jordan}\ and\ \citenamefont
  {Wang}(2003)}]{Jordan2003}%
  \BibitemOpen
  \bibfield  {author} {\bibinfo {author} {\bibfnamefont {K.~D.}\ \bibnamefont
  {Jordan}}\ and\ \bibinfo {author} {\bibfnamefont {F.}~\bibnamefont {Wang}},\
  }\bibfield  {title} {\enquote {\bibinfo {title} {{Theory of Dipole-Bound
  Anions}},}\ }\href {\doibase 10.1146/annurev.physchem.54.011002.103851}
  {\bibfield  {journal} {\bibinfo  {journal} {Annual Review of Physical
  Chemistry}\ }\textbf {\bibinfo {volume} {54}},\ \bibinfo {pages} {367--396}
  (\bibinfo {year} {2003})},\ \Eprint
  {http://arxiv.org/abs/https://doi.org/10.1146/annurev.physchem.54.011002.103851}
  {https://doi.org/10.1146/annurev.physchem.54.011002.103851} \BibitemShut
  {NoStop}%
\bibitem [{\citenamefont {Xu}\ and\ \citenamefont {Jordan}(2010)}]{Xu2010}%
  \BibitemOpen
  \bibfield  {author} {\bibinfo {author} {\bibfnamefont {J.}~\bibnamefont
  {Xu}}\ and\ \bibinfo {author} {\bibfnamefont {K.~D.}\ \bibnamefont
  {Jordan}},\ }\bibfield  {title} {\enquote {\bibinfo {title} {{Application of
  the Diffusion Monte Carlo Method to the Binding of Excess Electrons to Water
  Clusters}},}\ }\href {\doibase 10.1021/jp9066108} {\bibfield  {journal}
  {\bibinfo  {journal} {The Journal of Physical Chemistry A}\ }\textbf
  {\bibinfo {volume} {114}},\ \bibinfo {pages} {1364--1366} (\bibinfo {year}
  {2010})},\ \bibinfo {note} {pMID: 19788288},\ \Eprint
  {http://arxiv.org/abs/https://doi.org/10.1021/jp9066108}
  {https://doi.org/10.1021/jp9066108} \BibitemShut {NoStop}%
\bibitem [{\citenamefont {Hao}\ \emph {et~al.}(2018)\citenamefont {Hao},
  \citenamefont {Shee}, \citenamefont {Upadhyay}, \citenamefont {Ataca},
  \citenamefont {Jordan},\ and\ \citenamefont {Rubenstein}}]{Hao2018}%
  \BibitemOpen
  \bibfield  {author} {\bibinfo {author} {\bibfnamefont {H.}~\bibnamefont
  {Hao}}, \bibinfo {author} {\bibfnamefont {J.}~\bibnamefont {Shee}}, \bibinfo
  {author} {\bibfnamefont {S.}~\bibnamefont {Upadhyay}}, \bibinfo {author}
  {\bibfnamefont {C.}~\bibnamefont {Ataca}}, \bibinfo {author} {\bibfnamefont
  {K.~D.}\ \bibnamefont {Jordan}}, \ and\ \bibinfo {author} {\bibfnamefont
  {B.~M.}\ \bibnamefont {Rubenstein}},\ }\bibfield  {title} {\enquote {\bibinfo
  {title} {{Accurate Predictions of Electron Binding Energies of Dipole-Bound
  Anions via Quantum Monte Carlo Methods}},}\ }\href {\doibase
  10.1021/acs.jpclett.8b02733} {\bibfield  {journal} {\bibinfo  {journal} {The
  Journal of Physical Chemistry Letters}\ }\textbf {\bibinfo {volume} {9}},\
  \bibinfo {pages} {6185--6190} (\bibinfo {year} {2018})},\ \Eprint
  {http://arxiv.org/abs/https://doi.org/10.1021/acs.jpclett.8b02733}
  {https://doi.org/10.1021/acs.jpclett.8b02733} \BibitemShut {NoStop}%
\bibitem [{\citenamefont {Voora}\ \emph {et~al.}(2017)\citenamefont {Voora},
  \citenamefont {Kairalapova}, \citenamefont {Sommerfeld},\ and\ \citenamefont
  {Jordan}}]{Voora2017}%
  \BibitemOpen
  \bibfield  {author} {\bibinfo {author} {\bibfnamefont {V.~K.}\ \bibnamefont
  {Voora}}, \bibinfo {author} {\bibfnamefont {A.}~\bibnamefont {Kairalapova}},
  \bibinfo {author} {\bibfnamefont {T.}~\bibnamefont {Sommerfeld}}, \ and\
  \bibinfo {author} {\bibfnamefont {K.~D.}\ \bibnamefont {Jordan}},\ }\bibfield
   {title} {\enquote {\bibinfo {title} {Theoretical approaches for treating
  non-valence correlation-bound anions},}\ }\href {\doibase 10.1063/1.4991497}
  {\bibfield  {journal} {\bibinfo  {journal} {The Journal of Chemical Physics}\
  }\textbf {\bibinfo {volume} {147}},\ \bibinfo {pages} {214114} (\bibinfo
  {year} {2017})}\BibitemShut {NoStop}%
\bibitem [{\citenamefont {Raghavachari}\ \emph {et~al.}(1989)\citenamefont
  {Raghavachari}, \citenamefont {Trucks}, \citenamefont {Pople},\ and\
  \citenamefont {Head-Gordon}}]{Raghavachari1989}%
  \BibitemOpen
  \bibfield  {author} {\bibinfo {author} {\bibfnamefont {K.}~\bibnamefont
  {Raghavachari}}, \bibinfo {author} {\bibfnamefont {G.~W.}\ \bibnamefont
  {Trucks}}, \bibinfo {author} {\bibfnamefont {J.~A.}\ \bibnamefont {Pople}}, \
  and\ \bibinfo {author} {\bibfnamefont {M.}~\bibnamefont {Head-Gordon}},\
  }\bibfield  {title} {\enquote {\bibinfo {title} {{A fifth-order perturbation
  comparison of electron correlation theories}},}\ }\href {\doibase
  https://doi.org/10.1016/S0009-2614(89)87395-6} {\bibfield  {journal}
  {\bibinfo  {journal} {Chemical Physics Letters}\ }\textbf {\bibinfo {volume}
  {157}},\ \bibinfo {pages} {479--483} (\bibinfo {year} {1989})}\BibitemShut
  {NoStop}%
\bibitem [{\citenamefont {Stanton}\ and\ \citenamefont
  {Bartlett}(1993)}]{EOMCC1}%
  \BibitemOpen
  \bibfield  {author} {\bibinfo {author} {\bibfnamefont {J.~F.}\ \bibnamefont
  {Stanton}}\ and\ \bibinfo {author} {\bibfnamefont {R.~J.}\ \bibnamefont
  {Bartlett}},\ }\bibfield  {title} {\enquote {\bibinfo {title} {The equation
  of motion coupled‐cluster method. a systematic biorthogonal approach to
  molecular excitation energies, transition probabilities, and excited state
  properties},}\ }\href {\doibase 10.1063/1.464746} {\bibfield  {journal}
  {\bibinfo  {journal} {The Journal of Chemical Physics}\ }\textbf {\bibinfo
  {volume} {98}},\ \bibinfo {pages} {7029--7039} (\bibinfo {year}
  {1993})}\BibitemShut {NoStop}%
\bibitem [{\citenamefont {Stanton}\ and\ \citenamefont {Gauss}(1994)}]{EOMCC2}%
  \BibitemOpen
  \bibfield  {author} {\bibinfo {author} {\bibfnamefont {J.~F.}\ \bibnamefont
  {Stanton}}\ and\ \bibinfo {author} {\bibfnamefont {J.}~\bibnamefont
  {Gauss}},\ }\bibfield  {title} {\enquote {\bibinfo {title} {Analytic energy
  derivatives for ionized states described by the equation‐of‐motion
  coupled cluster method},}\ }\href {\doibase 10.1063/1.468022} {\bibfield
  {journal} {\bibinfo  {journal} {The Journal of Chemical Physics}\ }\textbf
  {\bibinfo {volume} {101}},\ \bibinfo {pages} {8938--8944} (\bibinfo {year}
  {1994})}\BibitemShut {NoStop}%
\bibitem [{\citenamefont {Nooijen}\ and\ \citenamefont
  {Bartlett}(1995)}]{EOMCC3}%
  \BibitemOpen
  \bibfield  {author} {\bibinfo {author} {\bibfnamefont {M.}~\bibnamefont
  {Nooijen}}\ and\ \bibinfo {author} {\bibfnamefont {R.~J.}\ \bibnamefont
  {Bartlett}},\ }\bibfield  {title} {\enquote {\bibinfo {title} {Equation of
  motion coupled cluster method for electron attachment},}\ }\href {\doibase
  10.1063/1.468592} {\bibfield  {journal} {\bibinfo  {journal} {The Journal of
  Chemical Physics}\ }\textbf {\bibinfo {volume} {102}},\ \bibinfo {pages}
  {3629--3647} (\bibinfo {year} {1995})}\BibitemShut {NoStop}%
\bibitem [{\citenamefont {Kucharski}\ \emph {et~al.}(2001)\citenamefont
  {Kucharski}, \citenamefont {W{\l}och}, \citenamefont {Musia{\l}},\ and\
  \citenamefont {Bartlett}}]{EOMCCSDT}%
  \BibitemOpen
  \bibfield  {author} {\bibinfo {author} {\bibfnamefont {S.~A.}\ \bibnamefont
  {Kucharski}}, \bibinfo {author} {\bibfnamefont {M.}~\bibnamefont {W{\l}och}},
  \bibinfo {author} {\bibfnamefont {M.}~\bibnamefont {Musia{\l}}}, \ and\
  \bibinfo {author} {\bibfnamefont {R.~J.}\ \bibnamefont {Bartlett}},\
  }\bibfield  {title} {\enquote {\bibinfo {title} {Coupled-cluster theory for
  excited electronic states: The full equation-of-motion coupled-cluster
  single, double, and triple excitation method},}\ }\href {\doibase
  10.1063/1.1416173} {\bibfield  {journal} {\bibinfo  {journal} {The Journal of
  Chemical Physics}\ }\textbf {\bibinfo {volume} {115}},\ \bibinfo {pages}
  {8263--8266} (\bibinfo {year} {2001})}\BibitemShut {NoStop}%
\bibitem [{\citenamefont {Dunning}(1989)}]{accd1}%
  \BibitemOpen
  \bibfield  {author} {\bibinfo {author} {\bibfnamefont {T.~H.}\ \bibnamefont
  {Dunning}},\ }\bibfield  {title} {\enquote {\bibinfo {title} {Gaussian basis
  sets for use in correlated molecular calculations. i. the atoms boron through
  neon and hydrogen},}\ }\href {\doibase 10.1063/1.456153} {\bibfield
  {journal} {\bibinfo  {journal} {The Journal of Chemical Physics}\ }\textbf
  {\bibinfo {volume} {90}},\ \bibinfo {pages} {1007--1023} (\bibinfo {year}
  {1989})}\BibitemShut {NoStop}%
\bibitem [{\citenamefont {Kendall}, \citenamefont {Dunning},\ and\
  \citenamefont {Harrison}(1992)}]{accd2}%
  \BibitemOpen
  \bibfield  {author} {\bibinfo {author} {\bibfnamefont {R.~A.}\ \bibnamefont
  {Kendall}}, \bibinfo {author} {\bibfnamefont {T.~H.}\ \bibnamefont
  {Dunning}}, \ and\ \bibinfo {author} {\bibfnamefont {R.~J.}\ \bibnamefont
  {Harrison}},\ }\bibfield  {title} {\enquote {\bibinfo {title} {Electron
  affinities of the first‐row atoms revisited. systematic basis sets and wave
  functions},}\ }\href {\doibase 10.1063/1.462569} {\bibfield  {journal}
  {\bibinfo  {journal} {The Journal of Chemical Physics}\ }\textbf {\bibinfo
  {volume} {96}},\ \bibinfo {pages} {6796--6806} (\bibinfo {year}
  {1992})}\BibitemShut {NoStop}%
\bibitem [{\citenamefont {Becke}(1993)}]{B3}%
  \BibitemOpen
  \bibfield  {author} {\bibinfo {author} {\bibfnamefont {A.~D.}\ \bibnamefont
  {Becke}},\ }\bibfield  {title} {\enquote {\bibinfo {title}
  {Density‐functional thermochemistry. iii. the role of exact exchange},}\
  }\href {\doibase 10.1063/1.464913} {\bibfield  {journal} {\bibinfo  {journal}
  {The Journal of Chemical Physics}\ }\textbf {\bibinfo {volume} {98}},\
  \bibinfo {pages} {5648--5652} (\bibinfo {year} {1993})}\BibitemShut {NoStop}%
\bibitem [{\citenamefont {Lee}, \citenamefont {Yang},\ and\ \citenamefont
  {Parr}(1988)}]{LYP}%
  \BibitemOpen
  \bibfield  {author} {\bibinfo {author} {\bibfnamefont {C.}~\bibnamefont
  {Lee}}, \bibinfo {author} {\bibfnamefont {W.}~\bibnamefont {Yang}}, \ and\
  \bibinfo {author} {\bibfnamefont {R.~G.}\ \bibnamefont {Parr}},\ }\bibfield
  {title} {\enquote {\bibinfo {title} {Development of the colle-salvetti
  correlation-energy formula into a functional of the electron density},}\
  }\href {\doibase 10.1103/PhysRevB.37.785} {\bibfield  {journal} {\bibinfo
  {journal} {Phys. Rev. B}\ }\textbf {\bibinfo {volume} {37}},\ \bibinfo
  {pages} {785--789} (\bibinfo {year} {1988})}\BibitemShut {NoStop}%
\bibitem [{\citenamefont {Vosko}, \citenamefont {Wilk},\ and\ \citenamefont
  {Nusair}(1980)}]{VWN}%
  \BibitemOpen
  \bibfield  {author} {\bibinfo {author} {\bibfnamefont {S.~H.}\ \bibnamefont
  {Vosko}}, \bibinfo {author} {\bibfnamefont {L.}~\bibnamefont {Wilk}}, \ and\
  \bibinfo {author} {\bibfnamefont {M.}~\bibnamefont {Nusair}},\ }\bibfield
  {title} {\enquote {\bibinfo {title} {Accurate spin-dependent electron liquid
  correlation energies for local spin density calculations: a critical
  analysis},}\ }\href {\doibase 10.1139/p80-159} {\bibfield  {journal}
  {\bibinfo  {journal} {Canadian Journal of Physics}\ }\textbf {\bibinfo
  {volume} {58}},\ \bibinfo {pages} {1200--1211} (\bibinfo {year}
  {1980})}\BibitemShut {NoStop}%
\bibitem [{\citenamefont {Brik}\ and\ \citenamefont
  {Srivastava}(2013)}]{Brik2013}%
  \BibitemOpen
  \bibfield  {author} {\bibinfo {author} {\bibfnamefont {M.~G.}\ \bibnamefont
  {Brik}}\ and\ \bibinfo {author} {\bibfnamefont {A.~M.}\ \bibnamefont
  {Srivastava}},\ }\bibfield  {title} {\enquote {\bibinfo {title} {{On the
  Optical Properties of the Mn$^{4+}$ Ion in Solids}},}\ }\href {\doibase
  10.1016/j.jlumin.2011.08.047} {\bibfield  {journal} {\bibinfo  {journal}
  {Journal of Luminescence}\ }\textbf {\bibinfo {volume} {133}},\ \bibinfo
  {pages} {69--72} (\bibinfo {year} {2013})}\BibitemShut {NoStop}%
\bibitem [{\citenamefont {Geschwind}\ \emph {et~al.}(1962)\citenamefont
  {Geschwind}, \citenamefont {Kisliuk}, \citenamefont {Klein}, \citenamefont
  {Remeika},\ and\ \citenamefont {Wood}}]{Geschwind1962}%
  \BibitemOpen
  \bibfield  {author} {\bibinfo {author} {\bibfnamefont {S.}~\bibnamefont
  {Geschwind}}, \bibinfo {author} {\bibfnamefont {P.}~\bibnamefont {Kisliuk}},
  \bibinfo {author} {\bibfnamefont {M.~P.}\ \bibnamefont {Klein}}, \bibinfo
  {author} {\bibfnamefont {J.~P.}\ \bibnamefont {Remeika}}, \ and\ \bibinfo
  {author} {\bibfnamefont {D.~L.}\ \bibnamefont {Wood}},\ }\bibfield  {title}
  {\enquote {\bibinfo {title} {{Sharp-Line Fluorescence, Electron Paramagnetic
  Resonance, and Thermoluminescence of Mn$^{4+}$ in -Al$_2$O$_3$}},}\ }\href
  {\doibase 10.1103/PhysRev.126.1684} {\bibfield  {journal} {\bibinfo
  {journal} {Physical Review}\ }\textbf {\bibinfo {volume} {126}},\ \bibinfo
  {pages} {1684--1686} (\bibinfo {year} {1962})}\BibitemShut {NoStop}%
\bibitem [{\citenamefont {Dunphy}\ and\ \citenamefont
  {Duley}(1990)}]{Dunphy1990}%
  \BibitemOpen
  \bibfield  {author} {\bibinfo {author} {\bibfnamefont {K.}~\bibnamefont
  {Dunphy}}\ and\ \bibinfo {author} {\bibfnamefont {W.~W.}\ \bibnamefont
  {Duley}},\ }\bibfield  {title} {\enquote {\bibinfo {title} {{Multiphoton
  Excitation of Mn$^{4+}$ and Cr$^{3+}$ Luminescence in MgO}},}\ }\href@noop {}
  {\bibfield  {journal} {\bibinfo  {journal} {Journal of Physics and Chemistry
  of Solids}\ }\textbf {\bibinfo {volume} {51}},\ \bibinfo {pages} {1077--1082}
  (\bibinfo {year} {1990})}\BibitemShut {NoStop}%
\bibitem [{\citenamefont {Brik}\ and\ \citenamefont
  {Srivastava}(2012)}]{Brik2012a}%
  \BibitemOpen
  \bibfield  {author} {\bibinfo {author} {\bibfnamefont {M.~G.}\ \bibnamefont
  {Brik}}\ and\ \bibinfo {author} {\bibfnamefont {A.~M.}\ \bibnamefont
  {Srivastava}},\ }\bibfield  {title} {\enquote {\bibinfo {title} {{Ab Initio
  Studies of the Structural, Electronic, and Optical Properties of K$_2$SiF$_6$
  Single Crystals at Ambient and Elevated Hydrostatic Pressure}},}\ }\href
  {\doibase 10.1149/2.071206jes} {\bibfield  {journal} {\bibinfo  {journal}
  {Journal of Electrochemical Society}\ }\textbf {\bibinfo {volume} {159}},\
  \bibinfo {pages} {J212--J216} (\bibinfo {year} {2012})}\BibitemShut {NoStop}%
\bibitem [{\citenamefont {Medvedev}\ \emph {et~al.}(2017)\citenamefont
  {Medvedev}, \citenamefont {Bushmarinov}, \citenamefont {Sun}, \citenamefont
  {Perdew},\ and\ \citenamefont {Lyssenko}}]{Medvedev2017}%
  \BibitemOpen
  \bibfield  {author} {\bibinfo {author} {\bibfnamefont {M.~G.}\ \bibnamefont
  {Medvedev}}, \bibinfo {author} {\bibfnamefont {I.~S.}\ \bibnamefont
  {Bushmarinov}}, \bibinfo {author} {\bibfnamefont {J.}~\bibnamefont {Sun}},
  \bibinfo {author} {\bibfnamefont {J.~P.}\ \bibnamefont {Perdew}}, \ and\
  \bibinfo {author} {\bibfnamefont {K.~A.}\ \bibnamefont {Lyssenko}},\
  }\bibfield  {title} {\enquote {\bibinfo {title} {{Density Functional Theory
  Is Straying from the Path Toward the Exact Functional}},}\ }\href {\doibase
  10.1126/science.aah5975} {\bibfield  {journal} {\bibinfo  {journal}
  {Science}\ }\textbf {\bibinfo {volume} {355}},\ \bibinfo {pages} {49--52}
  (\bibinfo {year} {2017})}\BibitemShut {NoStop}%
\bibitem [{\citenamefont {Lindner}(1977)}]{Lindner1977}%
  \BibitemOpen
  \bibfield  {author} {\bibinfo {author} {\bibfnamefont {P.}~\bibnamefont
  {Lindner}},\ }\bibfield  {title} {\enquote {\bibinfo {title} {{Theoretical
  Momentum Densities}},}\ }\href {\doibase 10.1088/0031-8949/15/2/004}
  {\bibfield  {journal} {\bibinfo  {journal} {Physica Scripta}\ }\textbf
  {\bibinfo {volume} {15}},\ \bibinfo {pages} {112--118} (\bibinfo {year}
  {1977})}\BibitemShut {NoStop}%
\bibitem [{\citenamefont {Williams}(1977)}]{Williams1977}%
  \BibitemOpen
  \bibfield  {author} {\bibinfo {author} {\bibfnamefont {B.~G.}\ \bibnamefont
  {Williams}},\ }\bibfield  {title} {\enquote {\bibinfo {title} {{The
  Experimental Determination of Electron Momentum Densities}},}\ }\href
  {\doibase 10.1088/0031-8949/15/2/003} {\bibfield  {journal} {\bibinfo
  {journal} {Physica Scripta}\ }\textbf {\bibinfo {volume} {15}},\ \bibinfo
  {pages} {92--111} (\bibinfo {year} {1977})}\BibitemShut {NoStop}%
\bibitem [{\citenamefont {Cooper}(1985)}]{Cooper1985}%
  \BibitemOpen
  \bibfield  {author} {\bibinfo {author} {\bibfnamefont {M.~J.}\ \bibnamefont
  {Cooper}},\ }\bibfield  {title} {\enquote {\bibinfo {title} {{Compton
  scattering and electron momentum determination}},}\ }\href {\doibase
  10.1088/0034-4885/48/4/001} {\bibfield  {journal} {\bibinfo  {journal}
  {Reports on Progress in Physics}\ }\textbf {\bibinfo {volume} {48}},\
  \bibinfo {pages} {415--481} (\bibinfo {year} {1985})}\BibitemShut {NoStop}%
\bibitem [{\citenamefont {Bell}\ and\ \citenamefont
  {Schneider}(2001)}]{Bell2001}%
  \BibitemOpen
  \bibfield  {author} {\bibinfo {author} {\bibfnamefont {F.}~\bibnamefont
  {Bell}}\ and\ \bibinfo {author} {\bibfnamefont {J.~R.}\ \bibnamefont
  {Schneider}},\ }\bibfield  {title} {\enquote {\bibinfo {title}
  {{Three-dimensional electron momentum densities of solids}},}\ }\href
  {\doibase 10.1088/0953-8984/13/34/327} {\bibfield  {journal} {\bibinfo
  {journal} {Journal of Physics: Condensed Matter}\ }\textbf {\bibinfo {volume}
  {13}},\ \bibinfo {pages} {7905--7922} (\bibinfo {year} {2001})}\BibitemShut
  {NoStop}%
\bibitem [{\citenamefont {Dugdale}(2016)}]{Dugdale2016}%
  \BibitemOpen
  \bibfield  {author} {\bibinfo {author} {\bibfnamefont {S.~B.}\ \bibnamefont
  {Dugdale}},\ }\bibfield  {title} {\enquote {\bibinfo {title} {{Life on the
  edge: a beginner's guide to the Fermi surface}},}\ }\href {\doibase
  10.1088/0031-8949/91/5/053009} {\bibfield  {journal} {\bibinfo  {journal}
  {Physica Scripta}\ }\textbf {\bibinfo {volume} {91}},\ \bibinfo {pages}
  {053009} (\bibinfo {year} {2016})}\BibitemShut {NoStop}%
\bibitem [{\citenamefont {Lock}, \citenamefont {Crisp},\ and\ \citenamefont
  {West}(1973)}]{Lock1973}%
  \BibitemOpen
  \bibfield  {author} {\bibinfo {author} {\bibfnamefont {D.~G.}\ \bibnamefont
  {Lock}}, \bibinfo {author} {\bibfnamefont {V.~H.~C.}\ \bibnamefont {Crisp}},
  \ and\ \bibinfo {author} {\bibfnamefont {R.~N.}\ \bibnamefont {West}},\
  }\bibfield  {title} {\enquote {\bibinfo {title} {{Positron annihilation and
  Fermi surface studies: a new approach}},}\ }\href {\doibase
  10.1088/0305-4608/3/3/014} {\bibfield  {journal} {\bibinfo  {journal}
  {Journal of Physics F: Metal Physics}\ }\textbf {\bibinfo {volume} {3}},\
  \bibinfo {pages} {561--570} (\bibinfo {year} {1973})}\BibitemShut {NoStop}%
\bibitem [{\citenamefont {Dugdale}\ \emph {et~al.}(2006)\citenamefont
  {Dugdale}, \citenamefont {Watts}, \citenamefont {Laverock}, \citenamefont
  {Major}, \citenamefont {Alam}, \citenamefont {Samsel-Czeka{\l}a},
  \citenamefont {Kontrym-Sznajd}, \citenamefont {Sakurai}, \citenamefont
  {Itou},\ and\ \citenamefont {Fort}}]{Dugdale2006}%
  \BibitemOpen
  \bibfield  {author} {\bibinfo {author} {\bibfnamefont {S.~B.}\ \bibnamefont
  {Dugdale}}, \bibinfo {author} {\bibfnamefont {R.~J.}\ \bibnamefont {Watts}},
  \bibinfo {author} {\bibfnamefont {J.}~\bibnamefont {Laverock}}, \bibinfo
  {author} {\bibfnamefont {Z.}~\bibnamefont {Major}}, \bibinfo {author}
  {\bibfnamefont {M.~A.}\ \bibnamefont {Alam}}, \bibinfo {author}
  {\bibfnamefont {M.}~\bibnamefont {Samsel-Czeka{\l}a}}, \bibinfo {author}
  {\bibfnamefont {G.}~\bibnamefont {Kontrym-Sznajd}}, \bibinfo {author}
  {\bibfnamefont {Y.}~\bibnamefont {Sakurai}}, \bibinfo {author} {\bibfnamefont
  {M.}~\bibnamefont {Itou}}, \ and\ \bibinfo {author} {\bibfnamefont
  {D.}~\bibnamefont {Fort}},\ }\bibfield  {title} {\enquote {\bibinfo {title}
  {{Observation of a Strongly Nested Fermi Surface in the Shape-Memory
  {AlloyNi}$_{0.62}$Al$_{0.38}$}},}\ }\href {\doibase
  10.1103/physrevlett.96.046406} {\bibfield  {journal} {\bibinfo  {journal}
  {Physical Review Letters}\ }\textbf {\bibinfo {volume} {96}} (\bibinfo {year}
  {2006}),\ 10.1103/physrevlett.96.046406}\BibitemShut {NoStop}%
\bibitem [{\citenamefont {Huotari}\ \emph {et~al.}(2010)\citenamefont
  {Huotari}, \citenamefont {Soininen}, \citenamefont {Pylkk{\"{a}}nen},
  \citenamefont {H{\"{a}}m{\"{a}}l{\"{a}}inen}, \citenamefont {Issolah},
  \citenamefont {Titov}, \citenamefont {McMinis}, \citenamefont {Kim},
  \citenamefont {Esler}, \citenamefont {Ceperley}, \citenamefont {Holzmann},\
  and\ \citenamefont {Olevano}}]{Huotari2010}%
  \BibitemOpen
  \bibfield  {author} {\bibinfo {author} {\bibfnamefont {S.}~\bibnamefont
  {Huotari}}, \bibinfo {author} {\bibfnamefont {J.~A.}\ \bibnamefont
  {Soininen}}, \bibinfo {author} {\bibfnamefont {T.}~\bibnamefont
  {Pylkk{\"{a}}nen}}, \bibinfo {author} {\bibfnamefont {K.}~\bibnamefont
  {H{\"{a}}m{\"{a}}l{\"{a}}inen}}, \bibinfo {author} {\bibfnamefont
  {A.}~\bibnamefont {Issolah}}, \bibinfo {author} {\bibfnamefont
  {A.}~\bibnamefont {Titov}}, \bibinfo {author} {\bibfnamefont
  {J.}~\bibnamefont {McMinis}}, \bibinfo {author} {\bibfnamefont
  {J.}~\bibnamefont {Kim}}, \bibinfo {author} {\bibfnamefont {K.}~\bibnamefont
  {Esler}}, \bibinfo {author} {\bibfnamefont {D.~M.}\ \bibnamefont {Ceperley}},
  \bibinfo {author} {\bibfnamefont {M.}~\bibnamefont {Holzmann}}, \ and\
  \bibinfo {author} {\bibfnamefont {V.}~\bibnamefont {Olevano}},\ }\bibfield
  {title} {\enquote {\bibinfo {title} {{Momentum Distribution and
  Renormalization Factor in Sodium and the Electron Gas}},}\ }\href {\doibase
  10.1103/physrevlett.105.086403} {\bibfield  {journal} {\bibinfo  {journal}
  {Physical Review Letters}\ }\textbf {\bibinfo {volume} {105}},\ \bibinfo
  {pages} {0860403} (\bibinfo {year} {2010})}\BibitemShut {NoStop}%
\bibitem [{\citenamefont {Holzmann}\ \emph {et~al.}(2011)\citenamefont
  {Holzmann}, \citenamefont {Bernu}, \citenamefont {Pierleoni}, \citenamefont
  {McMinis}, \citenamefont {Ceperley}, \citenamefont {Olevano},\ and\
  \citenamefont {Site}}]{Holzmann2011}%
  \BibitemOpen
  \bibfield  {author} {\bibinfo {author} {\bibfnamefont {M.}~\bibnamefont
  {Holzmann}}, \bibinfo {author} {\bibfnamefont {B.}~\bibnamefont {Bernu}},
  \bibinfo {author} {\bibfnamefont {C.}~\bibnamefont {Pierleoni}}, \bibinfo
  {author} {\bibfnamefont {J.}~\bibnamefont {McMinis}}, \bibinfo {author}
  {\bibfnamefont {D.~M.}\ \bibnamefont {Ceperley}}, \bibinfo {author}
  {\bibfnamefont {V.}~\bibnamefont {Olevano}}, \ and\ \bibinfo {author}
  {\bibfnamefont {L.~D.}\ \bibnamefont {Site}},\ }\bibfield  {title} {\enquote
  {\bibinfo {title} {{Momentum Distribution of the Homogeneous Electron
  Gas}},}\ }\href {\doibase 10.1103/physrevlett.107.110402} {\bibfield
  {journal} {\bibinfo  {journal} {Physical Review Letters}\ }\textbf {\bibinfo
  {volume} {107}},\ \bibinfo {pages} {110402} (\bibinfo {year}
  {2011})}\BibitemShut {NoStop}%
\bibitem [{\citenamefont {Silver}\ and\ \citenamefont {Sokol}(1989)}]{MomDist}%
  \BibitemOpen
  \bibinfo {editor} {\bibfnamefont {R.~N.}\ \bibnamefont {Silver}}\ and\
  \bibinfo {editor} {\bibfnamefont {P.~E.}\ \bibnamefont {Sokol}},\ eds.,\
  \href@noop {} {\emph {\bibinfo {title} {{Momentum Distributions}}}}\
  (\bibinfo  {publisher} {Plenum Press, New York},\ \bibinfo {year}
  {1989})\BibitemShut {NoStop}%
\bibitem [{\citenamefont {Hiraoka}\ and\ \citenamefont
  {Nomura}(2017)}]{Hiraoka2017}%
  \BibitemOpen
  \bibfield  {author} {\bibinfo {author} {\bibfnamefont {N.}~\bibnamefont
  {Hiraoka}}\ and\ \bibinfo {author} {\bibfnamefont {T.}~\bibnamefont
  {Nomura}},\ }\bibfield  {title} {\enquote {\bibinfo {title} {{Electron
  momentum densities near Dirac cones: Anisotropic Umklapp scattering and
  momentum broadening}},}\ }\href {\doibase 10.1038/s41598-017-00628-4}
  {\bibfield  {journal} {\bibinfo  {journal} {Scientific Reports}\ }\textbf
  {\bibinfo {volume} {7}},\ \bibinfo {pages} {765} (\bibinfo {year}
  {2017})}\BibitemShut {NoStop}%
\bibitem [{\citenamefont {Kyl{\"{a}}np{\"{a}}{\"{a}}}\ \emph
  {et~al.}(2019)\citenamefont {Kyl{\"{a}}np{\"{a}}{\"{a}}}, \citenamefont
  {Luo}, \citenamefont {Heinonen}, \citenamefont {Kent},\ and\ \citenamefont
  {Krogel}}]{Kylaenpaeae2019}%
  \BibitemOpen
  \bibfield  {author} {\bibinfo {author} {\bibfnamefont {I.}~\bibnamefont
  {Kyl{\"{a}}np{\"{a}}{\"{a}}}}, \bibinfo {author} {\bibfnamefont
  {Y.}~\bibnamefont {Luo}}, \bibinfo {author} {\bibfnamefont {O.}~\bibnamefont
  {Heinonen}}, \bibinfo {author} {\bibfnamefont {P.~R.~C.}\ \bibnamefont
  {Kent}}, \ and\ \bibinfo {author} {\bibfnamefont {J.~T.}\ \bibnamefont
  {Krogel}},\ }\bibfield  {title} {\enquote {\bibinfo {title} {{Compton profile
  of {VO}$_2$ across the metal-insulator transition: Evidence of a non-Fermi
  liquid metal}},}\ }\href {\doibase 10.1103/physrevb.99.075154} {\bibfield
  {journal} {\bibinfo  {journal} {Physical Review B}\ }\textbf {\bibinfo
  {volume} {99}},\ \bibinfo {pages} {075154} (\bibinfo {year}
  {2019})}\BibitemShut {NoStop}%
\bibitem [{\citenamefont {Zambelli}\ \emph {et~al.}(2000)\citenamefont
  {Zambelli}, \citenamefont {Pitaevskii}, \citenamefont {Stamper-Kurn},\ and\
  \citenamefont {Stringari}}]{Zambelli2000}%
  \BibitemOpen
  \bibfield  {author} {\bibinfo {author} {\bibfnamefont {F.}~\bibnamefont
  {Zambelli}}, \bibinfo {author} {\bibfnamefont {L.}~\bibnamefont
  {Pitaevskii}}, \bibinfo {author} {\bibfnamefont {D.~M.}\ \bibnamefont
  {Stamper-Kurn}}, \ and\ \bibinfo {author} {\bibfnamefont {S.}~\bibnamefont
  {Stringari}},\ }\bibfield  {title} {\enquote {\bibinfo {title} {{Dynamic
  structure factor and momentum distribution of a trapped Bose gas}},}\ }\href
  {\doibase 10.1103/physreva.61.063608} {\bibfield  {journal} {\bibinfo
  {journal} {Physical Review A}\ }\textbf {\bibinfo {volume} {61}},\ \bibinfo
  {pages} {063608} (\bibinfo {year} {2000})}\BibitemShut {NoStop}%
\bibitem [{\citenamefont {Lee}\ \emph {et~al.}(2017{\natexlab{b}})\citenamefont
  {Lee}, \citenamefont {Hippalgaonkar}, \citenamefont {Yang}, \citenamefont
  {Hong}, \citenamefont {Ko}, \citenamefont {Suh}, \citenamefont {Liu},
  \citenamefont {Wang}, \citenamefont {Urban}, \citenamefont {Zhang},
  \citenamefont {Dames}, \citenamefont {Hartnoll}, \citenamefont {Delaire},\
  and\ \citenamefont {Wu}}]{Lee2017}%
  \BibitemOpen
  \bibfield  {author} {\bibinfo {author} {\bibfnamefont {S.}~\bibnamefont
  {Lee}}, \bibinfo {author} {\bibfnamefont {K.}~\bibnamefont {Hippalgaonkar}},
  \bibinfo {author} {\bibfnamefont {F.}~\bibnamefont {Yang}}, \bibinfo {author}
  {\bibfnamefont {J.}~\bibnamefont {Hong}}, \bibinfo {author} {\bibfnamefont
  {C.}~\bibnamefont {Ko}}, \bibinfo {author} {\bibfnamefont {J.}~\bibnamefont
  {Suh}}, \bibinfo {author} {\bibfnamefont {K.}~\bibnamefont {Liu}}, \bibinfo
  {author} {\bibfnamefont {K.}~\bibnamefont {Wang}}, \bibinfo {author}
  {\bibfnamefont {J.~J.}\ \bibnamefont {Urban}}, \bibinfo {author}
  {\bibfnamefont {X.}~\bibnamefont {Zhang}}, \bibinfo {author} {\bibfnamefont
  {C.}~\bibnamefont {Dames}}, \bibinfo {author} {\bibfnamefont {S.~A.}\
  \bibnamefont {Hartnoll}}, \bibinfo {author} {\bibfnamefont {O.}~\bibnamefont
  {Delaire}}, \ and\ \bibinfo {author} {\bibfnamefont {J.}~\bibnamefont {Wu}},\
  }\bibfield  {title} {\enquote {\bibinfo {title} {{Anomalously low electronic
  thermal conductivity in metallic vanadium dioxide}},}\ }\href {\doibase
  10.1126/science.aag0410} {\bibfield  {journal} {\bibinfo  {journal}
  {Science}\ }\textbf {\bibinfo {volume} {355}},\ \bibinfo {pages} {371--374}
  (\bibinfo {year} {2017}{\natexlab{b}})}\BibitemShut {NoStop}%
\end{thebibliography}

\end{document}